\documentclass[fleqn,usenatbib]{mnras}

\usepackage{newtxtext,newtxmath}

\usepackage[T1]{fontenc}
\usepackage{ae,aecompl}
\usepackage{multirow}
\usepackage{tabularx}
\usepackage{supertabular}
\usepackage{longtable}
\usepackage[]{url}
\usepackage{graphicx}	
\usepackage{amsmath}	
\usepackage{amssymb}	

\usepackage{graphicx}	
\usepackage{amsmath}	
\usepackage{amssymb}	
\usepackage{multirow}
\usepackage[left]{lineno}
\usepackage{booktabs}
\usepackage{gensymb}
\usepackage{xcolor,color}
\usepackage[caption = false]{subfig}

\title[Observations of O3 alerts by GRANDMA]{GRANDMA Observations of Advanced LIGO's and Advanced Virgo's Third Observational Campaign}

\author[GRANDMA collaboration]{
S. Antier$^{1}$,
S. Agayeva$^{2}$, 
M. Almualla$^{3}$,
S. Awiphan$^{4}$, 
A. Baransky$^{5,6}$,
K. Barynova$^{5,6}$,
\newauthor
S. Beradze$^{7,8}$,
M. Bla\v{z}ek$^{9}$, 
M. Bo\"er$^{10}$,
O. Burkhonov$^{11}$,
N. Christensen$^{10}$,
A. Coleiro$^{1}$,
\newauthor
D. Corre$^{12}$,
M. W. Coughlin$^{13}$,
H. Crisp$^{14}$,
T. Dietrich$^{15}$, 
J.-G. Ducoin$^{12}$,
\newauthor
P.-A. Duverne$^{12}$,
G. Marchal-Duval$^{12}$,
B. Gendre$^{14}$,
P. Gokuldass$^{16}$
H. B. Eggenstein$^{17}$,
\newauthor
L. Eymar$^{10}$,
P. Hello$^{12}$,
E. J. Howell$^{14}$,
N. Ismailov$^{2}$,
D. A. Kann$^{9}$,
S. Karpov$^{18}$,
\newauthor
A. Klotz$^{19, 20}$,
N. Kochiashvili$^{7}$,
C. Lachaud$^{1}$,
N. Leroy$^{12}$,
W. L. Lin$^{21}$,
W. X. Li$^{21}$,
\newauthor
M. Ma\v{s}ek$^{18}$,
J. Mo$^{21}$,
R. Menard$^{1}$,
D. Morris$^{16}$,
K. Noysena$^{19, 20}$,
N. B. Orange$^{22}$,
\newauthor
M. Prouza$^{18}$,
R. Rattanamala$^{23}$,
T. Sadibekova$^{11, 24}$,
D. Saint-Gelais$^{1}$
M. Serrau$^{1}$,
\newauthor
A. Simon$^{25}$,
C. Stachie$^{10}$,
C. C. Th\"one$^{9}$, 
Y. Tillayev$^{11,26}$,
D. Turpin$^{24}$,
\newauthor
A. de Ugarte Postigo$^{9}$,
V. Vasylenko$^{25}$,
Z. Vidadi$^{2}$,
M. Was$^{27}$,
X. F. Wang$^{21}$,
\newauthor
J. J. Zhang$^{28}$
T. M. Zhang$^{29}$
X. H. Zhang$^{21}$
}
\date{Accepted XXX. Received YYY; in original form ZZZ}
\pubyear{2020}

\begin{document}
\label{firstpage}
\pagerange{\pageref{firstpage}--\pageref{lastpage}}
\maketitle

\begin{abstract}
 GRANDMA is a network of 25 telescopes of different sizes, including both photometric and spectroscopic facilities. The network aims to coordinate follow-up observations of gravitational-wave candidate alerts, especially those with large localisation uncertainties, to reduce the delay between the initial detection and the optical confirmation. In this paper, we detail GRANDMA's observational performance during Advanced LIGO/Advanced Virgo Observing Run 3 (O3), focusing on the second part of O3; this includes summary statistics pertaining to coverage and possible astrophysical origin of the candidates. To do so, we quantify our observation efficiency in terms of delay between gravitational-wave candidate trigger time, observations, and the total coverage. Using an optimised and robust coordination system, GRANDMA followed-up about 90\% of the gravitational-wave candidate alerts, i.e. 49 out of 56 candidates. This led to coverage of over 9000 deg$^2$ during O3. The delay between the gravitational-wave candidate trigger and the first observation was below 1.5 hour for 50\% of the alerts. We did not detect any electromagnetic counterparts to the gravitational-wave candidates during O3, likely due to the very large localisation areas (on average thousands of degrees squares) and relatively large distance of the candidates (above 200 Mpc for 60\% of BNS candidates). We derive constraints on potential kilonova properties for two potential binary neutron star coalescences (GW190425 and S200213t), assuming that the events' locations were imaged.
\end{abstract}




\section{Introduction}

Gravitational-wave (GW) signals detected by Advanced LIGO \citep{aLIGO} and Advanced Virgo \citep{adVirgo} since 2015 have revealed the existence of compact binary mergers which included neutrons stars (NSs) \citep{LSC_BNS_2017PhRvL,AbEA2019} and black holes (BHs) \citep{2019PhRvX...9c1040A}, providing unprecedented new insights into these objects. 

During the first and second Advanced LIGO and Advanced Virgo observing runs, O1 and O2 (2015-2017, Advanced Virgo joined at the end of O2), GWs from ten binary BH (BBH) systems and a binary NS system (BNS) were detected \citep{2019PhRvX...9c1040A,LSC_BNS_2017PhRvL}. Additional events produced by BBHs were also claimed by other groups analyzing the LIGO/Virgo data \citep{2019PhRvD.100b3007Z}. The GW observations of these compact binary sources have allowed gravity to be probed in the ultra-strong field regime \citep{PhysRevLett.116.221101}, and have yielded new ideas on the formation channels of the systems \citep{2019ApJ...882L..24A}. 

The first GW observation from a coalescing BNS, GW170817 \citep{LSC_BNS_2017PhRvL} and the joint observations of GRB 170817A by the \textit{Fermi}/GBM  and \textit{INTEGRAL}/SPI ACS gamma-ray detectors \citep{goldstein_ordinary_2017,savchenko_integral_2017} firmly established the association between BNS mergers and short duration gamma-ray bursts (GRBs). The estimation of the source location of GW170817 and its subsequent identification at a distance of 40\,Mpc~\citep{2017Sci...358.1556C} initiated a ground-breaking electromagnetic (EM) follow-up campaign from the X-ray to radio bands. These EM observations provided evidence that a successful jet was launched and that the initial gamma-ray emission were from a structured jet viewed approximately 20 degrees off-axis \citep{Mooley_2018Natur,2019Sci...363..968G}. 

Furthermore, optical, ultraviolet and infrared observations of the optical counterpart AT 2017gfo \citep{2017Sci...358.1556C} monitored the glow of a ``kilonova,'' produced by the radioactive decay of $r$-process elements in the neutron-rich ejecta \citep{LaSc1974,LiPa1998,KaMe2017} and showed that the source evolved from blue to red over a week \citep{LSC_MM_2017ApJ, 2018ApJ...855L..23A}. The impact of this discovery motivated a large number of small aperture telescopes to systematically scan the credible regions provided by LIGO/Virgo GW candidate event triggers during subsequent LIGO/Virgo observational runs. The most recent run, O3, which began in April 2019, was split into two parts to allow for a commissioning break: O3a (2019 April -- 2019 September) and O3b (2019 November -- 2020 March 27; somewhat more than a month earlier than planned because of the COVID-19 pandemic). 

Since the discovery of the EM counterpart of GW170817 no other counterpart associated with a GW candidate event has been observed despite significant coordinated efforts by numerous observing groups, e.g. \citealt{CoAh2019b,HoCo2019,GoAn2019,Andreoni2020,Ackley:2020qkz, 2019GCN25384}.

The potential BNS sources\footnote{Note that we make the assumption in this paper that any compact object component with mass $< 3M_{\mathrm{\odot}}$ is a NS.}reported so far by the LIGO/Virgo network in O3 (a sample of five with a BNS source probability of $>50\%$) have median estimated distances in the range $\sim157-241$ Mpc with only one source, GW190425 \citep{AbEA2019}, within 200\,Mpc.

Additionally, the $90\%$ confidence intervals of the sky localisations for this sample cover the range $1131-24226$\,deg$^{2}$. The combination of large credible regions coupled with the dimmer optical counterparts expected at such large distances have made searches for EM counterparts a highly challenging (and ultimately unsuccessful) endeavour during O3.

When conducting searches of large GW sky localisations with a single instrument, a telescope with a large field-of-view is optimal. However, there is a delicate balance; large credible regions are often the result of more distant sources observed at relatively lower signal-noise-ratio (SNR) -- but sources at greater distances are more accessible to larger diameter telescopes. It is both technically and financially challenging to have a large aperture telescope capable of covering a large field-of-view. This has motivated the use of multiple telescopes working in unison within global networks for the purpose of identification and characterisation of counterparts; see for example \citep{Coughlin2019}. In addition to GRANDMA (Global Rapid Advanced Network Devoted to the Multi-messenger Addicts), the heterogeneous telescope network described in this paper, other networks such as the Global Relay of Observatories Watching Transients Happen (GROWTH, \citealt{CoAh2019b}), MASTER-Net \citep{LiKo2010} and KMTNet\footnote{\url{https://kmtnet.kasi.re.kr/kmtnet-eng/}} have also been conducting counterpart searches during O3, supported by a host of individual facilities such as Pan-STARRS1 \citep{Pan-STARRS}, ATLAS \citep{ATLAS}, DECam \citep{DECAM}, MeerLICHT \citep{MeerLICHT}, DDOTI \citep{DDOTI}, and GOTO\footnote{Currently operating with one operational site, with funding approved for an additional node in Australia} \citep{Ob2018}. The O3 campaign has also seen the emergence of a variety of new tools and multi-messenger platforms to coordinate and optimise the follow-up: galaxy ranking can be optimised through the online catalogue MANGROVE \citep{2020MNRAS.492.4768D}; the GWEMOPT open software can be used to schedule GW follow-ups \citep{CoTo2018}; and the GW TreasureMap \citep{2020arXiv200100588W} is a system to report and visualise the progress of searches for EM counterparts to GW events.

In a previous paper \citep{GRANDMAO3A}, we published the GRANDMA results obtained during O3a. This current paper continues in a similar vein through the study of candidate events from O3b. The paper is organised as follows: Section \ref{GWalerts} outlines the GW candidate alerts received during O3, and Section \ref{GRANDMAtel} describes the new telescopes that joined the GRANDMA collaboration for O3b. Section \ref{strategy} provides details on the adopted observational strategies. Section 
\ref{O3asum} presents details on the GRANDMA observations, focuses on some specific events, and provides new constraints on the GW progenitors. We finish by presenting our conclusions in Section \ref{conclusions}.

\section{Overview on the third observational campaign O3}
\label{GWalerts}

The third observational campaign organized by LIGO and Virgo, O3, started 2019 April 1 at 15:00 UTC \citep{2019GCN.24045....1L}. After a month-long commissioning break in October, the second period of the campaign finished on 2020 March 27 at 17:00 UTC, for a total of 330 days, short of the goal of a full 365 days, cut short by the COVID-19 pandemic. For comparison, the second observational campaign lasted 269 days and the first observational campaign lasted 129 days \citep{PhysRevD.91.042003}.

\subsection{O3 alerts}

\begin{figure*}
\centering
 \includegraphics[width=0.8\textwidth]{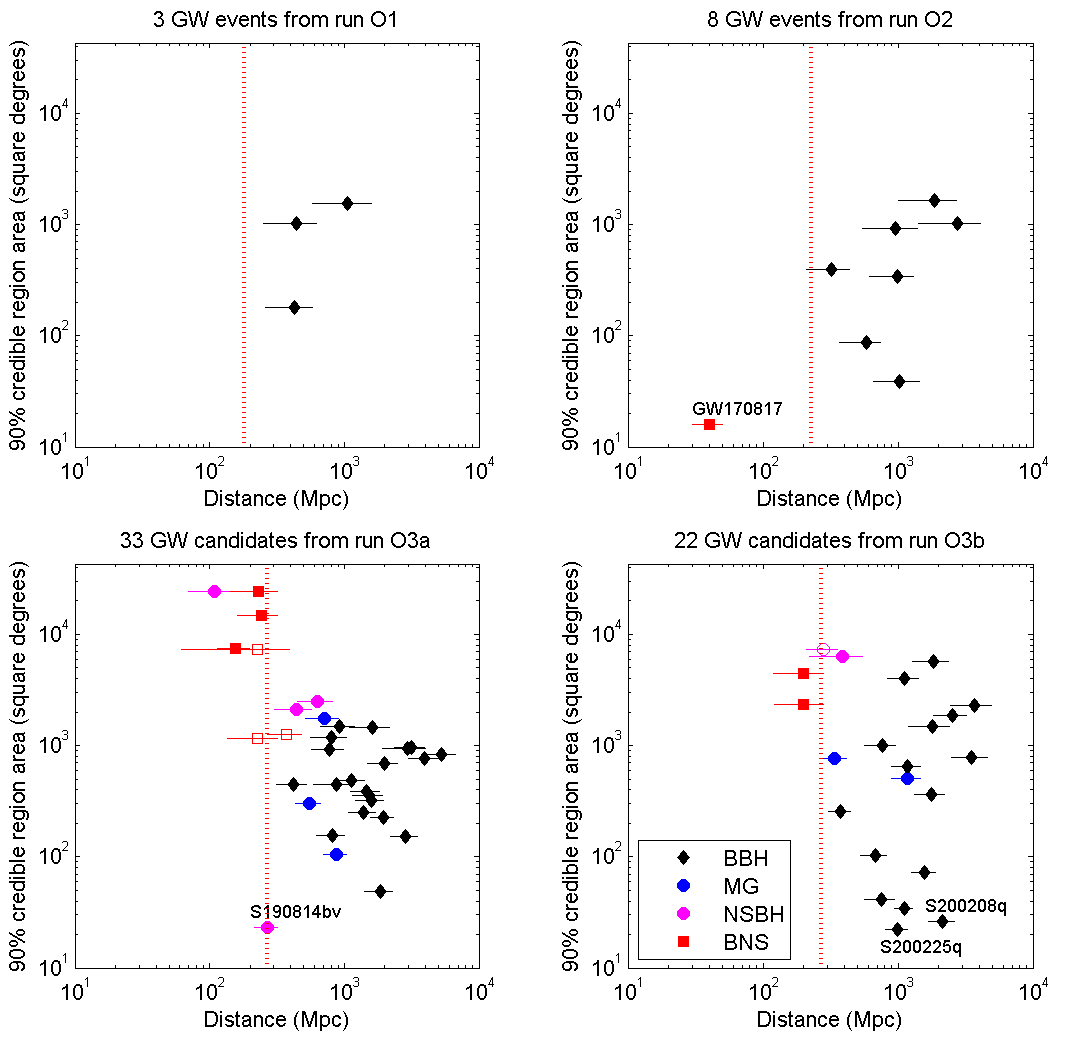}
 \caption{The most recently updated 90\% credible region area versus the most recently updated luminosity distance (posterior mean distance and posterior standard deviation of distance) for all LIGO/Virgo GW events/candidates of runs O1, O2, O3a and O3b. Vertical red dotted lines are the expected limiting distances to detect binary neutron star mergers (BNS) by the LIGO/Virgo detectors. BBH (binary black hole mergers), MG (mass gap mergers), NSBH (neutron star - black hole mergers) are represented in different colours. Four candidates for run O3 are indicated by open symbols corresponding to a predominantly terrestrial classification (above 50\%) and the classification indicated in these plots is the second most likely.}
\label{statistics_distances_omega}
\end{figure*}

\begin{figure}
\centering
 \includegraphics[width=\columnwidth]{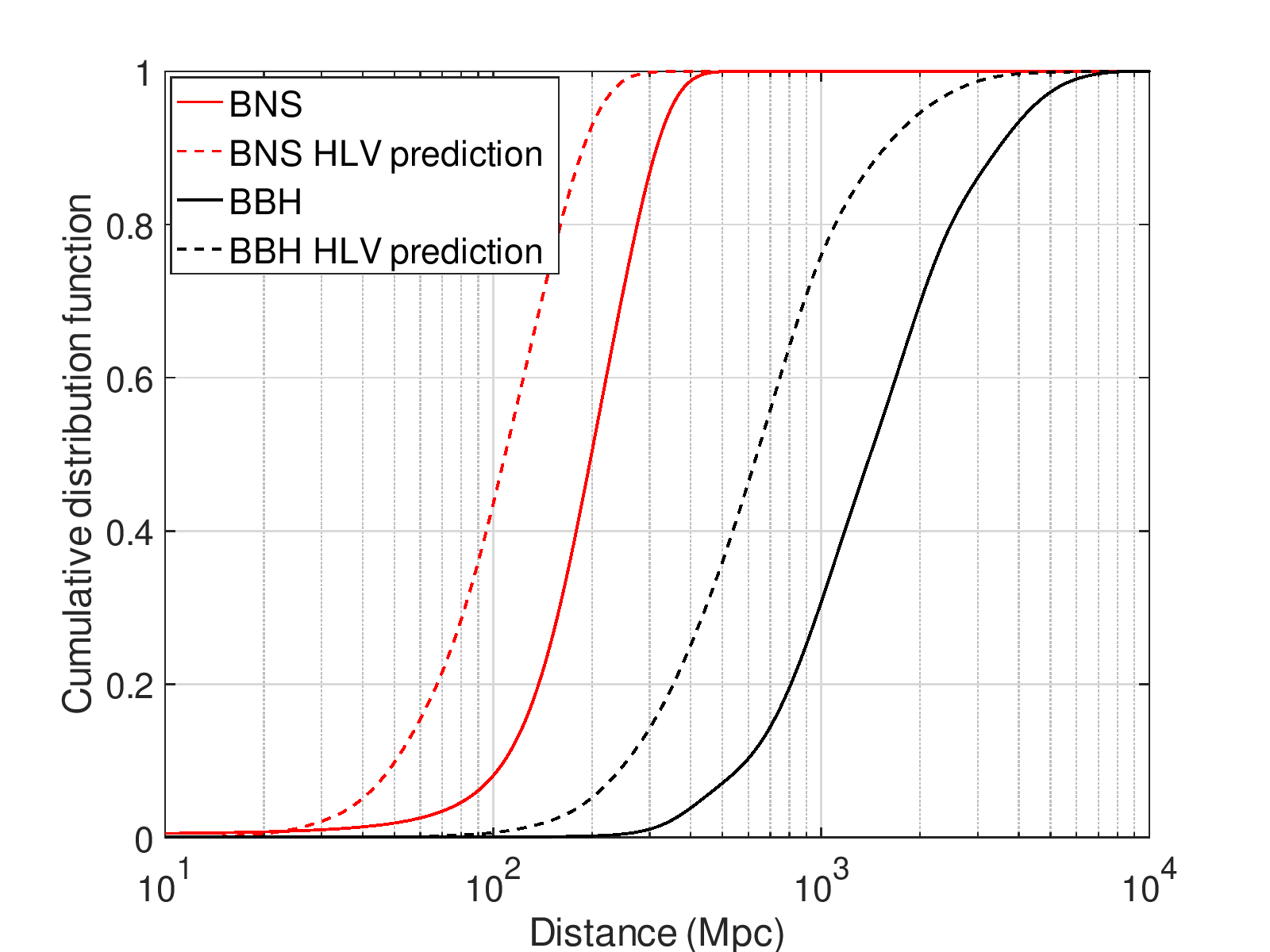}
 \caption{The cumulative distribution function of all LIGO/Virgo GW events and candidates in O3 with a Terrestrial indicator below 50\% versus the distance is shown in solid lines, compared to the predictions for these distribution from \citet{AbEA2020} shown in dashed lines.}
\label{statistics_runO3_distances_probability}
\end{figure}

\begin{figure}
\centering
 \includegraphics[width=\columnwidth]{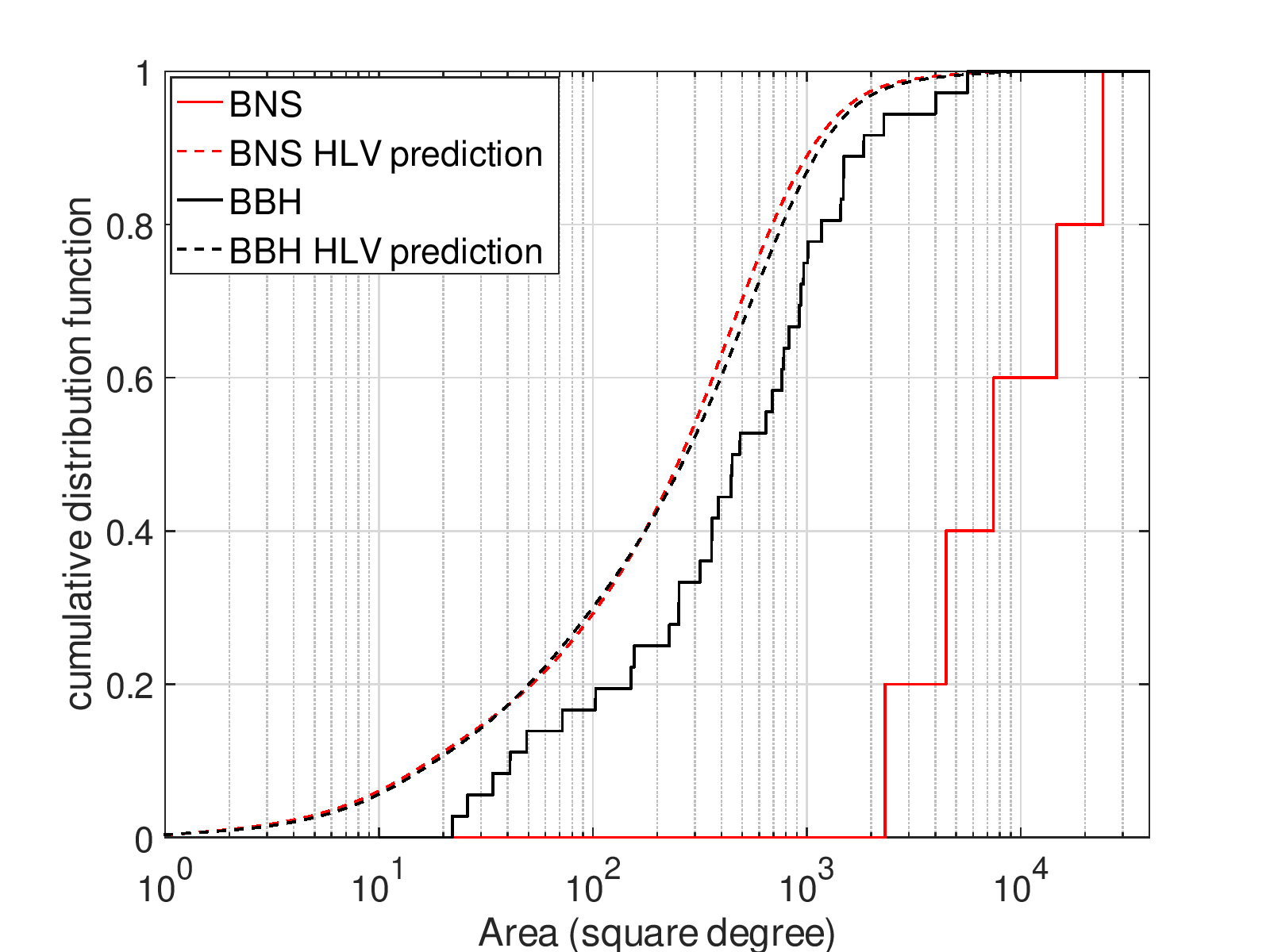}
 \caption{The cumulative distribution function of all LIGO/Virgo GW events and candidates in O3 with a Terrestrial indicator below 50\% versus the 90\% credible localisation area is shown in solid lines, compared to the predictions for these distributions from \citet{AbEA2020} shown in dashed lines.}
\label{statistics_runO3_area_probability}
\end{figure}

During O3, the LIGO Scientific Collaboration and the VIRGO Collaboration (LVC) sent 80 alerts to the astronomical community (and about 100 in total for the O1, O2 and O3 observational campaigns combined). Among them, 24 were false positives resulting in retractions due to data quality issues (see for instance \citealt{2019GCN.26413....1L,2020GCN.26665....1L,2020GCN.26785....1L}) and 56 are still considered as potential signals. 52 were classified as GW signals with a probability higher than 50\%\footnote{See \href{https://gracedb.ligo.org/superevents/public/O3/}{https://gracedb.ligo.org/superevents/public/O3/} for more information.} via the GCN platform (Gamma-ray burst Coordinates Network\footnote{see \href{https://gcn.gsfc.nasa.gov/}{https://gcn.gsfc.nasa.gov/} for more information}) thanks to a new dedicated alert system developed by the LVC \footnote{see the LIGO/Virgo userguide for more information \href{https://emfollow.docs.ligo.org/userguide/}{https://emfollow.docs.ligo.org/userguide/}}. The alert system was still being debugged for the first two months (April and May), producing a large delay of dozens of minutes between the alert delivery and the GW trigger time (the event UTC time). The delay was typically less than 10\,minutes for the remainder of the run, although some exceptional delays did occur for non-technical reasons, i.e. S200219ac \citep{2020GCN.27130....1L}, S200105ae \citep{2020GCN.26640....1L} and S191105ae \citep{2019GCN.26182....1L} during O3b.

The GW alerts were distributed with three different types of GCN/LVC notices with different distribution delays: preliminary (within 5 minutes), initial (within 30 minutes, produced after human vetting), and update notices (within 2 to 4 days, produced after a refined parameter estimation analysis). The notices contain information such as the trigger time, the online pipeline that generated the trigger, the event false alarm rate, a link to the sky localisation probability map (Bayestar, \citealt{PhysRevD.93.024013} or LALInference, \citealt{PhysRevD.91.042003}), and the estimate of the distance to the source. The alert also contains different source classification information if available for compact binary searches:  a BNS [p(BNS)], a BHNS [p(BHNS)], or terrestrial noise [p(terrestrial)] \citep{KaCa2019}, and an indicator to estimate the probability of producing an EM signature considering the candidate is of astrophysical origin [p(HasRemnant)] \citep{2019arXiv191100116C}. The GW trigger can be also classified as ``MassGap,'' completing the possible classifications. In the case of a Mass Gap event, the initial estimate is that at least one of the compact objects has a mass of $3-5M_\odot$, implying it is unclear whether it is a neutron star or a black hole. The other object can be a lighter neutron star or a more massive black hole. Up to the time when this paper was written, only the ``superevent'' S190425z has been confirmed as a real astrophysically produced GW event, GW190425 \citep{LVC2020ApJ190425}.

Out of the 52 GW candidates detected in O3, 36 have been classified as likely being emitted from a BBH merger, five from mass gap mergers, five from BHNS mergers, five from BNS mergers, and one was a GW ``burst'' type trigger (unmodeled search) of unidentified origin \citep{2020GCN.26734....1L,PhysRevD.93.042004}. Except for the case of the burst GW signal, the rest have information on their distance derived from the GW binary coalescence signal modeling. The final event classification can differ from the low latency analysis due to the further study subsequently conducted. This is especially the case for low-mass binaries with one object close to the lower or upper bound of the $3-5 M_\odot$ gap that can be reclassified ``NSBH'', ``BBH'', or ``Mass Gap'': S190814bv was given a final classification of 99\% ``NSBH'' \citep{2019GCN.25324....1L,2019GCN.25333....1L}, and S190728q as 52\% Mass Gap \citep{2019GCN.25187....1L,2019GCN.25208....1L}. In addition, the indication regarding probability to be of astrophysical origin has also evolved for some of the GW candidates: S190426c and S190510g now have a preferential origin of being ``Terrestrial'' \citep{2019GCN.24237....1L,2019GCN.24448....1L,2019GCN.25549....1L}; On the contrary, S190727h is now considered as an astrophysical event with 92\% ``BBH'' \citep{2019GCN.25164....1L}; S200105ae is still classified as 97\% ``Terrestrial'' but the LVC reported that the low false alarm rate for this trigger was due to only one detector (LIGO Livingston) responding to the event and offline analysis is likely to increase its significance \citep{2020GCN.26640....1L,2020GCN.26688....1L}.

 The localisation area for GW alerts also differs from the initial to the updated analysis for some events (for example, due to the inclusion of Virgo data in the analysis for S200225q, \citealt{2020GCN.27229....1L}). These issues will be discussed in section~\ref{BayestarvsLALInference}. Note that the reports for the distance estimate can also vary (e.g. S200219ac, \citealt{2020GCN.27130....1L}), but we will not discuss this distance analysis in this paper.
\smallbreak

Figure~\ref{statistics_distances_omega} displays the most recently issued 90\% credible localisation region produced for a GW candidate versus its most recently updated luminosity distance (posterior mean distance and posterior standard deviation of distance): all detailed numbers can be found in \citet{GRANDMAO3A} for O3a and Section~\ref{O3asum} for O3b. We split the diagram into four sections, one for each of the O1, O2, O3a, and O3b observational campaigns. We indicate the names of detected events and candidates if their error boxes are smaller than 30 square degrees containing one confirmed BNS event (GW170817, \citealt{LSC_BNS_2017PhRvL,2019PhRvX...9c1040A}), a NS-BH candidate (S190814bv, \citealt{2019GCN.25333....1L}), and two BBH candidate events (S200225q, \citealt{2020GCN.27229....1L}, and S200208q, \citealt{2020GCN.27036....1L}).

 For each kind of event, a cumulative distribution function on the distance can be constructed by approximating each measured distance and error bar with a Gaussian function. The combined distribution is shown in Figure~\ref{statistics_runO3_distances_probability}, along with the predictions provided by the LVC~\citep{AbEA2020}. There is a clear separation between the BNS and BBH distributions. This is due to the much higher rate of BNS mergers compared to BBH mergers\footnote{The GWTC-1 catalogue gives $110-3840$ events~Gp$c^{-3}y^{-1}$ for BNSs and $9.7-101$ events Gpc$^{-3}y^{-1}$ for BBHs} \citep{2019PhRvX...9c1040A}, and to the much smaller distance at which BNS mergers can be detected by GW detectors compared to BBH mergers. Hence, BBH mergers are detected often at large distances where the large volume compensates for the low rate. In particular, there is no BBH candidate or event detected below 200~Mpc.

The observed distances are larger than the prediction by approximately a factor 2 for BNS and BBH event types. This could be due to the conservative detection threshold on network SNR of 12 used in the prediction described in \citet{AbEA2020} and by masses for the binary system being larger than those used in the simulations. The study of this discrepancy is beyond the scope of this paper.
In particular, the median distance of BNS candidates in the run O3 is 200\,Mpc, 5 times further than GW170817. As a consequence, the apparent magnitudes of kilonovae associated with the BNS events of the run O3 are expected to be $\sim3.5$~mag fainter than for GW170817, assuming the same intrinsic lightcurve. 

The 90\% credible region areas can be combined to obtain a cumulative distribution function for the localisation area. This is shown in Figure~\ref{statistics_runO3_area_probability} and compared with the predictions provided by the LIGO/Virgo collaboration~\citep{AbEA2020}.
For BBH mergers, the localisation area is on average a factor 2 larger than predicted. Similar to the distance distribution, the difference could be due to larger masses and smaller network SNR of the signals detected as opposed to those from the simulations, as both properties are expected to reduce localisation performance. For BNS mergers the discrepancy with the prediction is larger by a factor of 10. This is more difficult to explain, but we note that three out of eight BNS candidates have a significant probability (above 50\%) of being Terrestrial. GW190425 is the only confident detection from O3 announced by the LVC so far \citep{LVC2020ApJ190425}, but it was observed by only two GW detectors and with a very low signal-to-noise ratio in one of them, which explains its poor sky localisation. In comparison with O2, the sky localisation areas of the eight BNS candidates produced during the run O3 are spread over a median value of 4500 deg$^2$ (i.e. 10\% of the sky), 300 times larger than for GW170817, which was observed by three GW detectors.

\smallbreak
The large GW alert rate in O3 provides motivation for a first estimate of the rate at which each classification of event (BNS, NS-BH, MassGap, and BBHs); see \cite{2019PhRvX...9c1040A,AbEA2020} for a discussion giving a more sophisticated approach.
Our method is as follows: for each type of event we calculate the mean distance for which they are detected.  Next, from this distance we can determine the enclosed volume that it represents, and estimate an event rate. To do this, we divide the number of GW candidates during the observational period by its duration, and divide that result by the volume determined for the type of event as presented in Table~\ref{rate}. A Monte Carlo study shows that this method yields, on average, the correct rate estimate if all events of the same type can be detected up to the same limiting distance (i.e. there is no large dispersion in the chirp masses). This is approximately true for BNS events, but for BBH systems there is a wide distribution in masses which will lead to a large systematic error \citep{2019PhRvX...9c1040A}. We cannot quantify this bias as the chirp mass of the systems are not yet publicly distributed.

\begin{table}
\caption{Statistics for each GW candidate/event source type during O3, see main text for further details.}
\begin{tabular}{ccccc}
\hline
Type    & Detections    & Mean D$_L$    & Volume    & Rate                          \\
        &               & Mpc           & Gpc$^3$   & events yr$^{-1}$ Gpc$^{-3}$   \\
\hline
BNS     & 5             & 206           & 0.037     & 152                           \\
NS-BH    & 5             & 366           & 0.205     &  27                           \\
Mass gap& 5             & 737           & 1.675     &   3.3                           \\
BBH     & 36            & 1649          & 18.77     &   2.1                         \\
\hline
\label{rate}
\end{tabular}
\end{table}

\subsection{Localisation and observational constraints}
\label{BayestarvsLALInference}

\begin{figure}
    \centering
    \includegraphics[width=\columnwidth]{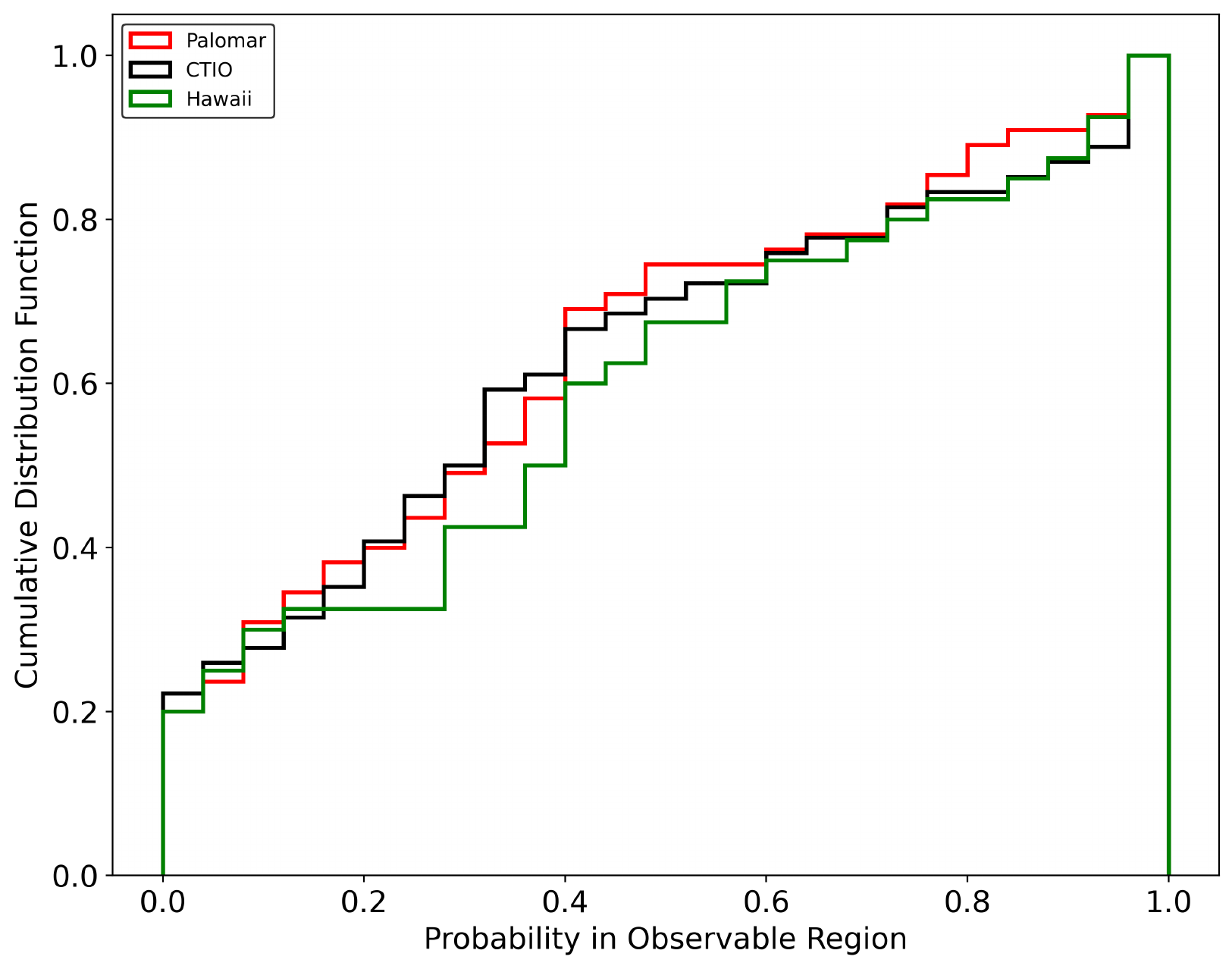}
    \caption{Probability contained within the observable region of the sky localisation for all GW candidates during O3; this is shown from three different locations (Palomar ($33.3563^\circ$, $-116.8648^\circ$), CTIO ($-30.1691^\circ$, $-70.8062^\circ$), and Hawaii ($19.8968^\circ$, $155.5828^\circ$); the LALInference sky localisation was used as long as it was available for the candidate/event in question.}
    \label{fig:observability}
\end{figure}

In order to give insight into the O3 campaign, it is useful to quantify the possibility of observing during the nights following the detection, as well as the benefit of observing after the initial sky localisation is released but before the updated sky localisation is available. Due to factors such as high airmass and the Sun's altitude, the possibility of observing from a certain location may be different than others, and so it is important to take this into consideration when attempting to follow up candidate events. Both initial and updated sky localisation areas are available for 41 candidates. Figure \ref{fig:observability} displays the cumulative distribution function for the observable probability within the localisation from three representative locations (Palomar, The Cerro Tololo Inter-American Observatory, and Hawaii) for all candidate events during O3. As can be seen in the figure, around 20\% of events were not observable at all from at least one of the three locations in the week following detection. We also find that a further $\sim16\%$ of candidates had $\lessapprox4\%$ of the probability observable from all three locations; all of these transients have 90\% credible regions spanning less than 300 square degrees (except for S190602aq, \citealt{2019GCN.24717....1T}, for which it spans $\sim750$ square degrees) and were classified to very likely be BBH mergers (aside from S190924h, \citealt{2019GCN.25861....1L}, which was classified as a mass gap candidate at $>99\%$). Additionally, $20\%$ of candidates had more than $40\%$ observability from only one of the three sites, demonstrating the utility of coordinating network-level telescope observations to obtain large coverage of the localisations. The importance of obtaining such observations at significant depth and coverage (regardless of whether an EM counterpart is ultimately discovered) cannot be overstated, as it has a direct impact on the ability to constrain the ejecta mass and binary parameters for the event if it is truly astrophysical in origin \citep{Coughlin2019}. 

\begin{figure}
    \centering
    \includegraphics[width=\columnwidth]{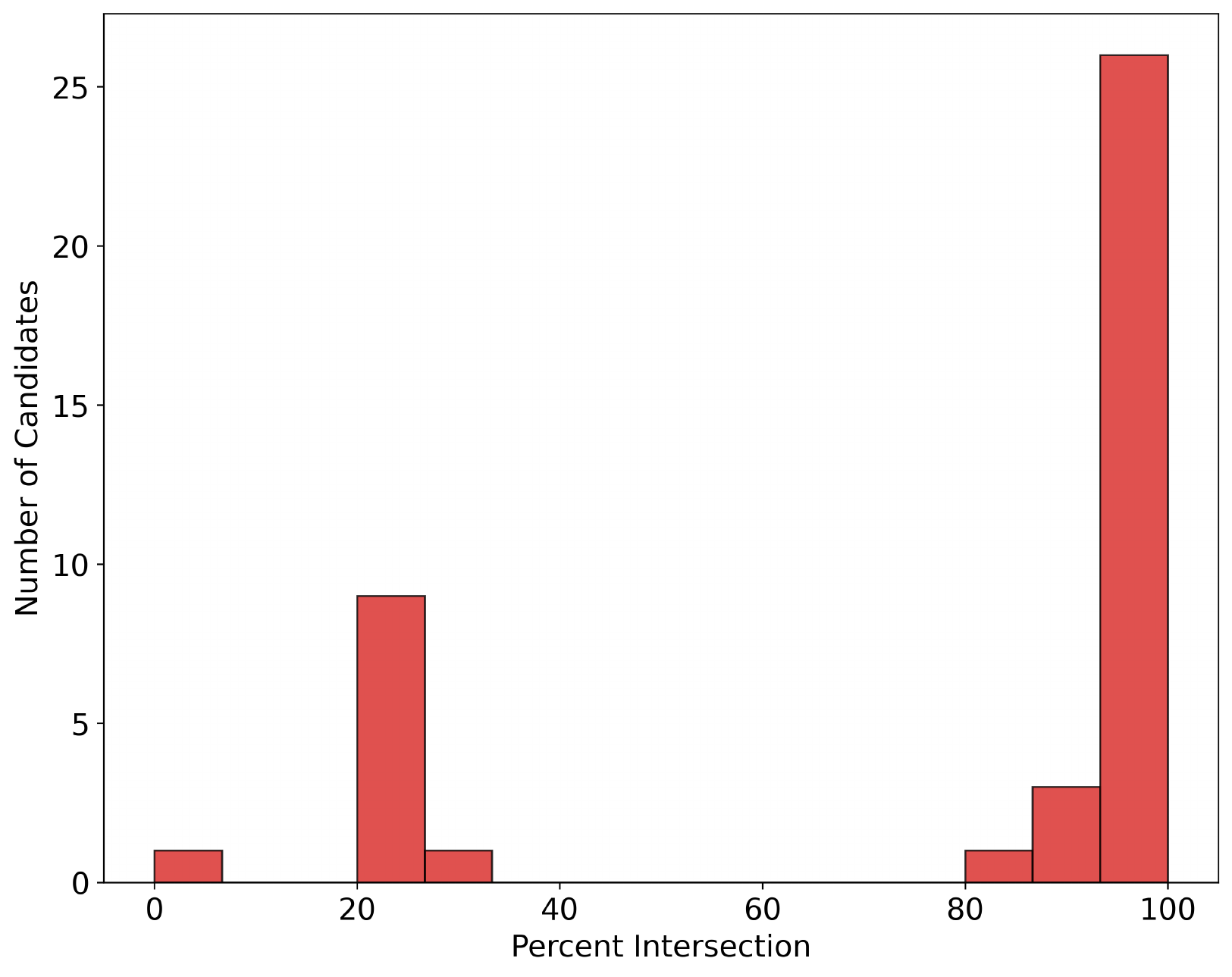}
    \caption{Overlap ratio between the initial and the most up-to-date 90\% credible regions of all GW candidate events during O3 (given that both sky localisations were available). Note that the overlapping regions can be discontinuous.}
    \label{fig:overlap}
\end{figure}

Figure \ref{fig:overlap} presents the distribution of the percent overlap between the 90\% credible regions of the initial Bayestar \citep{PhysRevD.93.024013} and the updated LALInference \citep{PhysRevD.91.042003} sky localisation areas for 41 candidate events from O3. To find the percent overlap, we compute the intersection in pixels between the two maps and then sum the probability contained in that region using the LALInference sky area. From the results, we find that the events fall into one of two possible categories: a low percentage overlap ($\sim$0\%-30\%) or a high percentage overlap (close to 100\%). Around 27\% of the events belong to the first category, showing that a significant number of candidates experience a large shift in their two-dimensional probability distribution. This shift is likely due to a variety of factors. For one, some multi-instrument detections did not factor in data from one of the GW detectors when generating the initial Bayestar skymap (e.g. S190720a, \citealt{2019GCN.25138....1L}), and for other events, the sky area decreased significantly between the two sky localisations (e.g. S200225q, \citealt{2020GCN.27229....1L}, with the inclusion of Virgo, and S190630ag, \citealt{2019GCN.25094....1L}). In addition, disagreement between the time-delays of the detected signals for the event and the results of the parameter estimation, as well as unanticipated issues (for example, the occurrence of scattered light glitches in some of the detectors in the case of S191213g, \citealt{gcn26402}, which possibly affected the sky position) may have led to the low percentage overlap for some events.

\section{Update of the GRANDMA collaboration}
\label{GRANDMAtel}

\begin{figure*}
\begin{center}
\includegraphics[scale=0.4]{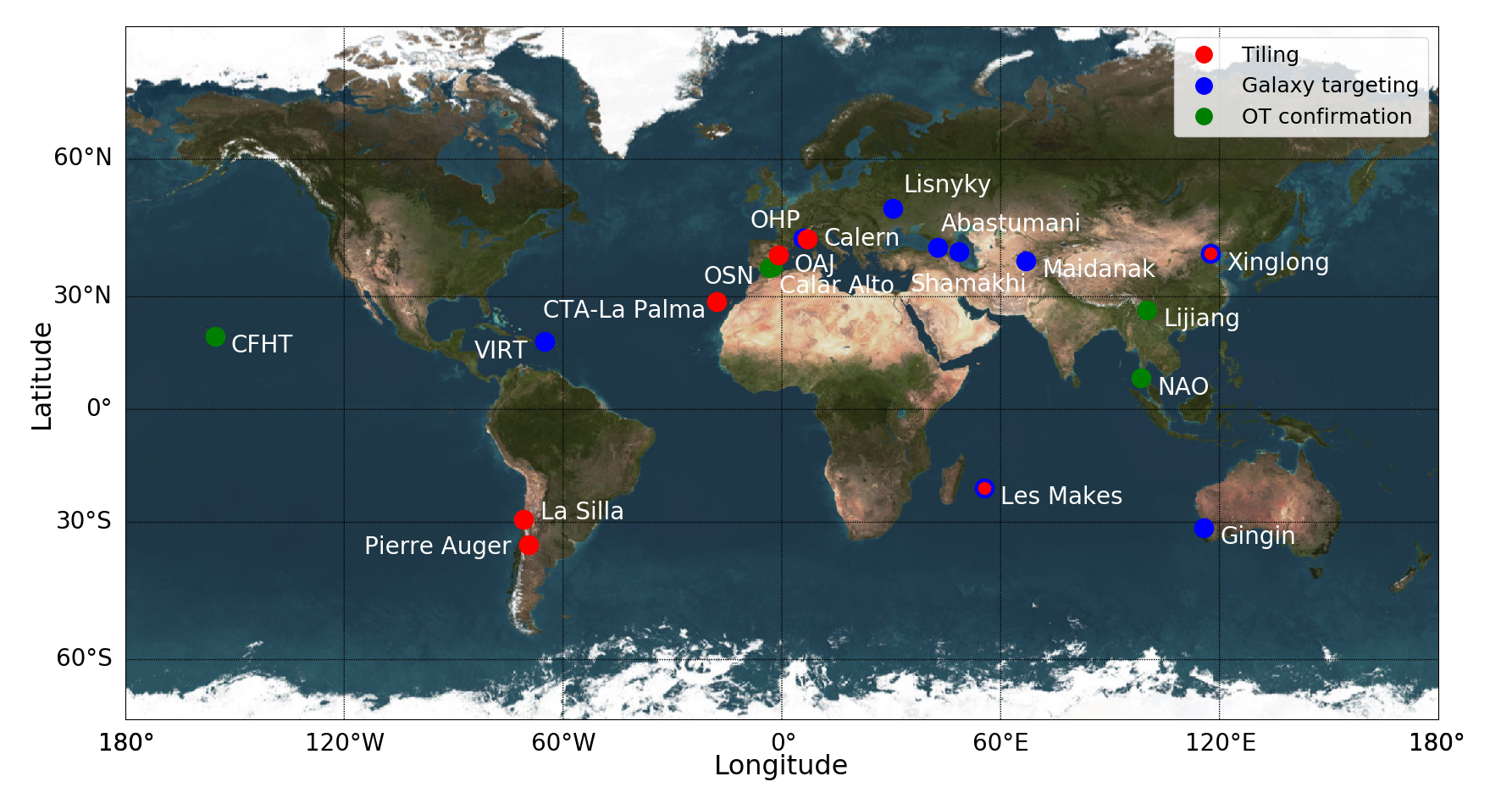}
\caption{Locations of the 20 observatories involved in the GRANDMA network. The colour encodes the observation strategy followed by telescopes at a given observatory: \textit{red} for tiling, \textit{blue} for targeting galaxies, \textit{green} for following-up candidates.}
\label{fig:GRANDMAnet2}
\end{center}
\end{figure*}

GRANDMA, a world-wide telescope network, was presented in \citet{GRANDMAO3A} in the context of O3a. Recently, new groups have joined the collaboration, as will be described below; also presented are updates on the observation strategies and results of certain subgroups. To date, GRANDMA is composed of 25 telescopes with both photometric and spectroscopic facilities, with a large amount of time allocated for observing transient alerts as a telescope network (see Figure~\ref{fig:GRANDMAnet2}, Table~\ref{tab:GRANDMAtelphoto} and Table~\ref{tab:GRANDMAtelspectro}). The GRANDMA collaboration includes 20 observatories, 29 institutions and groups from twelve countries.

For each of the observatories we provide a $5\sigma$ limiting magnitude as an estimate of the faintness of sources that can be reliably discovered by our system. The $5\sigma$ limit is the usual choice of most astronomical surveys (LSST, \citealt{LSST}; SDSS, \citealt{YorkSDSS}; PanSTARRS, \citealt{Chambers2016a}, etc.) as well as other GW follow-up efforts (e.g., GROWTH, \citealt{CoAh2019b}, and GOTO \citep{Ob2018}), and it implies that only one out of 1.7 million sources above this threshold would be a false detection. However, since in our case we are just looking for point-like sources in images with well-known point spread functions (PSFs), our system is capable of filtering out many of these sources (which do not have stellar PSFs) and significantly reducing the number of false positives. Our current neural network system (see Sect.~\ref{strategy}) has shown to be able to reduce the number of false positives by a factor of several thousand. This makes the $5\sigma$ limit provided here a conservative one.

\begin{table*}
\begin{tabular}{ccccccc}
Telescope      & Location        & Aperture & FOV            & Filters                                 & Typical lim mag     & Maximum Night slot\\
Name           &                 & (m)      &  (deg)     &                                         & (AB mag)             & (UTC) \\
\hline
\hline
TAROT/TCH      & La Silla Obs.   & 0.25 & $1.85\times1.85$   & Clear, $g^\prime r^\prime i^\prime $    &  18.0 in 60s (Clear)           & 00h-10h\\
FRAM-Auger     &   Pierre Auger Obs. &  0.30 &  $1.0\times1.0$  & $BVR_CI_C$, Clear   & 17.0 in 120s ($R_C$)          & 00h-10h\\
CFHT/WIRCAM    &    CFH Obs.             &    3.6  &      $0.35\times0.35$             &        $JH$                                &      22.0 in 200s ($J$)       & 10h-16h\\
CFHT/MEGACAM   &     CFH Obs.            &     3.6  &      $1.0\times1.0$               &  $g^\prime r^\prime i^\prime z^\prime$                                       &         23.0 in 200s ($r^\prime$ )    &  10h-16h \\
Thai National Telescope & Thai National Obs.  & 2.40 & $0.13\times0.13$ & Clear, $u^\prime g^\prime r^\prime i^\prime z^\prime$ & 22.3 in 3s ($g^\prime$) & 11h-23h \\
Zadko          & Gingin Obs.      & 1.00 & $0.17\times0.12$ & Clear, $g^\prime r^\prime i^\prime I_C$ & 20.5 in 40s (Clear)  & 12h-22h\\
TNT            & Xinglong Obs.   & 0.80 &  $0.19\times0.19$  & $BVg^\prime r^\prime i^\prime$          &    19.0 in 300s ($R_C$)  & 12h-22h\\
Xinglong-2.16  & Xinglong Obs.   &  2.16    &      $0.15\times0.15$              &     $BVRI$                                    &      21.0 in 100s ($R_C$)       & 12h-22h \\
GMG-2.4        & Lijiang Obs.    &  2.4    &     $0.17\times0.17$                &         $BVRI$                                &     22.0 in 100s ($R_C$)        & 12h-22h\\
UBAI/NT-60     & Maidanak Obs.   & 0.60 & $0.18\times0.18$   & $BVR_CI_C$                              &    18.0 in 180s ($R_C$) & 14h-00h\\
UBAI/ST-60     & Maidanak Obs.   & 0.60 & $0.11\times0.11$   & $BVR_CI_C$                              &    18.0 in 180s ($R_C$) & 14h-00h\\
TAROT/TRE      & La Reunion      & 0.18 &  $4.2\times4.2$  & Clear                                   &     16.0 in 60s (Clear)        & 15h-01h\\
Les Makes/T60  & La Reunion.     & 0.60 &  $0.3\times0.3$  & Clear, $BVR_C$                          &   19.0 in 180s ($R_C$)       & 15h-01h\\
Abastumani/T70 & Abastumani Obs. & 0.70 & $0.5\times0.5$   & $BVR_CI_C$                              &    18.2 in 60s ($R_C$)   & 17h-03h\\
ShAO/T60       & Shamakhy Obs.   & 0.60 & $0.28\times0.28$ & $BVR_CI_C$                              &    19.0 in 300s ($R_C$)  & 17h-03h \\
Lisnyky/AZT-8  & Kyiv Obs.       & 0.70 & $0.38\times0.38$ & $UBVR_CI_C$                             & 20.0 in 300s($R_C$) & 17h-03h\\
TAROT/TCA      & Calern Obs.     & 0.25 & $1.85\times1.85$ & Clear, $g^\prime r^\prime i^\prime$     &  18.0 in 60s (Clear)          & 20h-06h\\
FRAM-CTA    & ORM  & 0.25 &  $0.43\times0.43$   & Clear, $BVR_Cz^\prime$,    & 16.5 in 120s ($R_C$)          & 20h-06h\\
IRIS           & OHP             & 0.50  &  $0.4\times0.4$  &   Clear, $u^\prime g^\prime r^\prime i^\prime z^\prime $ & 18.5 in 60s ($r^\prime$)   & 20h-06h\\
T120           & OHP             & 1.20 &   $0.3\times0.3$     &     $BVRI$                                    &   20.0 in 60s ($R$)           & 20h-06h\\
OAJ/T80        & Javalambre Obs. & 0.80 &   $1.4\times1.4$                 & $r^\prime$                              & 21.0 in 180s ($r^\prime$) & 20h-06h\\
OSN/T150       & Sierra Nevada Obs. & 1.50 &  $0.30\times0.22$                  & $BVR_CI_C$                              & 21.5 in 180s ($R_C$) & 20h-06h\\
CAHA/2.2m      & Calar Alto Obs. & 2.20 &  $0.27\diameter$                & $ u^\prime g^\prime r^\prime i^\prime z^\prime $                              & 23.7 in 100s ($r^\prime$) & 20h-06h\\
VIRT           & Etelman Obs.   & 0.50 & $0.27\times0.27$ & $UBVRI, Clear$ &  19.0 in 120s (Clear) & 22h-04h\\

\hline
\end{tabular}
\caption{List of telescopes of the GRANDMA collaboration and their photometric performance when using their standard setup. }
\label{tab:GRANDMAtelphoto}
\end{table*}

\begin{table*}
\begin{tabular}{ccccccc}
Telescope/Instrument & Location & Wavelength range & Spectral resolution $\lambda/\Delta\lambda$ &  Limiting mag\\
\hline
\hline
2.2m CAHA/CAFOS   & Calar Alto Obs. & 3200-7000/6300-11000 & 400 & 20 in 1h \\
ShAO/T2m & Shamakhy Obs. & $3800-8000$ & 2000 & 17 in 1h \\
Xinglong-2.16/BFOSC & Xinglong Obs. & $3600-9600$ & 1000 & 18 in 1h \\
GMG-2.4/YFOSC & Lijiang Obs. & $3400-9100$ & 2000 & 19 in 1h \\
GTC & ORM & $3630-7500/7330-10000$ & 1018/2503 & 24 in 1h  \\
\hline
\end{tabular}
\caption{List of telescopes of the GRANDMA collaboration with spectroscopic capabilities.}
\label{tab:GRANDMAtelspectro}
\end{table*}

\subsection{Canada-France-Hawaii Telescope}

Twelve hours of observing time have been awarded (PI: A. Coleiro) on the Canada-France-Hawaii 3.6-m world-class telescope (CFHT) for period 2020A in order to trigger target-of-opportunity photometric observations. The telescope is located on top of the Mauna Kea summit in Hawaii and is equipped with two imaging instruments: a $1^\circ\times1^\circ$ field-of-view optical camera, MegaCam, complemented by a $20\arcmin\times20\arcmin$ field-of-view near-infrared detector, WIRCam. By providing high-sensitivity and wide field-of-view optical ($i^\prime$, $r^\prime$ and $z^\prime$) and near-infrared ($J$, $H$ and $K_s$) data, CFHT is an essential element of the GRANDMA network, allowing for a fast identification of the EM counterpart and/or its physical characterisation. To strengthen their scientific impact, we asked all data to be made freely available to the whole community when delivered to the GRANDMA team. However, due to the premature end of the scientific operations of both CFHT and the GW observatories at the end of March 2020, no observations were triggered on CFHT during period 2020A.

\subsection{FRAM network}
FRAM (F/(Ph)otometric Robotic Atmospheric Monitor) is a series of small robotic telescopes primarily designed for a continuous atmospheric monitoring of astronomical sites in order to measure atmospheric transparency with high spatial or temporal resolution. The original FRAM \citep{fram_auger} has been operated at the Pierre Auger Observatory in Argentina for more than a decade, while three more FRAMs \citep{fram_cta} are prepared to be used for a real-time atmospheric monitoring at Cherenkov Telescope Array (CTA) locations. Two of the FRAM sites are equipped, in addition to primary wide-field telephoto lenses, with larger diameter traditional telescopes, with larger diameter traditional telescopes intended for a follow-up and monitoring observations of a wide range of astrophysical transient objects (gamma-ray bursts, variable stars, comets, etc). 
These two telescopes are used for GRANDMA network observations. The first is a 30 cm f/6.8 telescope located at Pierre Auger observatory, Malargue, Argentina, and equipped with $B$, $V$, $R$ and $I$ filters, with a field-of-view of $60^\prime\times60^\prime$ and pixel scale of 0\farcs92/pix. 
The second is a 25 cm f/6.3 telescope located at Observatorio del Roque de los Muchachos, La Palma, Canary Islands, Spain, and is equipped with $B$, $V$, $R$ and $z^\prime$ filters, with a $26^\prime\times26^\prime$ field-of-view and 1\farcs52/pix pixel scale.

\subsection{HETH group at IAA}
The HETH (High-Energy Transients and their Hosts) group at the Instituto de Astrof\'isica de Andaluc\'ia (IAA) in Granada, Spain, continues to obtain competitive time at multiple telescopes.
The programmes at the Observatorio de Sierra Nevada (OSN, PI: Bla\v{z}ek) and the Centro Astron\'omico Hispano en Andalucía (CAHA, PI: Kann) continue as detailed in \cite{GRANDMAO3A}.

Competitive time was obtained at the Observatorio de Javalambre (OAJ) for further observations with the T80 telescope to obtain tiling observations (Proposal \#1900160, PI: Kann, see \citealt{GRANDMAO3A} for further details). Once again, the proposal was to observe $1.4^\circ\times1.4^\circ$ fields with an exposure time of 180\,s each, reaching a typical limiting magnitude of $r^\prime>21.0$ mag. In this proposal, based on the lessons learned from O3a, the requested time was increased to 12.46 hours (up to 100 tiles), including late-time re-observations for image-subtraction purposes. The second part of the proposal, multi-colour KN follow-up, was dropped as GRANDMA has access to facilities that are better suited for such deep, narrow-field, targeted observations.

The Gran Telescopio Canarias\footnote{\url{http://www.gtc.iac.es}} (GTC) is the world's largest single-aperture optical telescope, a 10.4m Ritchey–Chr\'etien telescope with a segmented mirror located at 2267 m altitude at Roque de los Muchachos observatory, La Palma, Canary Islands, Spain, at $28^\circ45^\prime24^{\prime\prime}$N, $17^\circ53^\prime31^{\prime\prime}$W. It is equipped with a large suite of instruments. The following three are part of the proposal.

The Optical System for Imaging and low-Intermediate-Resolution Integrated Spectroscopy (OSIRIS) is a multi-purpose imager and spectrograph mounted on the Nasmyth-B focus. It is equipped with SDSS $u^\prime g^\prime r^\prime i^\prime z^\prime$ filters and multiple grisms and volume-phased holographic (VPH) gratings allowing for different spectral resolutions. We generally employ the R1000B (wavelength range $3630-7500$ {\AA}, resolution 1018 at 5455 {\AA}) grism and the R2500I (wavelength range $7330-10000$ {\AA}, resolution 2503 at 8650 {\AA}) VPH grating, as well as $g^\prime r^\prime i^\prime z^\prime$ imaging. To obtain spectra with sufficient SNR, exposures of different integration times are employed: $4\times1200$~s if the target has $r^\prime >22$ mag, $3\times900$~s for $r^\prime\sim20-22$ mag and $2\times900$~s if $r^\prime<20$ mag. 

The Especrografo Multiobjeto Infra-Rojo (Infrared Multi-Object Spectrograph, EMIR), installed at the Nasmyth-A focus, is a multi-channel IR imager and medium-resolution spectrograph.
For spectra, it can achieve a SNR of 7 for a 21 mag source with 4800 s of exposure in the Y J grism. In imaging, limiting AB magnitudes at $5\sigma$ of $Y>24.2$, $J>23.6$, $H>23.1$ and $K_S>22.2$ mag can be achieved with exposure times of 180~s, 240~s, 420~s, and 600~s, respectively.

Finally, for the purpose of observing the host galaxy of a GW event in 3D, an application was made for time to use the Multi-Espectr\'ografo en GTC de Alta Resoluci\'on para Astronom\'ia (High-Resolution Multi-Spectrograph for Astronomy at GTC, MEGARA) in Integral Field Unit (IFU) mode, covering a field of $12\farcs5\times11\farcs3$ with a spaxel size of 0\farcs62. Two grisms would be used to cover the spectral lines H$\alpha$, [\ion{N}{II}], and [\ion{S}{II}] as well as H$\beta$ and [\ion{O}{III}].

An award was made (Proposal 118-MULTIPLE-3/20A, PI: Kann) 8\,hr of time with OSIRIS, 5\,hr with EMIR, and 2\,hr with MEGARA, ranked ``A''. The proposal was geared toward obtaining detailed observations of a single \textit{confirmed} EM counterpart of a GW event, and not meant for classifying candidates. The observing time was valid for the last two months of O3, in March and April 2020; however the worldwide COVID-19 pandemic caused GTC to be closed down already in mid-March, shortly thereafter followed by the end of O3b; no observations were ever triggered.

\subsection{Thai National Telescope}
 The Thai national telescope located at Doi Inthanon National Park is the main facility for the National Astronomical Research Institute of Thailand (NARIT). It is s 2.4 m diameter, Ritchy-Chr\'etien telescope with an $f/10$ ratio, providing a plate scale of 8\farcs6/mm at the two Nasmyth foci. The telescope system is an alt-az mount that can slew at a speed of 4\degree/s; therefore, it can track or acquire optical transient sources rapidly with a pointing accuracy at less than 3\arcsec, and can track targets without autoguiding for ten minutes or more with a pointing error of less than 0\farcs5. There are many scientific instruments that can be used upon request for specific observations such as ULTRASPEC~\citep{10.1093.mnras.stu1660} with SDSS filters ranging from 330 to 1000 nm, spectrographs varying from low to high resolution, and different types of scientific cameras.
 
The telescope was chosen to be built at the summit of the highest mountain in Thailand (Lat. 18.57372E\degree, Long. 98.48219E\degree, Alt. 2457 m). Its observational area and seeing conditions ensure dark skies for $B=21.5$ and $V=21.9$~mag/arcsec$^{2}$, and the median seeing is stable throughout the night at $\approx0\farcs9$. 

\subsection{VIRT}
The Virgin Islands Robotic Telescope (VIRT) is the primary research telescope at the University of the Virgin Islands and is located at the Etelman Observatory on the island of St. Thomas in the US Virgin Islands (USVI). The facility is at an elevation of 381 meters, latitude of 18.3N and longitude of 64.9W. The VIRT is a 0.5\,m Schmidt-Cassegrain telescope with a Finger Lakes Instrumentation (FLI) Proline 4240 back illuminated midband camera with a FLI Johnson/Bessel $UBVRI$ filter set. Typical seeing at the site is $1\farcs5-2\farcs0$.

\subsection{The \textit{kilonova-catcher} citizen science program}

\begin{figure*}
\begin{center}
\includegraphics[width=1.0\textwidth]{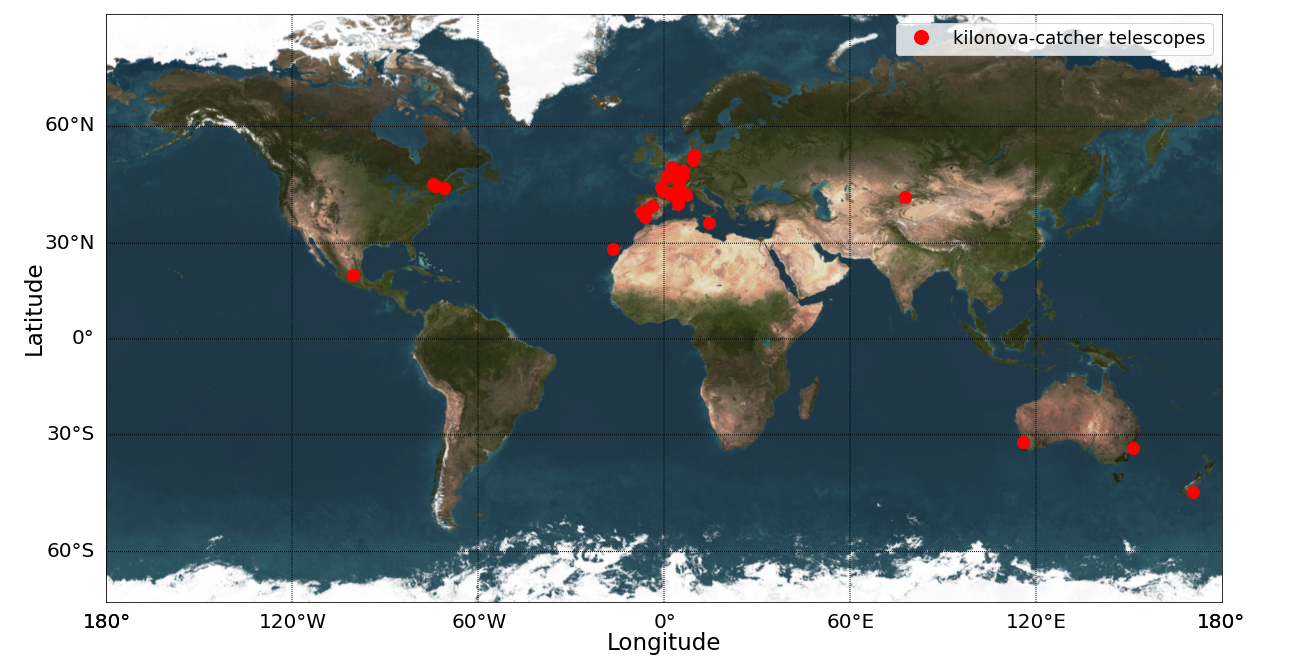}
\caption{Locations of the 34 telescopes involved in the GRANDMA citizen science program: \textit{kilonova-catcher}. Some citizen astronomers can operate several telescopes represented in this Earth map and also external remote facilities like the iTelescope Network (\url{https://www.itelescope.net/}), not represented here.}
\label{fig:KN_user_map}
\end{center}
\end{figure*}

Since the beginning of O3, GRANDMA has developed a citizen science program called \textit{kilonova-catcher} \citep{GRANDMAO3A}. It allows non-professional astronomers to add their observational capabilities to the GRANDMA network in order to perform optical follow-up of GW candidates on a best-effort basis. During the O3b run, 33 \textit{kilonova-catcher} users around the globe registered to the GRANDMA GW alert stream. In Figure \ref{fig:KN_user_map}, we show the locations of the telescopes they could operate during the O3b run.

During the O3b campaign, the kilonova-cather program was active during the follow-up of two BNS merger candidates \href{https://gracedb.ligo.org/superevents/S191213g/view/}{S191213g} and \href{https://gracedb.ligo.org/superevents/S200213t/view/}{S200213t}. Because of the narrow FoV of the amateur telescopes, we coordinated their observations with a strategy based on targeted searches for optical transients located in promising host galaxies. From these two citizen observational campaigns, we obtained 57 follow-up images in total. A detailed description of the amateur observation strategy and results can be found in Section~\ref{O3asum}. GRANDMA is unique with this global initiative of citizen observers. O3 illustrated the huge potential for GW astronomy to vitalise sky observations and EM counterpart candidates, while also highlighting the need of training and coordinated joint analysis.

\section{Update for the alert infrastructure, observation strategy and web interface and data analysis used in GRANDMA}
\label{strategy}

We use a similar infrastructure -- {\it ICARE} (Interface and Communication for Addicts of the Rapid follow-up in multi-messenger Era) -- for conducting coordinated observations of GW alerts to that discussed in \citet{GRANDMAO3A}. ICARE relies on the automatic reception of a GW alert, creation of an automatic observation plan that is then sent to the telescopes of the network, a central database, and cloud applications to monitor the full network. Similar tools and infrastructure have been developed by other teams for a similar purpose, i.e. the GROWTH Marshal \citep{2019PASP..131c8003K}. 
In this section, we will outline some improvements since the end of the O3a run, and in particular, during the break between O3a and O3b. We are going to focus on work done for the daily use version of ICARE that ran during O3b\footnote{There is a more ``universal version'' of ICARE under development now that O3 has ended}. These modifications introduced some minor negative impacts on the productivity of the collaboration at the level of the end-user communication (i.e. mostly with regard to the preparation of GCNs), which had no effect on observations, nor results reported in this manuscript.

The python-based alert distribution system used by GRANDMA during O3a remained mainly the same. Upon receiving a GW candidate alerts through a GCN, the coverage of the sky localisation region was generated for the GRANDMA collaboration telescopes (Table~\ref{tab:GRANDMAtelphoto}). For BBH type events, only large field-of-view instruments were included, while all instruments were involved for NSBH or BNS type events if the posterior mean distance was below 300 Mpc. The resulting observation plans were then sent via VOEvent as soon as the computation of a single plan was done using  Comet\footnote{\url{https://comet.transientskp.org/en/stable/}}. We improved this system for O3b by introducing a forwarding of the retraction notice to avoid unnecessary observations.

Our observation plan algorithm benefited from the
continuous improvements of the GWEMOPT\footnote{\url{https://github.com/mcoughlin/gwemopt}} telescope scheduling software \citep{CoTo2018}. Its latest version includes slicing sky localisations in right ascension to separately schedule different lobes in multi-lobed maps \citep{CoAn2019}, and automatically determining fields capable of being observed multiple times by several telescopes \citep{AlCo2020}. In doing so, the coverage of all our instruments, and especially the wide field-of-view telescopes, was improved. Our galaxy targeting strategy, i.e. when small field-of-view telescopes were utilised to survey known galaxies) was also revised using the dedicated MANGROVE galaxy catalogue presented in \citet{2020MNRAS.492.4768D}. We adopted the selection of preferential galaxies based on dependence on galaxy stellar mass. The most problematic issue in Galaxy Targeting Mode from O3a, however, that has since been addressed was the re-imaging of tiles when observations were executed. In short, this issue arose from an avoidance of considering whether or not a given galaxy requested for observation resided in the field-of-view of a preceding observation. As such, it resulted in an inefficient management of telescope observing time, and reduced the area covered by our collaboration. 
This improvement additionally benefited our strategy, as it led to extra time available for observing the most promising events. 

The start of the O3 run demonstrated the need for a standard data reduction pipeline in order to homogenise the photometry within the network. First, we still use individual online image processing developed by the various groups such as TAROT and FRAM \citep{2019ApJ...886...73N}. Secondly, we continue the  development of a common detection pipeline \textit{Gmadet}\footnote{\url{https://github.com/dcorre/gmadet}} adaptable for each telescope: it makes use of the popular AstrOmatic\footnote{\url{https://www.astromatic.net}} software (SCAMP, SWarp, PSFEx, SExtractor) to perform the astrometric calibration, align and stack images, estimate the PSF and extract sources. The detected sources are cross-matched to different catalogues (Pan-STARRS, Gaia, USNO-B1) using Xmatch, the CDS cross-match service, to discard already known objects. A crossmatch with known Solar system objects is also performed using SkyBoT \citep{skybot}. Particular attention is paid to the photometric calibration so that magnitudes are expressed in the AB system. Image substraction with a reference image is performed using HOTPANTS \citep{hotpants}, where Pan-STARRS stacked images can be used as reference. A Convolutional Neural Network algorithm is trained to identify stellar sources and discard most of the cosmics, bad pixels or artefacts coming from the subtraction process. In practice, during O3b \textit{Gmadet} was used to detect transient candidates in parallel to individual standard online image processing made by the groups. The results presented in this work correspond for \textit{Gmadet} only for the OAJ observations of S200213t. The other results use the individual image processing described in \citet{GRANDMAO3A} to allow for a direct comparison of our results against those for O3a. In case of gravitational-wave alert observations engaged by GRANDMA, each individual image was processed with a search for point sources detected at the 5-sigma or greater level. All the false positives which were deemed to be artifacts or cosmic rays, were eliminated by human inspection.

\section{GRANDMA electromagnetic follow-up campaign of O3a and O3b}
\label{O3asum}

\subsection{Observational summary of GRANDMA for O3a and O3b}

Over the eleven months of observations of the O3 campaign, GRANDMA followed-up 49/56 candidate events : 27/33 for O3a and 22/23 for O3b. The observation efficiency for GRANDMA is therefore $\sim$ 82\% for O3a and 96\% for O3b. This high cadence is possible due to the organization of the GRANDMA collaboration and the automated infrastructure discussed above. It is significantly above the reported observational rate of other ground-follow-up teams, aside from MASTER-Net \citep{LiKo2010} that reports GCNs for every alert based on their automated observations. 
Based on the GCN traffic, GRANDMA can be compared to GROWTH (about 30~\% of alerts, \citealt{CoAh2019b}), GOTO (about 15\% of alerts, \citealt{Ob2018}), ENGRAVE (about 5~\% of alerts, \citealt{Ackley:2020qkz}), GRAWITA (about 25\% of alerts, \citealt{2019GCN.25650....1S}), J-GEM (about 7\% of alerts, \citealt{2020GCN.27066....1K}), NOWT (about 4\% of alerts, \citealt{2019GCN.26488....1Z}), SAGUARO (about 11\% of alerts, \citealt{2019GCN.26473....1L}), and SVOM (about 30\% of alerts, \citealt{2020GCN.26948....1X}).

\begin{figure}
\centering
 \includegraphics[width=\columnwidth]{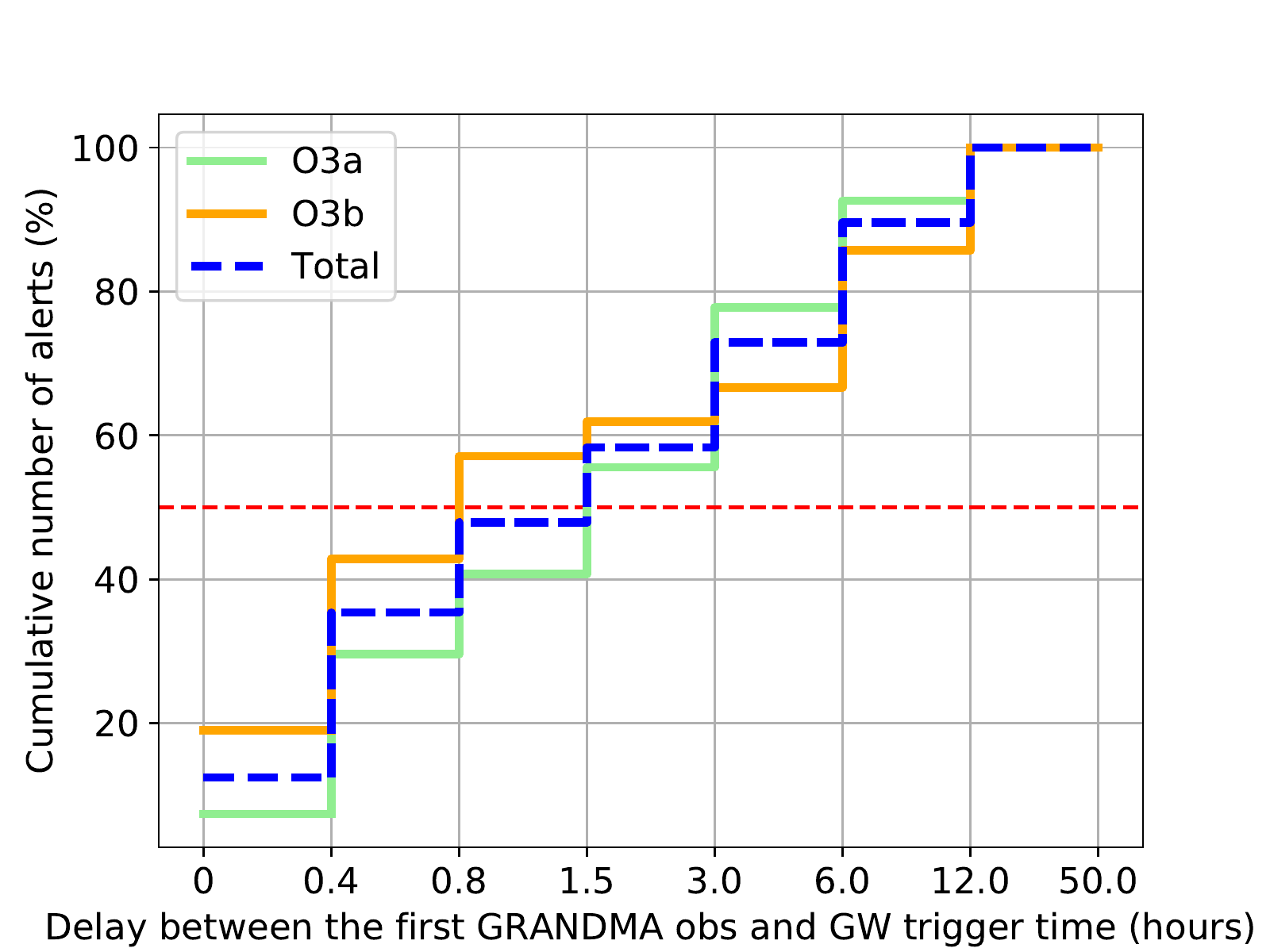}
 \caption{Cumulative distribution of the delay between the GW candidate/event trigger time and the first observation done by GRANDMA. The minimal delay is about 15 min obtained with both the TAROT and FRAM networks. About 50\% of the sky localisation areas of GW candidate/event alerts have been observed with a minimal delay of 1.5 hours.}
\label{DelayminimalGRANDMA}
\end{figure}

As shown in Figure~\ref{DelayminimalGRANDMA}, GRANDMA performed its first observation at most 1.5\,hours after the GW trigger time for more than half of the alerts, and at most 30 minutes for 15~\% of the alerts during O3. The minimal delay between the GW trigger time and the first GRANDMA observation is 15 min, which includes the $\sim$ 5\,min delay for sending the alert by LIGO/Virgo, $\sim$ 5\,min for computing the observation plan, and $\sim$ 5\,min for scheduling by the telescope. This has been achieved for both the TAROT and FRAM networks. Moreover, we see some improvement in terms of the rapid response by GRANDMA between O3a and O3b: this is due to improvements not only from the LIGO/Virgo infrastructure system \citep{2019arXiv190103310T} but also the GRANDMA infrastructure improvements discussed above. This demonstrates once more the need for a heterogenous network of telescopes sharing joint tools to face the challenge of follow-up of GW triggers and their requirement of rapid EM observational responses. Indeed, \citet{Metzger:2019zeh} and references therein, predict different light curve evolutions for the first hours after the merger of two compact objects~\citep{2018ApJ...855L..23A} depending on the properties of the initial binary system. Early time observations after the merger (less than two hours and in different passbands if possible) might have provided important constraints on properties of the fast ejecta components, the composition of the ejected material, and the heating rate for the unbound material. Other contribution channels or kilonova-precursors powered by free neutron decay can also be ruled out by early observations~\citep{Metzger:2014yda}.

\begin{figure}
\centering
 \includegraphics[width=\columnwidth]{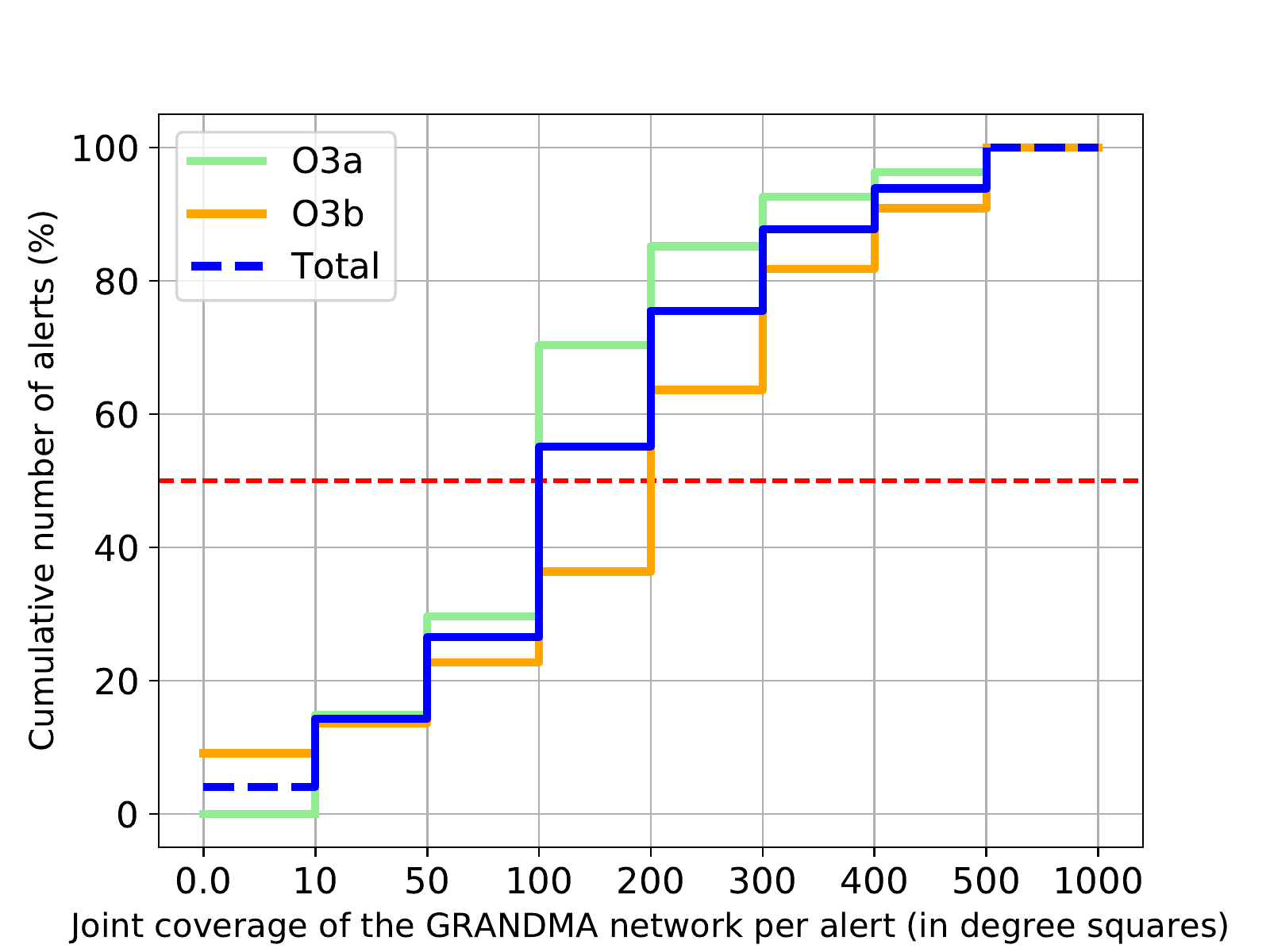}
 \caption{Cumulative distribution of the total sky coverage of the GRANDMA network (in deg$^2$) engaged for each GW candidate/event alert. The distribution depends on the sky localisation area size of the alerts and observability constraints, as discussed in Section~\ref{GWalerts}.}
\label{CoverageGRANDMA}
\end{figure}

\begin{figure*}
\centering
 \includegraphics[width=1.0\textwidth]{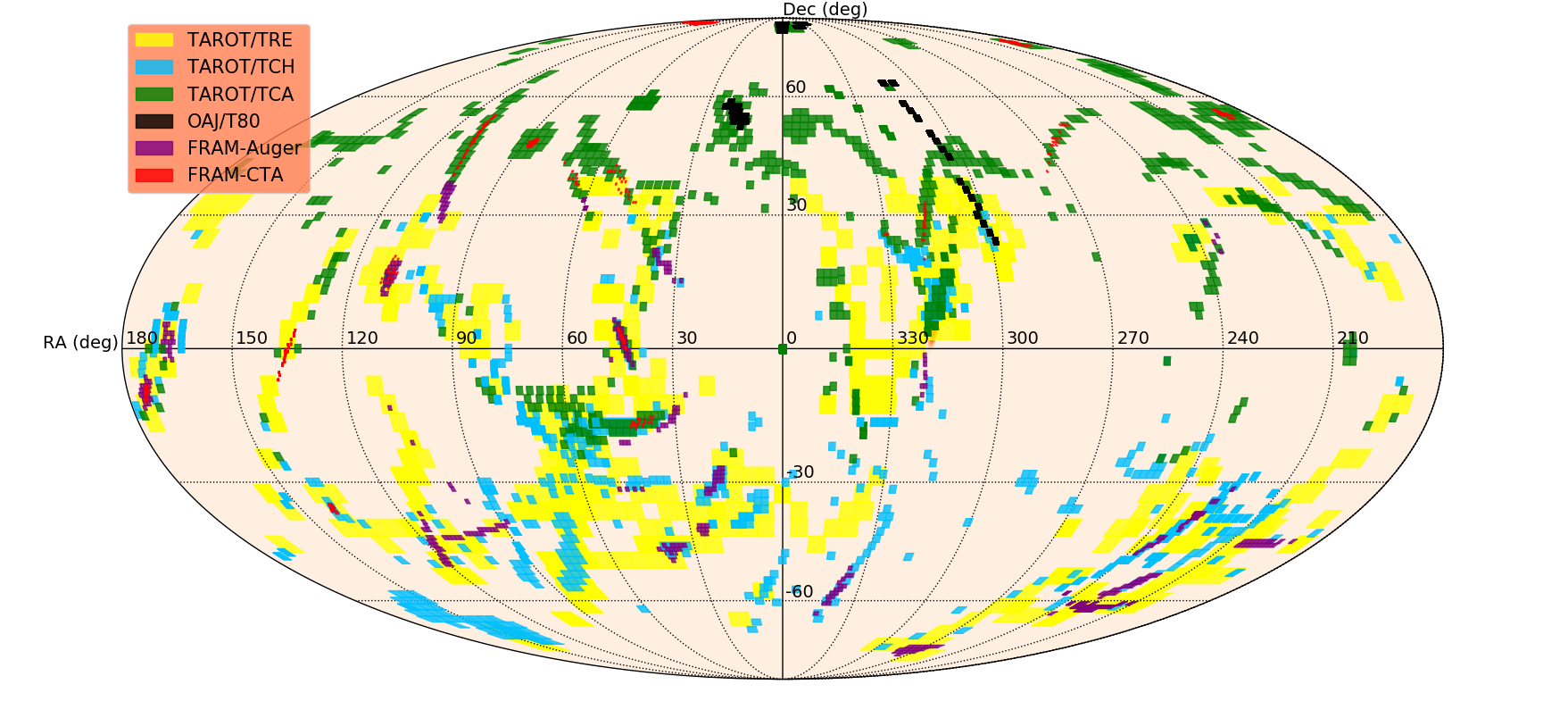}
\caption{Total sky coverage of the wide field-of-view telescopes of GRANDMA during O3. Note the distribution of observations due to the GRANDMA coordinated program. Also note that the heterogeneity of the maps is imprinted from observing constraints and the distribution of the alerts over the sky.}
\label{CoveragetotGRANDMA}
\end{figure*}

GRANDMA covered $213 \pm 173$ deg$^2$ on average for the GW alerts during O3. As shown in Figure~\ref{CoverageGRANDMA}, more than 100 deg$^2$ of the sky has been observed by the GRANDMA network for more than half of the alerts. We do not see major differences in terms of sky coverage between O3a and O3b; however, improvements in the network-level scheduling as discussed in \cite{CoAn2019} in June 2019 led to a significant increase of the maximum coverage reachable by GRANDMA (typically by a factor of 2). The total maximal coverage for a single alert was for S191204r, a BBH candidate event, at 550 deg$^2$, representing 90\% of the sky localisation area. The lowest coverage was only 10~\% of the candidate S200225q's most updated sky localisation area; although the event was well localised, with a 90\% credible region of 22 deg$^2$ as seen in Appendix~\ref{followup_plots} and \citet{2020GCN.27229....1L}, the vast majority of this region was behind the sun at the time of the event. In Figure~\ref{CoveragetotGRANDMA}, we can see the total sky coverage of GRANDMA over the full O3 campaign. There is a clear distribution of northern and southern hemispheres among the different telescopes due to the global coordination (see Section~\ref{strategy}). In total, we covered 9218 deg$^2$ with six wide field-of-view telescopes, with multi-epoch observations for most of the fields. Our near real-time analysis did not find any interesting kilonova candidates (see Section~\ref{O3asum} and \citealt{GRANDMAO3A}).

\begin{figure}
\centering
 \includegraphics[width=0.47\textwidth]{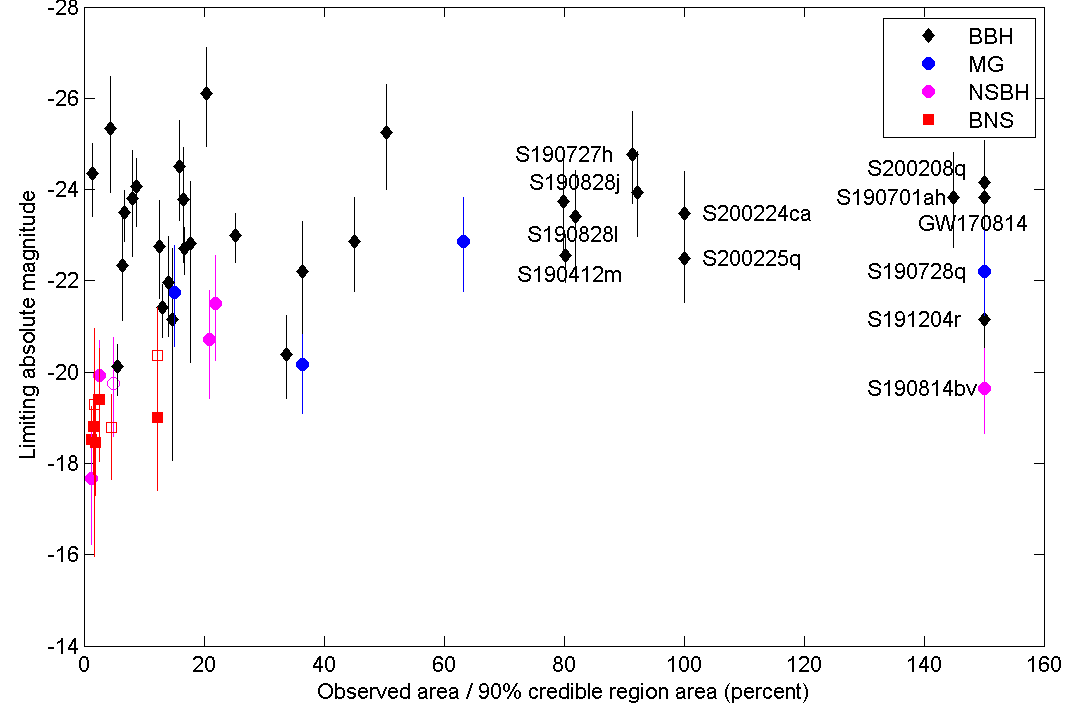}
 \caption{All LIGO/Virgo GW candidates/events observed by GRANDMA telescopes are plotted as limiting absolute magnitude versus the percentage of the 90\% credible region that was observed. We added the TAROT/TRE limit of GW~170814 published in \citet{2019ApJ...886...73N}.
 candidates/events having a percentage of the 90\% credible region greater than 150\% are plotted at 150\% to keep clarity of the plot.
 }
\label{absmag_omega_obs}
\end{figure}

\begin{figure}
\centering
 \includegraphics[width=0.47\textwidth]{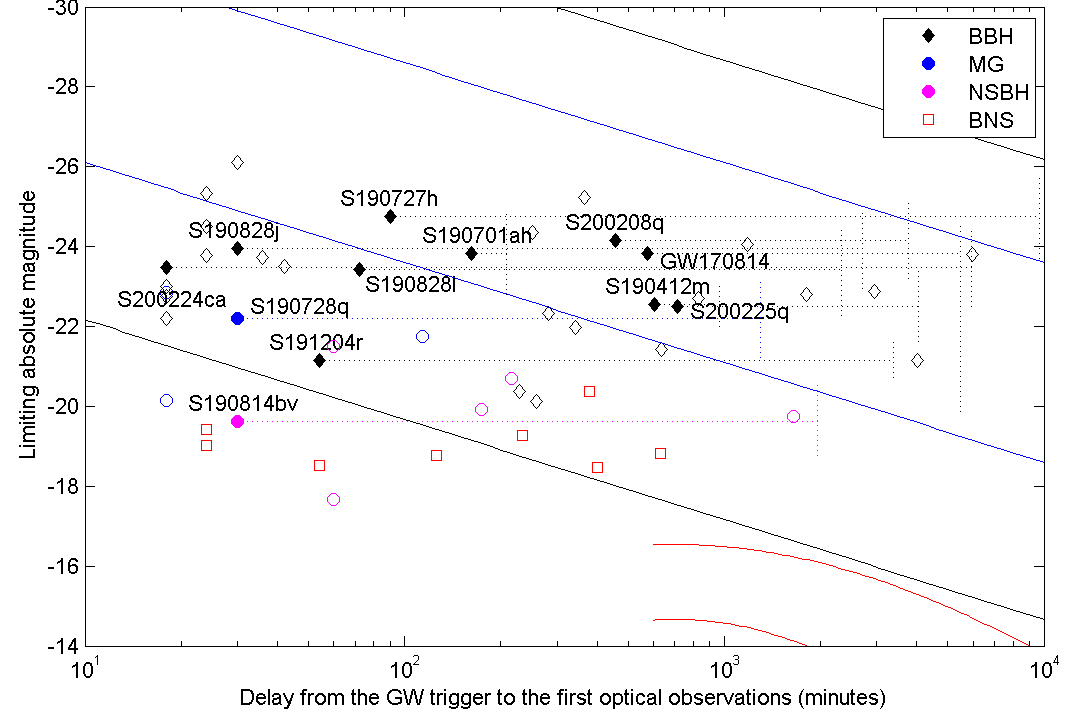}
 \caption{All LIGO/Virgo GW candidates/events observed by GRANDMA telescopes are plotted as limiting absolute magnitude versus the delay between the GW trigger and the first GRANDMA observations. Filled symbols are used for candidates/events for which more than 80\% of the 90\% credible region have been observed. We added the TAROT/TRE limit of GW170814 published in \citet{2019ApJ...886...73N}. The dashed horizontal bars indicate the duration of the first visit for the best candidates/events (selected from the Figure \ref{absmag_omega_obs}). The black lines are the typical limits of LGRB light curves, blue lines for SGRBs and red lines for the GW170817 kilonova.
 }
\label{light_curve_obs}
\end{figure}

According to Figure~\ref{statistics_runO3_distances_probability}, the median distance of O3 BNS merger candidates is $\sim$\,200~Mpc, corresponding to a distance modulus $\textnormal{DM}=36.5$. The median distance of BBHs is $\sim1.3$~Gpc ($\textnormal{DM}=40.6$).  Considering an apparent limiting magnitude of 17, this corresponds to approximate absolute magnitude non-detection limits of $-$19.5 for BNSs and $-$23.6 for BBHs. Figure \ref{absmag_omega_obs} displays the absolute magnitude limit based on the distance of each GW event versus the observed area, and in Figure \ref{light_curve_obs}, versus the time of the first optical observations. It
shows that only 12 GW events were observed by GRANDMA covering more than 80\% of the error box. The names of these events are indicated in the Figure \ref{absmag_omega_obs}.
Essentially only BBH events have full coverage of the credible region; this is due to a combination of the statistical rate of observability (see section~\ref{BayestarvsLALInference}), GRANDMA coverage capacity ($\sim200$ deg$^2$), and the poor localisation of NSBH or BNS events. S190814bv, with a current classification of NSBH, remains the exception. In conclusion, the disparity of alerts during O3 in terms of localisation and their observational accessibility, the evolution of the localisation over time, as well as the possible nature of candidates, shows a clear need for monitoring of routine observations of the alerts and calls for rapid reactions at all times. In this way, by observing all alerts during O3, GRANDMA has gained training and experience and is ready to observe a large number of alerts during the subsequent GW observation campaigns, starting with O4.

\subsection{Report summary of GRANDMA observations during O3b}

Specifically for O3b, we report our observations for the BBH candidates in Table~\ref{tab:GRANDMArecapO3BBH}, and for the systems potentially containing at least one NS in Table~\ref{tab:GRANDMArecapO3BNS}. The unmodeled GW burst candidate S200114f is listed in Table~\ref{tab:GRANDMArecapO3Burst}. All the sky localisation coverage are shown in Appendix~\ref{skycoveragelink}.

We have similar observational results during O3b as in O3a in terms of coverage; at least 30\% of the error boxes of the GW candidate events (using the most recently updated sky localisation areas) were observed in 63\% of the cases. The TAROT network followed and reported search results~\citep{2019ApJ...886...73N} for all of the GW alerts sent by GRANDMA, and FRAM only missed one of them. We note also the participation of the OAJ telescope for the BNS candidate S200213t \citep{gcn27042,gcn27116}. The participation of the narrow field-of-view telescopes has been reduced in O3b to host galaxy targeting searches due to a low number of GW candidates with a most probable luminosity distance below 300\,Mpc; this consisted of three events \citep{gcn27042,gcn26402,2020GCN.26640....1L}. However, we can highlight the participation of the Lisnyky and Tingshua groups for S191213g, the UBAI group for S191213t, and the VIRT group for S200213t. We also highlight the participation of amateur astronomers for S191213g and S200213t.

\begin{table*}
\caption[Summary of the GRANDMA observations]{Summary of the GRANDMA observations during the last five months of O3 for BBH candidates. Observations are not necessarily continuous during the time interval.
\href{https://gracedb.ligo.org/superevents/S191110x/view/}{S191110x} \href{https://gracedb.ligo.org/superevents/S191213ai/view/}{S191213ai},
\href{https://gracedb.ligo.org/superevents/S191120at/view/}{S191120at},
\href{https://gracedb.ligo.org/superevents/S191124be/view/}{S191124be},
\href{https://gracedb.ligo.org/superevents/S191212q/view/}{S191212q},
\href{https://gracedb.ligo.org/superevents/S191225aq/view/}{S191225aq},
\href{https://gracedb.ligo.org/superevents/S200108v/view/}{S200108v},
\href{https://gracedb.ligo.org/superevents/S200303ba/view/}{S200303ba},
were retracted by the LVC due to data quality issues in the detectors.  90\% c.r. corresponds to the 90\% credible region of the latest sky localisation area sent by LIGO-Virgo ($^a$, when only the ``Bayestar sky localisation'' is available), $\delta \, t$ to the delay with respect to the trigger time, $\Delta \, \mathrm{T}$ to the duration of the observations, Prob (\%), Area (${\mathrm{deg^2}}$) to the coverage of GRANDMA compared to the latest revision of the sky localisation area in percentage and in squares degrees. Limiting magnitudes are computed from 5$\sigma$ threshold for source extraction.}
\label{tab:GRANDMArecapO3BBH}
\begin{tabular}{ccccc|cccccc}
Alert & Time & Type & Dist & 90\% c.r. & Telescope       & $\delta \, t$ & $\Delta \, \mathrm{T}$ & Lim. mag &  Prob & Area   \\
 & (UTC) &  & (Mpc) & (deg$^2$) &  & (h) & (h) &  &  (\%) & (${\mathrm{deg^2}}$)   \\
\hline
\multirow{6}{*}{\href{https://gracedb.ligo.org/superevents/S191105e/view/}{S191105e}} & \multirow{6}{*}{14:35:21} & \multirow{6}{*}{BBH (95\%)} & \multirow{6}{*}{$1183 \pm 281$} & \multirow{6}{*}{643} & GRANDMA & 49.0  & 16.2  & 17-18  & 64.1  &  289 \\
      &  &    & &  & FRAM-Auger & 57.6 & 1.6 & 16.8  &  3.6   &  19  \\
      &  &    & &  & FRAM-CTA &58.7  & 1.5   & 15.9  &  0.2   & 4   \\
 &  &  &  &   & TCA  & 56.1 & 6.4 &  18 &  1.1  & 25  \\
   &  &    & &  & TCH  &57.4 & 7.9  & 18  &  33.6   & 46   \\
      &  &    & &  & TRE &49.0  & 8.47   & 17  &  57.3   &  242  \\
\hline
\multirow{6}{*}{\href{https://gracedb.ligo.org/superevents/S191109d/view/}{S191109d}$^a$} &
\multirow{6}{*}{01:07:17} & \multirow{6}{*}{BBH (>99\%)} & \multirow{6}{*}{$1810 \pm 604$} & 
\multirow{6}{*}{1487} & GRANDMA & 0.4 &  31.2 & 17-18  & 18.0  & 263  \\
      &  &    & &  & FRAM-Auger & 5.7 &  1.8  & 16.0 & 2.8  &    19 \\
      &  &    & &  & FRAM-CTA & 0.4 & 2.7   & 16.0  &  0.5  & 5  \\
 &  &  &  &   & TCA  & 1.0 &  26.0  &  18  &  2.1  & 43 \\
   &  &    & &  & TCH  & 29.5 & 2.2  & 18   & 1.0   & 11  \\
      &  &    & &  & TRE & 20.2 &  2.5  & 17 & 12.3  &  191  \\
\hline
\multirow{4}{*}{\href{https://gracedb.ligo.org/superevents/S191129u/view/}{S191129u}}  &
\multirow{4}{*}{13:40:29} & \multirow{4}{*}{BBH (100\%)} & \multirow{4}{*}{$742 \pm 180$} &
\multirow{4}{*}{852} & GRANDMA & 10.6 &  49.1 & 18   & 27.9  & 130  \\
      &  &    & &  & FRAM-Auger & 11.3 & 2.2   & 16.2  & 6.1  & 17   \\
 &  &  &  &   & TCA  & 10.6 & 5.5  & 18   &  4.6 & 63 \\
   &  &    & &  & TCH  & 10.7 & 49.0  &  18  &  21.9  & 63  \\
\hline
\multirow{5}{*}{\href{https://gracedb.ligo.org/superevents/S191204r/view/}{S191204r}$^a$} &
\multirow{5}{*}{17:15:26} & \multirow{5}{*}{BBH (100\%)} & \multirow{5}{*}{$678 \pm 149$} &
\multirow{5}{*}{103} & GRANDMA & 0.9 & 58.6  & 18  & 89.1  & 550 \\
      &  &    & &  & FRAM-Auger & 7.4 &  1.7  & 16.7 & 0.1  &  20  \\
 &  &  &  &   & TCA  & 0.9 & 57.1  &  18  &  0.8 & 68 \\
   &  &    & &  & TCH  & 7.5 & 30.4  &  18  &  18.2  & 68  \\
      &  &    & &  & TRE & 1.0 &   47.8 & 17 & 72.9  & 417   \\
\hline
\multirow{6}{*}{\href{https://gracedb.ligo.org/superevents/S191215w/view/}{S191215w}} &
\multirow{6}{*}{22:30:52} & \multirow{6}{*}{BBH (>99\%)} & \multirow{6}{*}{$1770 \pm 455$} &
\multirow{6}{*}{361} & GRANDMA & 0.6 & 56.3  & 17-18  & 37.3  & 288 \\
      &  &    & &  & FRAM-Auger & 5.0 &  1.4  & 16.7 &  2.1 &   20 \\
     &  &    & &  & FRAM-CTA & 46.1 &  0.2  & 17.5 &  0.7 &    1 \\
   &  &    & &  & TCA  & 42.6 &  5.1 &  18  &  18.6  & 25  \\
      &  &    & &  & TCH  & 2.2 & 54.7  &  18  &  13.0  &   36 \\
      &  &    & &  & TRE & 0.6 &  47.3  & 17 &  8.6 &  226  \\    
\hline
\multirow{6}{*}{\href{https://gracedb.ligo.org/superevents/S191216ap/view/}{S191216ap}} &
\multirow{6}{*}{21:33:38} & \multirow{6}{*}{BBH (99\%)} & \multirow{6}{*}{$376 \pm 70$} &
\multirow{6}{*}{253} & GRANDMA & 3.8 & 162.1  & 17-18  & 42.8  &  85 \\
      &  &    & &  & FRAM-Auger & 3.8 &  1.1  & 15.1 &  1.8 & 8   \\
     &  &    & &  & FRAM-CTA & 21.6 &  1.5  & 17.4 &  4.4 & 5   \\
   &  &    & &  & TCA  & 19.4 &  146.4 &  18  &  40.1  & 67  \\
      &  &    & &  & TCH  & 27.6 & 122  &  18  &  0.7  & 11  \\
\hline
\multirow{6}{*}{\href{https://gracedb.ligo.org/superevents/S191222n/view/}{S191222n}} &
\multirow{6}{*}{03:35:37} & \multirow{6}{*}{BBH (>99\%)} & \multirow{6}{*}{$2518 \pm 679$} &
\multirow{6}{*}{1850} & GRANDMA &  0.4 & 51  & 17-18  & 23.8  & 292 \\
      &  &    & &  & FRAM-Auger & 0.4 & 28.4  & 16.3  & 3.0  & 20    \\
     &  &    & &  & FRAM-CTA & 0.7 & 1.5 & 16.6 & 0.3  &  5  \\
   &  &    & &  & TCA  & 0.7 &  1.2 & 18  & 3.6  & 71  \\
      &  &    & &  & TCH  & 3.0 &  48.5 & 18  &  8.8  & 56  \\
      &  &    & &  & TRE  & 12.5 & 30.5  & 17  & 10.0 & 157   \\
\hline
\multirow{6}{*}{\href{https://gracedb.ligo.org/superevents/S200112r/view/}{S200112r}} &
\multirow{6}{*}{15:58:38} & \multirow{6}{*}{BBH (>99\%)} & \multirow{6}{*}{$1125 \pm 289$} &
\multirow{6}{*}{4004} & GRANDMA & 0.3  & 58.8   & 17-18  &  16.9 & 500 \\
      &  &    & &  & FRAM-Auger & 9.5 & 5.2  & 15.4   & 0.5   &  15  \\
     &  &    & &  & FRAM-CTA & 13.5  & 25.6 & 16.6 &  0.7 &  5  \\
   &  &    & &  & TCA  & 1.5 & 48.8  & 18  &  5.8 &  71 \\
      &  &    & &  & TCH  & 8.6 & 50.5  & 18  & 2.4   &  63 \\
      &  &    & &  & TRE  & 0.3 & 47.9  & 17 & 8.7  &  379  \\
\hline
\multirow{6}{*}{\href{https://gracedb.ligo.org/superevents/S200115j/view/}{S200115j}} &
\multirow{6}{*}{04:23:09} & \multirow{6}{*}{MassGap (>99\%)} & \multirow{6}{*}{$340 \pm 79$} &
\multirow{6}{*}{765} & GRANDMA & 0.3  & 166.3  & 17-18  & 14.8  & 278 \\
      &  &    & &  & FRAM-Auger & 20.8 & 1.7  & 15.8 & 1.6  &  19  \\
     &  &    & &  & FRAM-CTA & 15.3 & 1.5 &  17.4 & 0.4   & 5    \\
   &  &    & &  & TCA  & 12.9 & 153.8  & 18  & 3.8  & 71  \\
      &  &    & &  & TCH  & 0.3  &  164.9 & 18  &  6.5  &  68 \\
      &  &    & &  & TRE  & 11.7  & 28.8  & 17 & 7.4 & 174  \\
\hline
\multirow{6}{*}{\href{https://gracedb.ligo.org/superevents/S200128d/view/}{S200128d}} &
\multirow{6}{*}{02:20:11} & \multirow{6}{*}{BBH (97\%)} & \multirow{6}{*}{$3702 \pm 1265$} &
\multirow{6}{*}{2293} & GRANDMA & 0.4  & 54.7  & 17-18  & 11.2  & 97 \\
      &  &    & &  & FRAM-Auger & 0.45 & 5.6  & 17.8 & 2.6  &  20  \\
     &  &    & &  & FRAM-CTA & 0.4 & 4.8 & 16.8 &  0.4  & 4    \\
   &  &    & &  & TCA  & 19.1 & 3.2  & 18  & 1.5  & 22  \\
      &  &    & &  & TCH  & 2.6 &  52.5 & 18  &  8.7  &  67 \\
\hline
\end{tabular}
\end{table*}

\begin{table*}
\contcaption{GRANDMA observations during the last five months for BBH candidates.}
\label{tab:GRANDMArecapO3BBH2}
\begin{tabular}{ccccc|cccccc}
Alert & Time & Type & Dist & 90\% c.r. & Telescope       & $\delta \, t$ & $\Delta \, \mathrm{T}$ & Lim. mag &  Prob & Area   \\
 & (UTC) &  & (Mpc) & (deg$^2$) &  & (h) & (h) &  &  (\%) & (${\mathrm{deg^2}}$)   \\
\hline
\multirow{2}{*}{\href{https://gracedb.ligo.org/superevents/S200129m/view/}{S200129m}} &
\multirow{2}{*}{06:54:58} & \multirow{2}{*}{BBH (100\%)} & \multirow{2}{*}{$755 \pm 194$} &
\multirow{2}{*}{41} & GRANDMA & 11.0 & 12.2 & 18  & 8.3  & 7 \\
   &  &    & &  & TCA  & 11.0 & 12.2 & 18  & 8.3  & 7  \\
\hline
\multirow{6}{*}{\href{https://gracedb.ligo.org/superevents/S200208q/view/}{S200208q}} &
\multirow{6}{*}{13:01:17} & \multirow{6}{*}{BBH (>99\%)} & \multirow{6}{*}{$2142 \pm 459$} &
\multirow{6}{*}{26} & GRANDMA & 7.6  & 104.0  & 17-18  & 95.1  & 488 \\
      &  &    & &  & FRAM-Auger & 18.3 & 0.12  & 16 & 8.6  &  2  \\
     &  &    & &  & FRAM-CTA & 9.3 & 0.98 &  15.3 & 11.3   & 3    \\
   &  &    & &  & TCA  & 7.6 & 104.0  & 18  & 0.1  & 69  \\
      &  &    & &  & TCH  & 11.4  &  56.5 & 18  &  43.6  &  69 \\
      &  &    & &  & TRE  & 27.0  & 8.9  & 17 & 91.1 & 364  \\
\hline
\multirow{6}{*}{\href{https://gracedb.ligo.org/superevents/S200219ac/view/}{S200219ac }}&
\multirow{6}{*}{09:44:15} & \multirow{6}{*}{BBH (96\%)} & \multirow{6}{*}{$3533 \pm 1031$} &
\multirow{6}{*}{781} & GRANDMA & 6.1  & 53.1  & 17-18  & 39.2  & 392 \\
      &  &    & &  & FRAM-Auger & 14.7  & 1.6  & 16.5 & 4.8  & 17    \\
     &  &    & &  & FRAM-CTA & 13.3 & 1.5 & 17.3 &  0.8  &  4  \\
   &  &    & &  & TCA  & 8.5 &  50.7 & 18 & 8.6  &  70 \\
      &  &    & &  & TCH  & 14.6  & 27.6  &  18  & 11.2  &  63 \\
      &  &    & &  & TRE  & 6.1  &  31.9 & 17 & 23.8 &  277 \\
\hline
\multirow{6}{*}{\href{https://gracedb.ligo.org/superevents/S200224ca//view/}{S200224ca}}&
\multirow{6}{*}{22:22:34} & \multirow{6}{*}{BBH (>99\%)} & \multirow{6}{*}{$1575 \pm 322$} &
\multirow{6}{*}{72} & GRANDMA & 0.3  & 99.3 & 17-18  & 92.5 & 169 \\
      &  &    & &  & FRAM-Auger & 6.9  &  1.8 & 16.9 & 51.9  & 18    \\
     &  &    & &  & FRAM-CTA & 0.3 & 1.5 & 16.5 & 19.2   & 4.0   \\
   &  &    & &  & TCA  & 0.3 & 99.3 & 18 & 1.4  & 26  \\
      &  &    & &  & TCH  & 5.0 & 52.9  &  18 & 20.9  & 32 \\
      &  &    & &  & TRE  & 0.8 & 98.2  & 17 &  90.6 & 139  \\
\hline
\multirow{6}{*}{\href{https://gracedb.ligo.org/superevents/S200225q/view/}{S200225q}}&
\multirow{6}{*}{06:04:21} & \multirow{6}{*}{BBH (96\%)} & \multirow{6}{*}{$995 \pm 188$} &
\multirow{6}{*}{22} & GRANDMA & 12.0  & 58.2  & 17-18  & 10.2  & 277 \\
      &  &    & &  & FRAM-Auger & 18.1  & 3.8  & 16.4 & 0.1  & 20     \\
     &  &    & &  & FRAM-CTA & 13.8 & 28.7 & 17.2 &  9.7  & 9   \\
   &  &    & &  & TCA  & 12.0 & 58.1 & 18 & 0.4  &  81 \\
      &  &    & &  & TCH  &  18.4 & 51.8  & 18  &  $<0.1$  & 74.0 \\
      &  &    & &  & TRE  & 11.9 & 26.7  & 17 & $<0.1$ & 106  \\
\hline
\multirow{4}{*}{\href{https://gracedb.ligo.org/superevents/S200302c/view/}{S200302c}}&
\multirow{4}{*}{01:58:11} & \multirow{4}{*}{BBH (89\%)} & \multirow{4}{*}{$1820 \pm 536$} &
\multirow{4}{*}{5656} & GRANDMA & 1.0  & 71.5 & 17-18  & 18.0  & 450  \\
      &  &    & &  & FRAM-Auger &  1.0 & 1.9  & 16.1 & 1.6  &  19  \\
   &  &    & &  & TCA  & 17.5 & 54.9 & 18 & 6.3  &  92 \\
      &  &    & &  & TRE  & 17.0 & 28.7  & 17 & 10.9 & 349  \\
\hline
\multirow{1}{*}{\href{https://gracedb.ligo.org/superevents/S200311bg/view/}{S200311bg}} &
\multirow{1}{*}{11:58:53} & \multirow{1}{*}{BBH (100\%)} & \multirow{1}{*}{$1115 \pm 175$} &
\multirow{1}{*}{34} & \multicolumn{6}{c}{no observations, Sun-constrained}  \\
\hline
\multirow{5}{*}{\href{https://gracedb.ligo.org/superevents/S200316bj/view/}{S200316bj}} &
\multirow{5}{*}{21:57:26} & \multirow{5}{*}{MassGap (>99\%)} & \multirow{5}{*}{$1178 \pm 283$} &
\multirow{5}{*}{508} & GRANDMA & 0.3 &  119.9  & 17-18  & 73.3   & 321   \\
  &  &    & &  & FRAM-Auger  & 1.8 & 1.6  & 16.1  & 1.3  & 16 \\ 
 &  &    & &  & FRAM-CTA  & 0.4 & 1.4  & 17.1  & 15.0  & 4 \\
   &  &    & &  & TCA  & 0.3 & 52.5  & 18  & 68.0  & 69  \\
   &  &    & &  & TRE  & 41.7 & 6.4  & 17  & 4.9  & 244 \\
\hline
\end{tabular}
\end{table*}
\begin{table*}
\caption{Summary of the GRANDMA observations during the last five months for BNS or NS - BH merger candidates, using the latest versions of the LALInference sky localisations ($^a$, when only the ``Bayestar sky localisation'' is available). Observations are not necessarily continuous during the time interval. \href{https://gracedb.ligo.org/superevents/S191117j/view/}{S191117j},
\href{https://gracedb.ligo.org/superevents/S191120aj/view/}{S191120aj},
\href{https://gracedb.ligo.org/superevents/S191213ai/view/}{S191213ai},
\href{https://gracedb.ligo.org/superevents/S191220af/view/}{S191220af},
\href{https://gracedb.ligo.org/superevents/S200106au/view/}{S200106au},
\href{https://gracedb.ligo.org/superevents/S200116ah/view/}{S200116ah},
\href{https://gracedb.ligo.org/superevents/S200308e/view/}{S200308e},
were retracted by the LIGO-Virgo collaboration due to data quality issues in the detectors.  90\% c.r. corresponds to the 90\% credible region of the latest sky localisation area sent by LIGO-Virgo, $\delta \, t$ to the delay with respect to the trigger time, $\Delta \, \mathrm{T}$ to the duration of the observations, Prob (\%), Area (${\mathrm{deg^2}}$) to the coverage of GRANDMA compared to the latest revision of the sky localisation area in percentage and in squares degrees. Limiting magnitudes are computed from 5$\sigma$ threshold for source extraction.
}
\label{tab:GRANDMArecapO3BNS}
\begin{tabular}{ccccc|cccccc}
Alert & Time & Type & Dist & 90\% c.r. & Telescope       & $\delta \, t$ & $\Delta \, \mathrm{T}$ & Lim. mag &  Prob & Area   \\
 & (UTC) &  & (Mpc) & (deg$^2$) &  & (h) & (h) &  &  (\%) & (${\mathrm{deg^2}}$)   \\\hline
\multirow{4}{*}{\href{https://gracedb.ligo.org/superevents/S191205ah/view/}{S191205ah}$^a$} &\multirow{4}{*}{21:52:08} 
 &\multirow{4}{*}{NSBH (93\%)}  & \multirow{4}{*}{$385\pm 164$} & \multirow{4}{*}{6378} & GRANDMA & 2.9 & 151.6 &  18 & 4.8  & 158 \\
      &  &    & &  & FRAM-Auger & 8.2 &  1.4  & 16.5 & 0.5  &  20  \\
 &  &  &  &   & TCA  & 18.9 &  133.2 & 18   & 3.0  & 71 \\
   &  &    & &  & TCH  & 2.9 &  151.6 & 18   &  1.2  &  68 \\
\hline
\multirow{4}{*}{\href{https://gracedb.ligo.org/superevents/S191213g/view/}{S191213g}} &\multirow{4}{*}{04:34:08} 
 &\multirow{4}{*}{BNS (77\%)}  & \multirow{4}{*}{$201\pm 81$} & \multirow{4}{*}{4480} & GRANDMA & 0.9 & 120.4 &  18 & 0.6  & 43 \\
      &  &    & &  & FRAM-CTA & 0.9 &  0.3  & 15.2 & $<0.1$  & 1    \\
 &  &  &  &   & TCA  & 47.6 &  73.3 & 18   & 0.5  &  42 \\
   &  &  &  &   & Kilonova-catcher & 39.8 & 13 & 18  & $<1$ & $<1$ \\
 \hline
\multirow{4}{*}{\href{https://gracedb.ligo.org/superevents/S200105ae/view/}{S200105ae}} & \multirow{4}{*}{16:24:26} & \multirow{4}{*}{NSBH (3\%)} & \multirow{4}{*}{$282 \pm 74$} & \multirow{4}{*}{7373} & GRANDMA &27.5  & 130.2  & 17-18  & 12.5  &  356 \\
      &  &    & &  & FRAM-Auger & 60.0 & 1.7 & 15.8  &  0.6   &  20  \\
      &  &    & &  & FRAM-CTA &28.1  & 1.5   & 16.3  &  0.2   & 5   \\
 &  &  &  &   & TCA  &27.5 & 118.2 &  18 &  3.2  & 70  \\
   &  &    & &  & TCH  &59.0 & 98.7  & 18  &  3.3   & 70   \\
      &  &    & &  & TRE  &48.0 & 26.5  & 17  &  9.9   & 295   \\
\hline
\multirow{7}{*}{\href{https://gracedb.ligo.org/superevents/S200213t/view/}{S200213t}} &
\multirow{7}{*}{04:10:40} & \multirow{7}{*}{BNS (63\%)} & \multirow{7}{*}{$201 \pm 80$} &
\multirow{7}{*}{2326} & GRANDMA & 0.4 & 164.8  & 16-21  & 32.8  & 281 \\
      &  &    & &  & FRAM-Auger & 20.3 & 1.3 & 16.0 & 0.1  &  15  \\
     &  &    & &  & FRAM-CTA &  15.3 & 1.5 & 17.3 & 4.0   & 4    \\
   &  &    & &  & TCA  & 0.4 & 163.3  & 18  & 30.4  & 70  \\
      &  &    & &  & TCH  & 45.1  &  120.1 & 18  &  $<0.1$  &  4 \\
      &  &    & &  & TRE  & 12.0 & 31.0  & 17 & 0.8 & 193  \\
            &  &    & &  & OAJ  & 15.0 & 1.5  & 20.1 & 17.5 & 25  \\
 &  &  &  &   & UBAI/NT-60 & 161.8 & 73.9  & 17  & $<1$ & $<1$ \\
  &  &  &  &   & UBAI/ST-60 & 161.0  & 73.1  & 17  & $<1$ & $<1$ \\
 &  &  &  &   & VIRT & 187.3 & 0.5 & 17  & $<1$ & $<1$ \\
  &  &  &  &   & Kilonova-catcher & 22.0 & 120  & 16-19  & $<1$ & $<1$ \\
\hline
\end{tabular}
\end{table*}

\subsection{The binary neutron star merger candidate S191213g}
\label{sec:S191213g}

The GW candidate S191213g was detected on 2019-12-13 04:34:08 UTC \citep{gcn26402}. The trigger was identified by Virgo, LIGO-Hanford, and LIGO-Livingston with a reported false alarm rate of 1.12 per year. It has a non-negligible probability of 23\% of being of terrestrial origin, but assuming its origin is astrophysical, the GWs were likely emitted from a BNS (BNS 77\%).  Using the low-latency analysis \citep{KaCa2019}, if it is astrophysical, it has more than 99\% chance to have a remnant. The low-latency localisation analysis \citep{PhysRevD.93.024013} gave an estimated distance of  $195 \pm 59$ Mpc, and a 90\% credible sky area localisation of 1393 $\text{deg}^2$ \citep{gcn26402}. The next day, the off-line analysis performed by LALInference \citep{PhysRevD.91.042003} gave a distance of $201 \pm 81$ Mpc and a poorer localisation with a 90\% sky area of 4480 $\text{deg}^2$ \citep{gcn26417}. 
As S191213g is considered a potential BNS event, several teams performed follow-up in both neutrino and EM domains. Concerning neutrinos, no candidates were found by IceCube \citep{gcn26399}, ANTARES \citep{gcn26404}, or the Pierre Auger Observatory \citep{gcn26423}. Concerning photons, at high energies, \textit{INTEGRAL} \citep{gcn26401}, \textit{MAXI} \citep{gcn26403}, \textit{AGILE} \citep{gcn26407,gcn26408}, \textit{Fermi}/GBM and LAT \citep{gcn26409,gcn26412}, \textit{Swift} \citep{gcn26410}, \textit{CALET} \citep{gcn26419}, \textit{AstroSat} \citep{gcn26425} and \textit{Insight}-HXMT\citep{gcn26433} did not detect any transients in the 90\% credible region. In the optical, ZTF published two lists of nine and ten candidates, respectively \citep{gcn26424,gcn26437}. All of them were ruled out after spectroscopic vetting or looking at archival data from wide field-of-view telescopes.

\subsubsection{GRANDMA follow-up of the GW alert for S191213g}

Observation plans were sent to 17 GRANDMA telescopes as the event distance enables the search for counterparts located close to interesting galaxies. Unfortunately, the time of the event was relatively close to the end of the European night, and combined with the minimal coverage possible, mostly constrained to the Northern hemisphere; rapid observations by GRANDMA were not an easy task. Due to the low significance of the event, we preferred not to engage our ToO time and instead concentrated our effort on following up some possible EM candidates reported by others. Two of our robotic telescopes performed observations to follow up the event: FRAM-CTA-N and TCA. FRAM-CTA-N started observations 53\,min after the GW trigger in the R$_c$ band and TCA started 2858\,min after the GW trigger with no filter. 
These observations covered about 3.4\% of the cumulative probability of the Bayestar sky localisation area created at 2019-12-13 04:36:53 (UTC), representing only 1\% of the most recently updated LALInference localisation \citep{2019GCN.26417....1L}. 
No significant transient candidate was found during our real time analysis.
We also note that none of the candidates found by other teams were contained in fields observed by FRAM or TCA.

We also conducted follow-up observations via the citizen-scientist programme. As soon as the alert was received, we ranked the 200 most promising host galaxies spatially compatible with the GW skymap following the method described in \cite{2020MNRAS.492.4768D}. This list was promptly distributed to the \textit{kilonova-catcher} community to perform targeted searches for optical transients. The citizen-astronomers were asked to favour the best-ranked galaxies for their observations and to use a clear filter in order to maximise the chances of detection. The idea was to maximise their discovery potential before performing multi-band follow-up observations to confirm an optical transient candidate and study its physical properties. From this citizen observational campaign, we obtained 13 follow-up images, with 16 galaxies localised in the 90\% C.L. probability sky area. The amateur observations started between $1.87-2.2$~days after the GW trigger. All the observations are reported in the Appendix Table \ref{tab:KN_observations}. No optical transient was found at any galaxy position and constraints on their underlying optical flux were derived, resulting in typical unfiltered limiting magnitude of $\mathrm{m_{Lum}=17.6\pm 0.2}$ at 5$\sigma$ confidence.

\subsubsection{GRANDMA follow-up of the electromagnetic counterpart candidate AT2019wxt}

At 2019-12-18, Pan-STARRS published a GCN about a potential counterpart, AT2019wxt/PS19hgw (now named SN 2019wxt), located in the galaxy KUG 0152+311 at a distance of 144 Mpc, and compatible with the GW localization \citep{gcn26485}. Its place in the localisation area and its blue and quite featureless spectrum made SN 2019wxt an interesting kilonova candidate. Eventually, the transient was ruled out by the VLT \citep{gcn26504} due to the discovery of broad helium emission lines which are usually found in type Ib supernovae; such a blue continuum had already been observed in the early phases of type IIb supernovae. The final classification of AT2019wxt as a Type IIb SN was done by LBT \citep{gcn26508}.

On the GRANDMA side, Lisnyky/AZT-8 observed the transient location on 2019-12-18 between 20:39:20 and 21:19:07 (UTC). Stacking the 40 images, the OT was not detected up to an upper limit of 19.0~mag in the R$_C$ band.

The Tingshua-NAOC Telescope (TNT) \citep{2008ApJ...675..626W,2012RAA....12.1585H} was also observing on 2019-12-18 between 16:53:05 and 17:03:05 (UTC) in the $g^\prime$ band, on 2019-12-19 in the $BVg^\prime r^\prime i^\prime$ bands, and on 2019-12-26 in the $g^\prime r^\prime i^\prime$ bands. The source was barely visible because of bad seeing. The upper limit results are reported in the Vega system for the $BV$ bands and in the AB system for the $g^\prime r^\prime i^\prime$ bands in Appendix~\ref{tab:2019wxt_observations}.

The 2.16m telescope located in Xing\-long observed on 2019-12-27. The source was very faint in the $V$ band, making the measurements inaccurate, whereas it was readily visible in the $R$~band. The techniques in photometry employed for both TNT and Xinglong/2.16m were image substraction, with templates that were created by interpolating from the surrounding pixels using the Laplace method.

The Thai National Telescope, equipped with ULTRASPEC, the high speed camera, obtained images in $u^\prime g^\prime r^\prime i^\prime z^\prime $ on 2019-12-20 UTC. After a month it revisited the position again, on 2020-01-16 UTC. The transient magnitude was measured by subtracting the galaxy background using the image of the galaxy obtained after the transient had faded, before doing photometry, with PSF fitting. We obtained detections with $u^\prime=20.72\pm0.29$ mag, $g^\prime=19.60\pm0.18$ mag, $r^\prime =19.58\pm0.30$ mag, $i^\prime=19.34\pm0.23$ mag, and $z^\prime =19.30\pm0.35$ mag, with a median SNR and limiting magnitude of 4.0 and 20.0 mag, respectively.

The citizen astronomers performed observations, and the results are presented in the Appendix in Table \ref{tab:2019wxt_observations}. Three astronomers, M. Serrau, D. St-Gelais and H.-B. Eggenstein sent images of the transient location. The data reduction was done by using references images taken by the same astronomers to perform host subtraction, and the photometric calibration was done using the Pan-STARRS catalogue. For M. Serrau and H.-B. Eggenstein, the transient was not visible to a limiting magnitude of 17.7 in the luminance filter and 18.10 in the $I_C$ band. In D. St-Gelais' image, the transient is visible with a magnitude of 18.8 in the luminance filter. As this is a point source with a well-known position, all limits are given at $3\sigma$ confidence.

\begin{figure*}
\centering
 \includegraphics[width=0.75\textwidth]{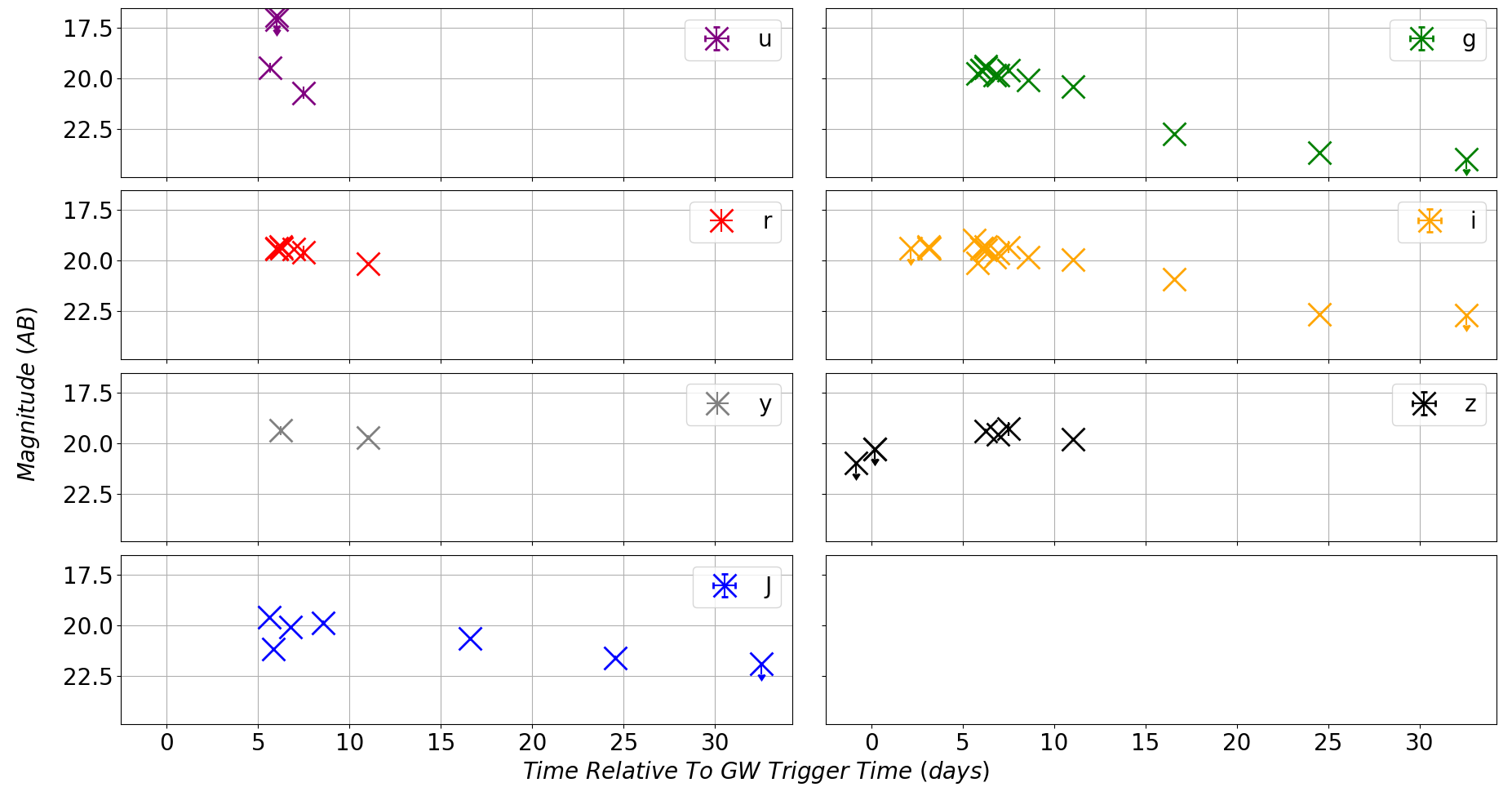}
 \caption{Light curves of SN 2019wxt for the seven bands used for comparison to kilonovae models. $T_0$ corresponds to 2019-12-13 04:33:36.000 UTC \citep{gcn26402}, GW trigger time. Note that all bands are not shown in this figure but are available in Appendix~\ref{tab:2019wxt_observations}, with measurements from both GRANDMA (including the \textit{kilonova-catcher} contributions) and GCNs.}
\label{fig:SN2019wx_lc}
\end{figure*}

A lightcurve of SN2019wxt with photometric points reported in GCNs and with the GRANDMA measurements can be seen in Figure~\ref{fig:SN2019wx_lc}. We conducted an independent investigation for determining the possible nature of the AT2019wxt using the photometric evolution of the light curve in the $g$- and $r$-bands. We use the method proposed in \citet{Stachie:2019nae}; it relies on a deep-learning algorithm \citep{Muthukrishna:2019wgc}, which was previously trained on ZTF-like transient lightcurves (see a more extended description of the training data set in these references). Each data was classified in different categories: ``SN'' (supernovae such as SNIa, SNIb, SNIc, SNII), ``KN'' (kilonova) and ``Others'' (regrouping various types of transients such as active galactic nuclei, tidal disruption events). The algorithm also reports as ``Indistinguishable'' cases when either there is not enough data (few observations) or the classifier does not recognize the lightcurve. As demonstrated in \citet{Stachie:2019nae}, the algorithm is efficient for classify a wide variety of publicly reported ZTF supernovae during the testing sequence. By extension, the simulated kilonovae that can be observed by ZTF are well-classified by this procedure, with an efficiency higher than 50\% after around 10 observations. In the case of AT2019wxt, the situation is more complex due to a heterogeneous dataset made by various observatories different than ZTF (whereas training was only done with ZTF), with different photometric measurement techniques, irregular time of observations and the lack of further measurements after 20 days. However, about 20 data points in $g$- and $r$-bands (see in the Appendix~\ref{tab:2019wxt_observations}) have been used; the algorithm classified the transient as ``Indistinguishable'' after each new observation, and does not favor a kilonova classification. Given the extensive observations, the non-identification of the transient likely indicates that AT2019wxt is a peculiar transient.

We also compare this lightcurve to expected kilonovae models. To do so, we employ a Gaussian process regression based surrogate model~\citep{DoFa2017,CoDi2018,CoDi2018b,CoDi2019} that employs pre-computed kilonova light curves modelled with the full radiative-transfer Monte-Carlo code of~\cite{KaMe2017}. This allows us to compute arbitrary light curves for generic ejecta properties/configurations. This analysis finds that at the distance of the host galaxy, the ejecta masses required approach 0.1\,$M_\odot$ and beyond, higher than expected in BNS mergers. In addition, even a 2-component model, for the parameter space considered, is unable to reproduce the $u$- and $g$-band excesses seen in this particular transient, although the evolution in the redder bands is broadly consistent. In this way, a kilonova explanation for this transient is disfavored within the expected parameter space for these types of transients. 

\begin{figure}
\centering
 \includegraphics[width=\columnwidth]{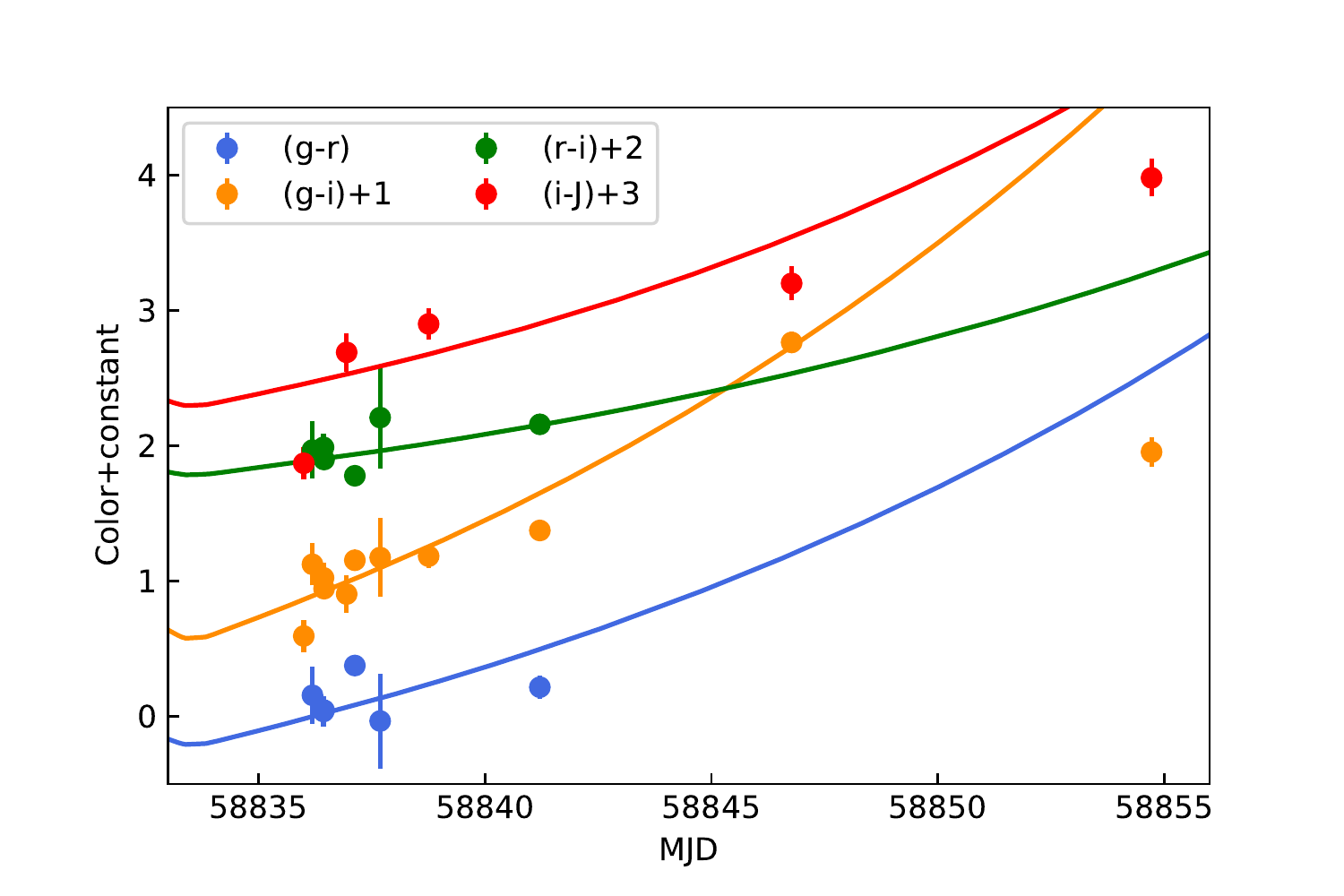}
 \caption{The colour evolution of AT 2019wxt. Note that the GW trigger time corresponds to MJD 58830.190. The circles correspond to the observed colours that are corrected for Galactic extinction. The solid lines correspond to the fitting results based on the Interaction model. }
\label{fig:SN2019wxt_lcfit}
\end{figure}

The rapidly-declining light curves of SN 2019wxt suggest a possibility that this supernova could be powered by the interaction of the ejecta with circumstellar materials (CSM; e.g. \citealp{2012ApJ...746..121C, 2013ApJ...773...76C, 2019MNRAS.489.1110W}). Assuming that the CSM has a stellar-wind density profile (i.e. $\rho\propto r^{-2}$) and that the density is uniform for the inner ejecta and follows $\rho\propto r^{-7}$ for the outer ejecta, a fit to the multi-band data (corrected for Galactic extinction) of this supernova is obtained with $\chi^2$/d.o.f.$=43$. The corresponding parameters include $T_0=$58830.190, $M_\mathrm{ej}=0.87M_\odot$, $v_\mathrm{ej}=9.33\times10^{8}$ cm~s$^{-1}$, $M_\mathrm{CSM}=0.16M_\odot$, $\rho_\mathrm{CSM,in}=8.25\times10^{-12}$ g~cm$^{-3}$, $r_\mathrm{CSM,in}=1.8\times10^{14}$ cm and $\epsilon=0.1$, where $T_0$ is the explosion date, $M_\mathrm{ej}$ is the mass of ejecta, $v_\mathrm{ej}$ is the velocity of ejecta, $M_\mathrm{CSM}$ is the mass of CSM, $\rho_\mathrm{CSM,in}$ is the density at the inner radius of the CSM ($r_\mathrm{CSM,in}$) and $\epsilon$ is the radiation efficiency. As seen in Figure~\ref{fig:SN2019wxt_lcfit}, we compare the fit to the results with the evolution in different colours, which could minimise the effect of the difference in observed brightnesses from different telescopes. If the CSM results from a mass ejection of the progenitor star before explosion, the mass-loss rate can be estimated as $0.05/(v_\mathrm{CSM}/10$~km~s$^{-1}$) $M_\odot$~yr$^{-1}$, where $v_\mathrm{CSM}$ is the velocity of the CSM. Wolf-Rayet stars or red supergiant stars have difficulty in producing such intense ejections; while binary interaction can possibly be responsible for mass ejection of $10^{-3}-10^{-1} M_\odot$ yr$^{-1}$ (see \citealp{2014ARA&A..52..487S, 2019MNRAS.488.3783B} and references therein). Therefore, SN 2019wxt could possibly originate from an explosion of a binary star.

\subsection{The unmodeled candidate S200114f}

The GW candidate S200114f, an unmodeled transient trigger, occurred on 2020-01-14 02:08:18 UTC \citep{gcn26734}. The candidate was found by the Coherent Wave Burst analysis pipeline with a version of the search tagged as ``IMBH'' \citep{2020GCN.26734....1L,PhysRevD.93.042004}. This event was observed by Virgo, LIGO-Livingston and LIGO-Hanford. Its false alarm rate was about one in 25 years. The 90\% credible region of the sky localisation area is 403 $\text{deg}^2$; no further update on this trigger has been circulated to date. No distance estimate was provided. It is the only ``burst'' alert transmitted during O3. By comparison, 5 ``burst'' alerts were sent during O2, but these triggers were either retracted or found to be consistent with noise \citep{2019arXiv190103310T}. The event S200114f is interesting since none of the modeled search pipelines produced triggers during the same time period. Since no neutrinos (high or low energy) were recorded around the time of the trigger \citep{gcn26732,gcn26742,2020GCN.26751....1C}, a galactic supernova is not likely to be the origin. If the event is astrophysical, a possible scenario for the GW emission is a massive BBH coalescence for which the inspiral phase is at too low of frequency and cannot be detected by LIGO and Virgo \citep{Salemi:2019ovz}. A GW signal from a cosmic string is a more exotic possibility \citep{Abbott:2017mem}, as is GW emission from an unknown source \citep{Abbott:2019prv}.
 
Several teams followed the GW burst candidate S200114f and about 30 GCN circulars were sent. According to the follow-up observations, no neutrino events have been reported in spatial coincidence with S200114f by IceCube, which collected good quality data during the time the event occurred \citep{gcn26732}, or by ANTARES, which had only a 6.0\% chance that the GW source was in its field-of-view \citep{gcn26742}; nor has any counterpart been reported by a gamma-ray space instrument, such as \textit{Fermi}/LAT \citep{gcn26745}. SAGUARO observed 180 deg$^2$ of the S200114f sky localisation area starting from 2.1\,hr after the trigger time with a typical limiting magnitude of 21. Five candidates were reported but they are unlikely to be related to the trigger.  All of these are possibly variable stars or AGNs \citep{gcn26753}. \textit{Swift} reported a possible X-ray transient source, S200114f-X2, which is coincident with a candidate AGN, and therefore this rise in flux may indicate nothing more than AGN activity and variability \citep{gcn26787,gcn26791}.

\subsubsection{GRANDMA observations}

Five GRANDMA telescopes participated in the follow-up of S200114f, as shown in Table~\ref{tab:GRANDMArecapO3Burst}, as soon as they received the GRANDMA observation plan (see section~\ref{strategy}). Time delays for FRAM between the first observation and the trigger time were as follows: FRAM-Auger -– 19 min, FRAM-CTA-N -– 20 min; observations were obtained in the $R_C$ filter. The TAROT network observed the most probable region about 15~hr after the alert due to technical problems during the first night. The respective time delays are: TCA -- 934 min, TRE -- 901 min, TCH -- 908 min; images were obtained without a filter. These observations covered about 77\% of the sky localisation area in the following 90~hr for a total sky coverage of 351 deg$^2$, as shown in Figure \ref{skycoveragelink} in the Appendix. Note that FRAM-Auger observed 28\% of the sky localisation area in the first two hours after the GW trigger time for a total coverage of 20 deg$^2$. GRANDMA observations were reported in \citet{gcn26875} and no significant transient candidates were found with our low-latency analysis. We also conducted further analysis on our FRAM images on the two lists of ZTF candidates mentioned above. In particular, ZTF20aafemdh and ZTF20aafeogg were observed by FRAM-Auger 29~hr and 0.5~hr before ZTF observations in $g^\prime r^\prime $ bands to a depth of 21 mag. We did not detect any point source at $5\sigma$ confidence in the $R_C$ band to 16.2~mag. ZTF20aafemxx was also not detected to 16.2~mag in the $R_C$ band by FRAM-CTA 1.5~hr before the ZTF detection at 22 mag in the $r^\prime $-band.

\begin{table}
\caption{Summary of the GRANDMA observations of the \href{https://gracedb.ligo.org/superevents/S200114f/view/}{S200114f} unmodeled (``burst'') GW candidate detected on 14, January, 2020 at 02:08:18 UT. The 90\% credible region of the event spans 403 deg$^2$. An addition burst candidate \href{https://gracedb.ligo.org/superevents/S191110af/view/}{S191110af} was retracted by the LVC due to a data quality issue in the detectors.  $\delta \, t$ corresponds to the delay with respect to the trigger time, $\Delta \, \mathrm{T}$ to the duration of the observations, Prob (\%), Area (${\mathrm{deg^2}}$) to the coverage of GRANDMA compared to the sky localisation area in percentage and in squares degrees. Limiting magnitudes are computed from 5$\sigma$ threshold for source extraction.
}
\label{tab:GRANDMArecapO3Burst}
\begin{tabular}{cccccc}
 Telescope       & $\delta \, t$ & $\Delta \, \mathrm{T}$ & Lim. mag &  Prob & Area   \\
 & (h) & (h) &  &  (\%) & (${\mathrm{deg^2}}$)   \\\hline
 GRANDMA & 0.3 & 90.7 &  17-18 & 76.4  & 351  \\
 FRAM-Auger & 0.3 &  1.7  & 16.2 &   28.5 &   19 \\
 FRAM-CTA & 0.3  & 1.7  & 16.2  &     7.4 & 5  \\
 TCA  & 15.6 &  76.1  & 18   & 13.4  & 11.2  \\
 TCH  & 15.1 & 35.6  & 18   &  14.0 &   39 \\
 TRE  & 15.0 & 31.4  & 17   & 68.2  & 311  \\
\hline
\end{tabular}
\end{table}

\subsection{The binary neutron star merger candidate S200213t}
\label{sec:S200213t}
The GW candidate S200213t was identified on 2020-02-13 04:10:40.328 UTC \citep{gcn27042}. It is a three-detector trigger event, being observed by Virgo and both LIGO-Livingston and LIGO-Hanford. S200213t was reported to have a false-alarm rate of about one in 1 year and 9 months. It has a non-negligible probability of 37\% to be a terrestrial artifact, whereas the probability of being a BNS is 63\%.

Nevertheless, assuming the candidate is astrophysical in origin, the probability of the presence of remnant matter is higher than 99\%. The first evaluation of the distance and the first estimate of 90\% credible sky localisation area provided by the realtime analysis \citep{KaCa2019} were $224\pm90$ Mpc and 2587 $\text{deg}^2$. These two quantities were updated three days later as follows \citep{gcn27096}: the luminosity distance is $201\pm80$ Mpc and the 90\% credible region to 2326 $\text{deg}^2$. The most updated sky localisation using LALInference is shown in Figure~\ref{fig:GRANDMAS200213tfull}, both the full sky coverage and a zoomed-in panel. 

Given the possible BNS nature of S200213t, the follow-up campaign performed by multiple groups was extensive, producing more than 50 GCNs. A muon neutrino candidate was detected by IceCube 175.94 seconds prior to GW trigger, which overlapped with the sky localisation of S200213t \citep{gcn27043}. No gamma-ray transients were reported by \textit{MAXI} \citep{2020GCN.27048....1K}, \textit{INTEGRAL} \citep{2020GCN.27050....1R}, and Insight-\textit{HXMT} \citep{2020GCN.27071....1Z}. \textit{Swift}/BAT and Fermi were not in science mode \citep{2020GCN.27058....1S, 2020GCN.27062....1C, 2020GCN.27056....1V}. In the optical bands, galaxy targeted observations were performed by KAIT, observing 108 galaxies within the 90\% probability region \citep{2020GCN.27064....1V}, J-Gem, observing a total of 37 galaxies \citep{2020GCN.27066....1K}, GECKO, observing four galaxies \citep{2020GCN.27067....1P}, and NAOC/CAS, observing 16 galaxies \citep{2020GCN.27070....1X} compatible with the sky localisation volume of S200213t. ZTF covered almost 80\% of the Bayestar sky localisation area in one night to a magnitude limit of 21~mag in the $g^\prime r^\prime $ bands. ZTF also reported, in three different circulars \citep{gcn27051, gcn27065, gcn27068}, a total of 15 candidates. ZTF provided a re-evaluation of the candidates after the delivery of an updated sky localisation area, leaving seven candidates \citep{gcn27095}. \cite{gcn27118, gcn27154, gcn27140} showed that four of these candidates were unrelated to S200213t and offered some non-detection upper-limits on a fifth candidate. MASTER also claimed a candidate in \cite{gcn27072} and another one in \cite{gcn27077}. The last candidate was proven to be unrelated to S200213t by \cite{gcn27085}. DDOTI/OAN covered $\approx50$\% of the Bayestar sky localisation area down to a limiting magnitude of 19.2 mag in the $w$ band and less than an hour after the trigger time \citep{2020GCN.27061....1W}. GOTO also immediately covered $\approx54$\% of the Bayestar sky localisation area, down to a limiting magnitude of 18.4 mag in the $L$ band \citep{2020GCN.27069....1C}.

\subsubsection{GRANDMA observations}

\begin{figure*}
\begin{center}
    \centering
    \subfloat[]{{\includegraphics[width=0.92\textwidth]{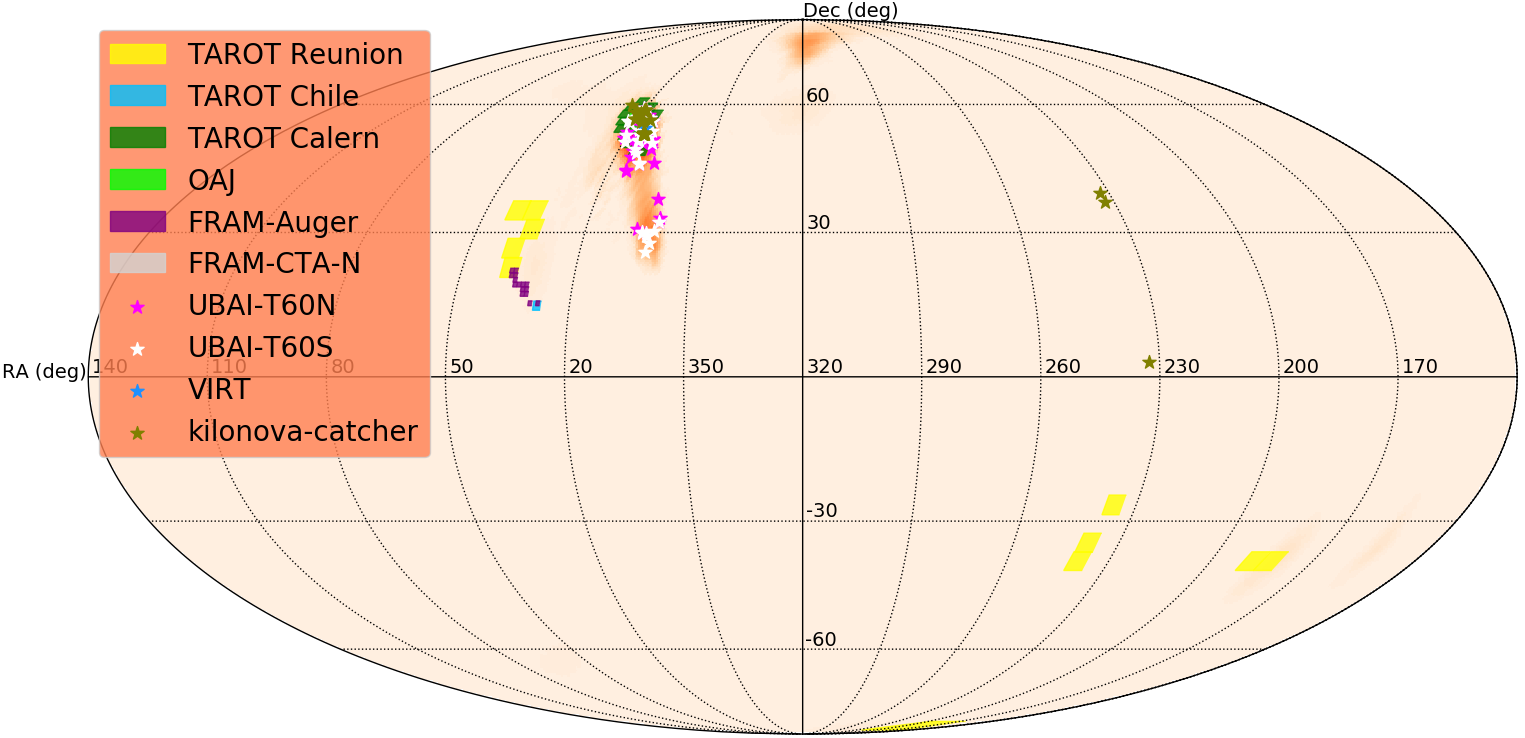} }}%
    \qquad
    \subfloat[]{{\includegraphics[scale=0.63]{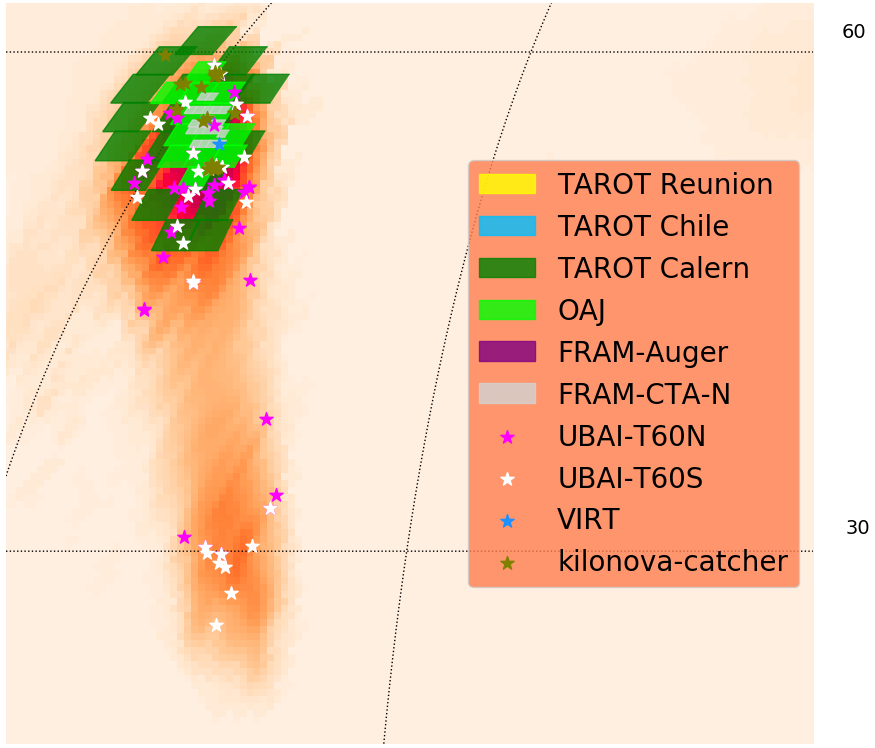} }}%
\caption{GRANDMA follow-up of the GW candidate S200213t, a BNS merger candidate. Yellow, blue and dark green tiled areas represent observational tiles obtained by the TAROT network. Purple and grey tiled areas represent observational tiles obtained by the FRAM network. Light green tiled areas represent observational tiles obtained by OAJ. In red, the LALInference sky localisation area of S200213t is shown. We note that TRE tiles covered 193 deg$^2$ yet only 1\% of the final sky area localisation. Stars represent galaxy-targeting fields obtained by UBAI and VIRT (several days after the GW trigger time) and the citizen science program \textit{kilonova-catcher} (with first images taken a few hours after the GW candidate trigger time).}
\label{fig:GRANDMAS200213tfull}
\end{center}
\end{figure*}



The first notice was released 25 min after the trigger. Five wide field-of-view telescopes participated in the follow-up : TCA, TRE, OAJ, FRAM-CTA-N, FRAM-Auger and TCH. At the time of the trigger, the FRAM network, TCA and TCH had the capacity to observe but due to bad weather or technical issues did not; OAJ requested to trigger the target of opportunity program, but it would not have been performed before the end of the night. Only TCA responded quickly to the alert. The first image was taken 26~min after the alert for TCA, 717 min for TRE, 898 min for OAJ, 932 min for FRAM-CTA-N (second night), 1220 min for FRAM-Auger (second night), and 2703 min for TCH (third night). These observations account for about 33\% of the cumulative probability of the LALInference sky localisation area and are shown in Figure~\ref{fig:GRANDMAS200213tfull}. This sky localisation area was covered by TCA in less than 24\,hr. No significant transient candidates were discovered \citep{gcn27116}.

The observational campaign was complemented with three days of optical follow-up from two 60\,cm UBAI telescopes (UBAI-T60S and UBAI-T60N). The observations were performed in the galaxy targeting mode 6.7 days after the GW alert, with a delay due to bad weather conditions. We obtained photometric observations in the $R_C$ filter reaching a limiting magnitude of 17~mag at $5\sigma$ confidence, using the list of fields provided for the most updated version of the sky localisation area. The 16 fields observed contained a total of 47 galaxies observed by UBAI-T60N and 39 galaxies observed by UBAI-T60S, but no significant transient candidate was found. VIRT also observed two fields containing galaxies of interest located in the sky localisation area starting at 2020-02-21 00:00:37, e.g. 188\,hr after the trigger time. The delay was due to poor weather conditions as well. The photometric analysis of the two fields gave an upper limit in the $R_C$-band at $5\sigma$ confidence of 17\,mag .

From the GRANDMA \textit{kilonova-catcher} community, we obtained 34 images containing 29 galaxies localised in the 90\% C.L. sky localisation area. The imaging latencies with respect to the GW trigger times were $0.91-4.96$ days. We summarise the \textit{kilonova-catcher} observations in Table \ref{tab:KN_observations}. No optical transient was found at any galaxy position and constraints on their underlying optical flux were derived, resulting in unfiltered limiting magnitudes spanning $\mathrm{m_{Lum} = 17.7\pm 0.7}$ mag at $5\sigma$ confidence. 

No significant candidates were found by our initial analysis when crosschecking with databases of known objects or transients (see section~\ref{strategy}). GRANDMA also observed some interesting counterpart candidates mentioned above. The independent observations of GRANDMA covered most of the candidates presented in \cite{gcn27051, gcn27065, gcn27068,gcn27077}.  For example, ZTF20aapvtip and ZTF20aamvoxx were observed by TAROT-Reunion but at a lower sensitivity and hours after ZTF. VIRT performed targeted observations of ZTF20aamvnth (AT2020cjb) and ZTF20aamvmzj (AT2020cja) and reported upper limits on the two sources: 17.8 mag (in the $Rc$ band at 5$\sigma$ confidence) in data taken from 2020-02-13 23:31:11 to 2020-02-14 00:46:03 and 17.8 mag (in the $R_C$ band) on data taken from  2020-02-14 00:48:27 to 2020-02-14 01:11:05. Two months after the event, we conducted further analysis on OAJ observations with Gmadet (see section~\ref{strategy}). Among the $\sim$\,1800 transient candidates, we selected the most 200 promising events. We found an interesting transient which we classified as a moving object based on a near-in-time upper limit from the Zwicky Transient Facility.
However, the upper limits obtained with TAROT (with data taken the first night) and OAJ (with data taken the second night) can be used to constrain the ejecta properties of potential kilonova that may have occurred in the vicinity of the observed galaxies. These are described in more details in the following Sec.~\ref{sec:kilonova_properties}.

\subsubsection{Constraining kilonova properties}

\label{sec:kilonova_properties}
\begin{figure}
\centering
 \includegraphics[width=\columnwidth]{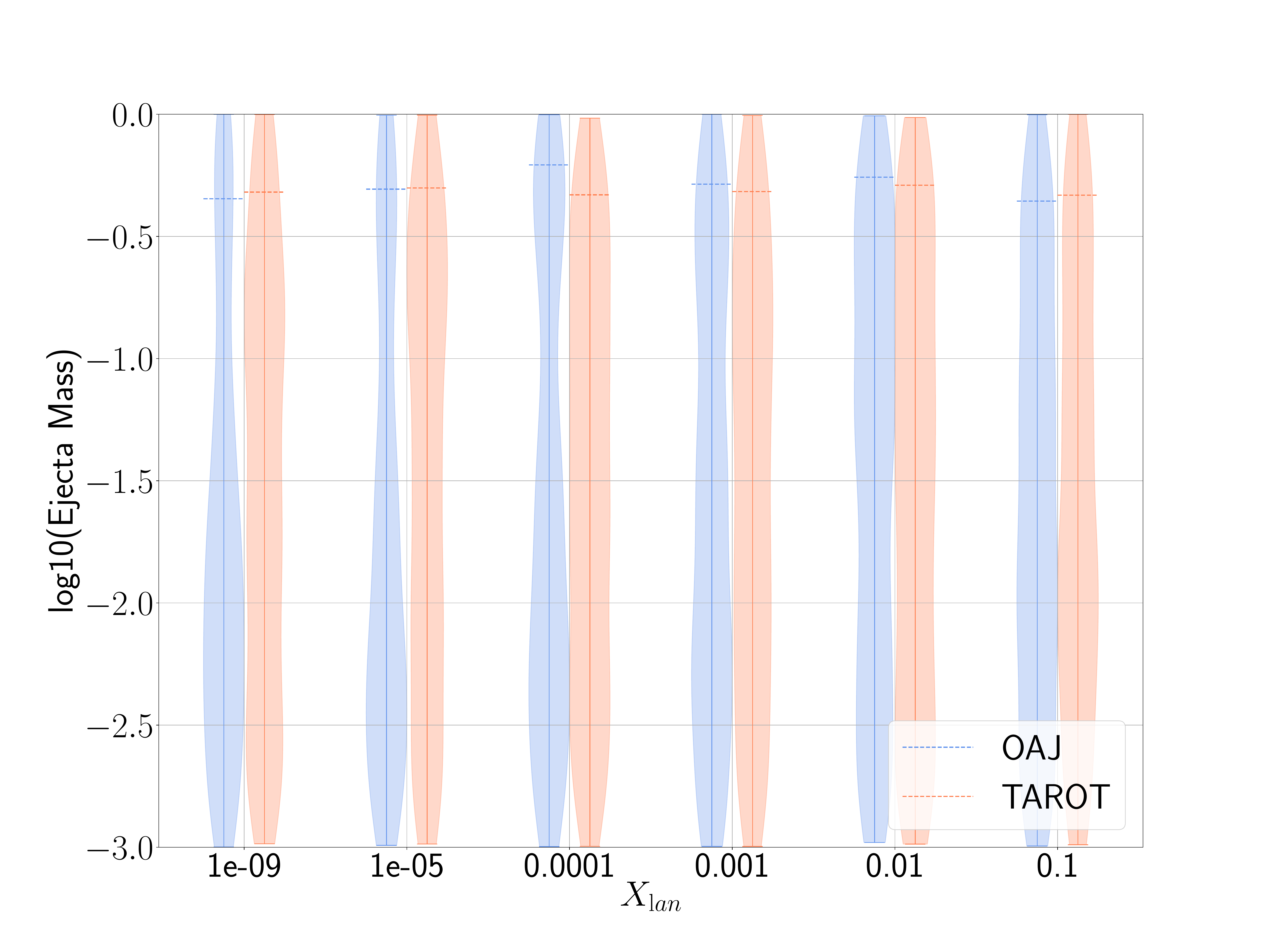}
 \caption{Constraints on the ejecta mass in terms of lanthanide fractions $X_{\rm lan}$ for the BNS candidate S200213t based on the OAJ and TAROT observations. The thickness of the vertical bands represents the probability density for this particular ejecta mass. Thick regions mark more probable regions of the parameter space, thin regions less probable ejecta masses. Overall, we find that for lower lanthanide fractions $X_{\rm lan}$ we are able to disfavour high ejecta masses, while for large lanthanide fractions no real constraints can be extracted. 
 We mark the 90\% upper limits as horizontal dashed lines, where we clearly see that for 
 most scenarios our upper bounds are reflecting the prior.}
\label{fig:S200213t_ejecta}
\end{figure}

We seek now to derive constraints on the observed GW transients from GRANDMA observations. As an example, we will focus in particular on the BNS candidate S200213t. 
Within the GRANDMA network, a number of telescopes searched for a kilonova counterpart to this event, among these OAJ which covered part of the sky localisation area to a magnitude $r^\prime>20.1$ mag (at 5$\sigma$ detection level),TAROT to a limiting magnitude of $CR>18$ mag, and a number of individual galaxy-targeted searches that also reached limiting magnitudes around 18~mag (see Table \ref{tab:KN_observations} in the Appendix). 
Under the assumption that TAROT (which achieved 30\% coverage of the relevant sky area) or OAJ (which achieved 18\% coverage of the relevant sky area) covered the relevant sky location, we derive constraints on the ejecta mass consistent with the non-detection of any kilonova.

For a quantitative interpretation, we follow~\cite{GRANDMAO3A} and \cite{Coughlin2019}, we derive generic ejecta masses $M_{\rm ej}$, ejecta velocities $v_{\rm ej}$, and lanthanide fractions $X_{\rm lan}$ that are consistent with the non-observation of a counterpart in our observational searches. Fig.~\ref{fig:S200213t_ejecta} summarises our results. Unfortunately, independent of the lanthanide fraction of the ejected material, we are not able to derive a constraint on the ejecta mass. 
Even very large ejecta masses of the order of $~\sim 0.4M_\odot$, that are typically disfavoured by numerical-relativity simulations~\cite{Dietrich:2016fpt}, cannot be ruled out. For comparison, the ejecta mass of GW170817 is estimated to be $\sim0.05M_\odot$, i.e., about one order of magnitude smaller. 
Furthermore, the exact 90\% upper bound on the ejecta mass (vertical lines in Fig.~\ref{fig:S200213t_ejecta}) seems to be prior-dominated, which is another indicator that no reliable information can be extracted.
Therefore, it seems likely that just from our observations, sources compatible to GW170817 would have been missed due to the significantly larger distance of S200213t, $\sim$ five times further away.




\section{Conclusions}
\label{conclusions}

In conclusion, the coordination of joint observations across a telescope network, above and beyond just recording independent observations from various groups, is an important goal for deriving multi-messenger science with EM and GW data. Within GRANDMA, we have performed such joint observations whereby we obtain a combination of repeated and split observations over the telescopes in our network, including scheduling them to perform observations in different bands at different times in order to make it possible to potentially determine colour and luminosity evolution of any transients that we might discover. GRANDMA has demonstrated the coordination of observations from a very diverse array of facilities; it has also achieved significant success in terms of coverage of the GW sky localisation areas in a variety of bands and to a variety of limiting magnitudes. In this paper, we particularly focused on the second part of O3 and the two BNS merger candidates S191213g and S200213t. While we derived constraints on potential kilonova properties of S200213t, thanks to the active follow-up by six wide field of view telescopes of the network, we also targeted follow-up of a candidate optical counterpart which turned out to be a Type IIb SN with possible evidence of CSM interaction.

GRANDMA has followed-up about 90\% of the GW alerts during O3, covering about 50 events, including BBHs, to demonstrate our global coordination system. GRANDMA is composed of 25 telescopes inspired to contribute to this global effort, and remains a flexible and open collaboration. We have also highlighted the GRANDMA's unique citizen science program, involving, for the first time, amateur astronomers following-up potential GW events. Given the number of GW triggers, optical transients, and the necessary data reduction procedures, such a collaboration needs dedicated human organisation, data analysis, and protocol communication to manage targets of observations obtained by many different telescopes. In this sense, GRANDMA is well-prepared for the next GW observing run O4 with possible rates of one transient per day.



\section*{Acknowledgements}
Parts of this research were conducted by the Australian Research Council Centre of Excellence for Gravitational Wave Discovery (OzGrav), through project number CE170100004. EJH acknowledges support from an Australian Research Council DECRA Fellowship (DE170100891).
AdUP and CCT acknowledge support from Ram\'on y Cajal fellowships RyC-2012-09975 and RyC-2012-09984 and the Spanish Ministry of Economy and Competitiveness through project AYA2017-89384-P. DAK acknowledges Spanish research project RTI2018-098104-J-I00 (GRBPhot).  MB acknowledges funding as ``personal tecnico de apoyo'' under fellowship number PTA2016-13192-I. SA is supported by the CNES Postdoctoral Fellowship at Laboratoire AstroParticule et Cosmologie. SA and CL acknowledge the financial support of the Programme National Hautes Energies (PNHE). DT acknowledges the financial support of CNES post-doctoral program. UBAI acknowledges support from the Ministry of Innovative Development through projects FA-Atech-2018-392 and VA-FA-F-2-010. SB acknowledges Shota Rustaveli National Science Foundation (SRNSF) grant No - PHDF/18-1327. TAROT has been built with the support of the Institut National des Sciences de l'Univers, CNRS, France. TAROT is funded by the CNES and thanks the help of the technical staff of the Observatoire de Haute Provence, OSU-Pytheas. MP, SK and MM are supported by European Structural and Investment Fund and the Czech Ministry of Education, Youth and Sports (Projects CZ.02.1.01/0.0/0.0/16\_013/0001402, CZ.02.1.01/0.0/0.0/16\_013/0001403 and CZ.02.1.01/0.0/0.0/15\_003/0000437). NBO, DM, and PG acknowledge financial support from NASA-MUREP-MIRO grant NNX15AP95A, NASA-EPSCoR grant NNX13AD28A, and NSF EiR AST Award 1901296. The GRANDMA collaboration thank the amateur participants to the \textit{kilonova-catcher} program. The \textit{kilonova-catcher} program is supported by the IdEx Université de Paris, ANR-18-IDEX-0001. This research made use of the cross-match service provided by CDS, Strasbourg. We thank Ulrich Hopp to provide the precise date of observations for AT2019wxt Wendelstein optical observations.

\bibliographystyle{mnras}
\bibliography{references}

\section{Sky Coverage Links}
\label{skycoveragelink}

\begin{table*}
\caption{Sky coverage observations of the GRANDMA collaboration during the last five months.}
\begin{tabular}{ccc}
\hline
GW trigger & GRANDMA sky localisation coverage Link & GCN report \\
\hline
S191105e &  \href{https://grandma-owncloud.lal.in2p3.fr/index.php/s/FnfqEwZtgcSxeLl/download}{https://grandma-owncloud.lal.in2p3.fr/index.php/s/FnfqEwZtgcSxeLl/download} & \citet{gcn26215} \\ 
S191109d &  \href{https://grandma-owncloud.lal.in2p3.fr/index.php/s/utGS2Zigv2qGp8S/download}{https://grandma-owncloud.lal.in2p3.fr/index.php/s/utGS2Zigv2qGp8S/download} & \citet{gcn26219} \\ 
S191129u &  \href{https://grandma-owncloud.lal.in2p3.fr/index.php/s/YMk334EKRjSfnGD/download}{https://grandma-owncloud.lal.in2p3.fr/index.php/s/YMk334EKRjSfnGD/download} & \citet{gcn26362}  \\ 
S191204r &  \href{https://grandma-owncloud.lal.in2p3.fr/index.php/s/PqlwSoGEclSr2VJ/download}{https://grandma-owncloud.lal.in2p3.fr/index.php/s/PqlwSoGEclSr2VJ/download} & \citet{gcn26361}  \\ 
S191205ah &  \href{https://grandma-owncloud.lal.in2p3.fr/index.php/s/4AJrPrBKtZRWSyA}{https://grandma-owncloud.lal.in2p3.fr/index.php/s/4AJrPrBKtZRWSyA} & \citet{gcn26384}  \\ 
S191213g &  \href{https://grandma-owncloud.lal.in2p3.fr/index.php/s/mK004QVEHlLkh5V}{https://grandma-owncloud.lal.in2p3.fr/index.php/s/mK004QVEHlLkh5V} & \citet{gcn26558}  \\ 
S191215w & \href{https://grandma-owncloud.lal.in2p3.fr/index.php/s/FdAcqmoFJ8U3dAS}{https://grandma-owncloud.lal.in2p3.fr/index.php/s/FdAcqmoFJ8U3dAS} & \citet{gcn26469}  \\ 
S191216ap & \href{https://grandma-owncloud.lal.in2p3.fr/index.php/s/YYGlFk1mRHMzCTr}{https://grandma-owncloud.lal.in2p3.fr/index.php/s/YYGlFk1mRHMzCTr} & \citet{gcn26528}  \\ 
S191222n & \href{https://grandma-owncloud.lal.in2p3.fr/index.php/s/gVU51ziBqxCVYpK}{https://grandma-owncloud.lal.in2p3.fr/index.php/s/gVU51ziBqxCVYpK} & \citet{gcn26575} \\ 
S200105ae & \href{https://grandma-owncloud.lal.in2p3.fr/index.php/s/xrQjLfudAshgdYt}{https://grandma-owncloud.lal.in2p3.fr/index.php/s/xrQjLfudAshgdYt} & \citet{gcn26687} \\ 
S200112r & \href{https://grandma-owncloud.lal.in2p3.fr/index.php/s/cR4lmCGjlUjiQOJ}{https://grandma-owncloud.lal.in2p3.fr/index.php/s/cR4lmCGjlUjiQOJ} & \citet{gcn26789} \\ 
S200114f & \href{https://grandma-owncloud.lal.in2p3.fr/index.php/s/xA8XBol3EVWckpq}{https://grandma-owncloud.lal.in2p3.fr/index.php/s/xA8XBol3EVWckpq} & \citet{gcn26875} \\ 
S200115j & \href{https://grandma-owncloud.lal.in2p3.fr/index.php/s/0zzUckHMfewfW4I}{https://grandma-owncloud.lal.in2p3.fr/index.php/s/0zzUckHMfewfW4I} & \citet{gcn26820} \\  
S200128d & \href{https://grandma-owncloud.lal.in2p3.fr/index.php/s/LNfiDNPraReH0nM}{https://grandma-owncloud.lal.in2p3.fr/index.php/s/LNfiDNPraReH0nM} & \citet{gcn26947} \\  
S200129m & \href{https://grandma-owncloud.lal.in2p3.fr/index.php/s/OYkJW42IL6QXF67}{https://grandma-owncloud.lal.in2p3.fr/index.php/s/OYkJW42IL6QXF67} & $-$ \\  
S200208q & \href{https://grandma-owncloud.lal.in2p3.fr/index.php/s/INCbvLAPi6SQqyO}{https://grandma-owncloud.lal.in2p3.fr/index.php/s/INCbvLAPi6SQqyO} & \citet{gcn27054} \\  
S200213t & \href{https://grandma-owncloud.lal.in2p3.fr/index.php/s/AwWEotlQ2Erf282}{https://grandma-owncloud.lal.in2p3.fr/index.php/s/AwWEotlQ2Erf282} & \citet{gcn27116} \\ 
S200219ac & \href{https://grandma-owncloud.lal.in2p3.fr/index.php/s/mSj3R8GlO28kI4v}{https://grandma-owncloud.lal.in2p3.fr/index.php/s/mSj3R8GlO28kI4v} & \citet{gcn27168} \\  
S200224ca & \href{https://grandma-owncloud.lal.in2p3.fr/index.php/s/SJ1bZVKqoTkD52K}{https://grandma-owncloud.lal.in2p3.fr/index.php/s/SJ1bZVKqoTkD52K} & \citet{gcn27238} \\  
S200225q & \href{https://grandma-owncloud.lal.in2p3.fr/index.php/s/CKWTsw6dUdAEgdB}{https://grandma-owncloud.lal.in2p3.fr/index.php/s/CKWTsw6dUdAEgdB} & \citet{gcn27275} \\  
S200302c & \href{https://grandma-owncloud.lal.in2p3.fr/index.php/s/M2iREd4e61fkXCt}{https://grandma-owncloud.lal.in2p3.fr/index.php/s/M2iREd4e61fkXCt} & \citet{gcn27322} \\  
S200316bj & \href{https://grandma-owncloud.lal.in2p3.fr/index.php/s/ejyjwHrAGJNp0eY}{https://grandma-owncloud.lal.in2p3.fr/index.php/s/ejyjwHrAGJNp0eY} & \citet{gcn27409} \\  
\hline
\end{tabular}
\end{table*}

\begin{figure*}
\subfloat[S191105e]{\includegraphics[width = 3.4in]{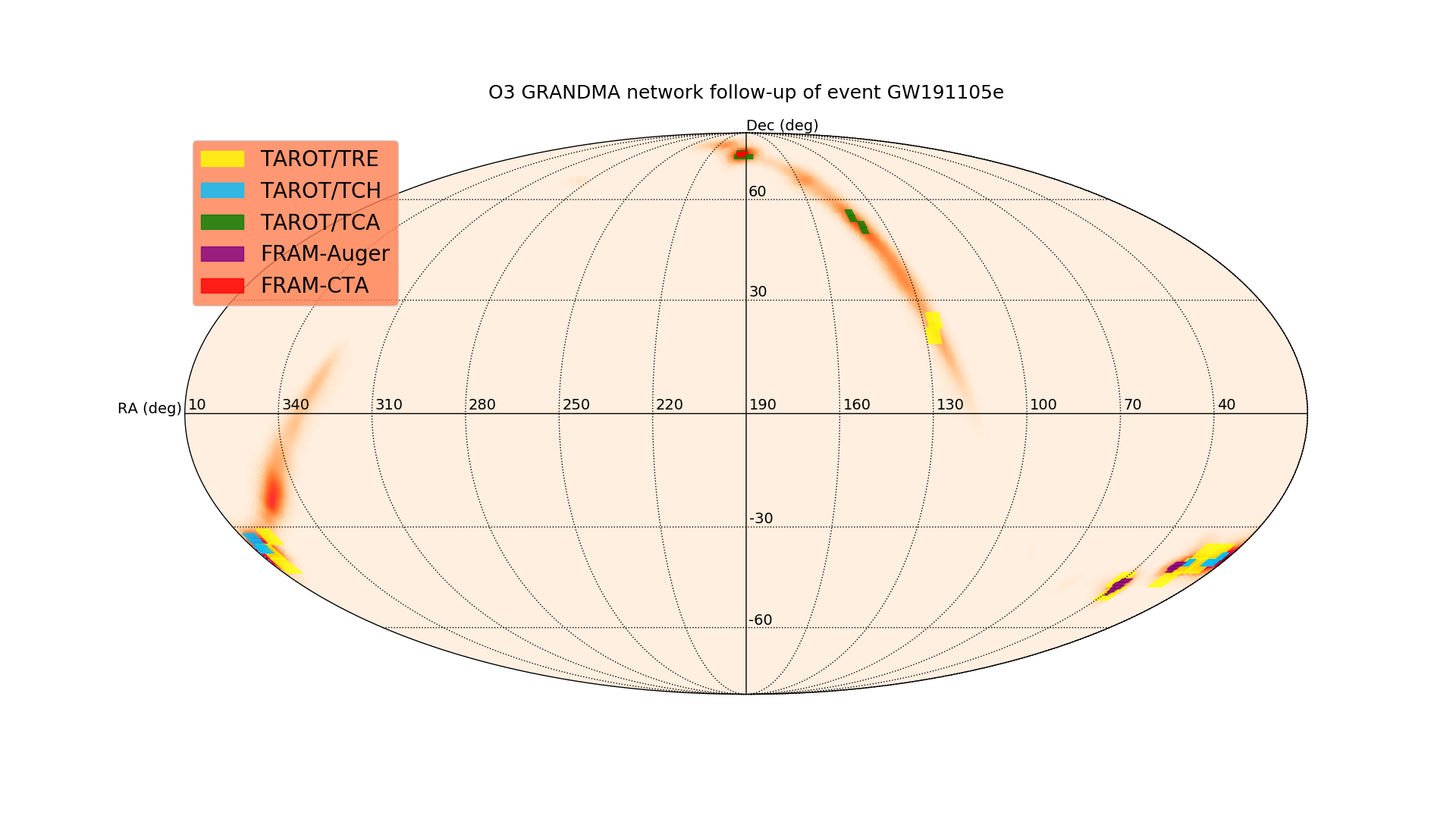}}
\subfloat[S191109d]{\includegraphics[width = 3.4in]{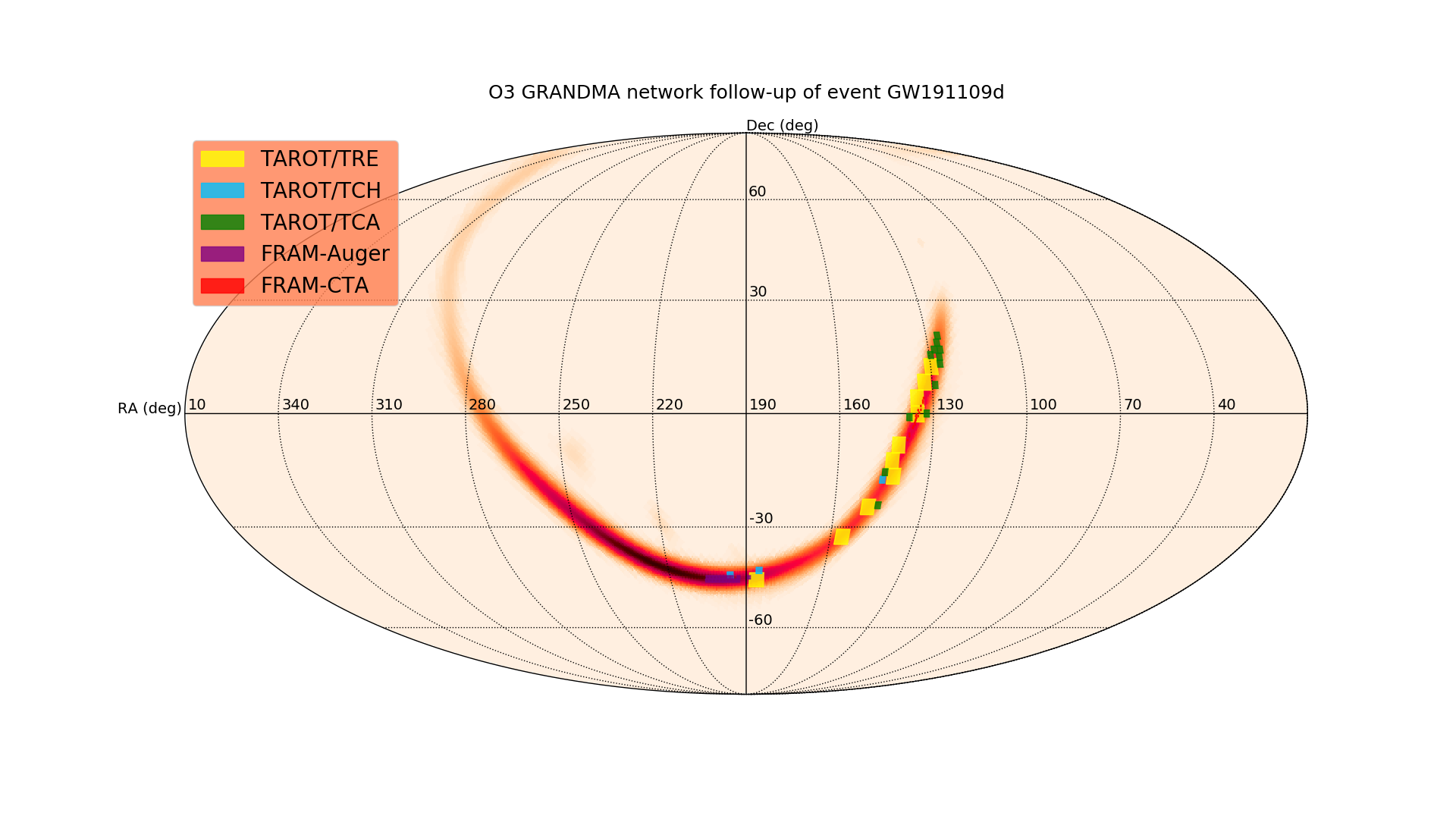}}\\
\subfloat[S191129u]{\includegraphics[width = 3.4in]{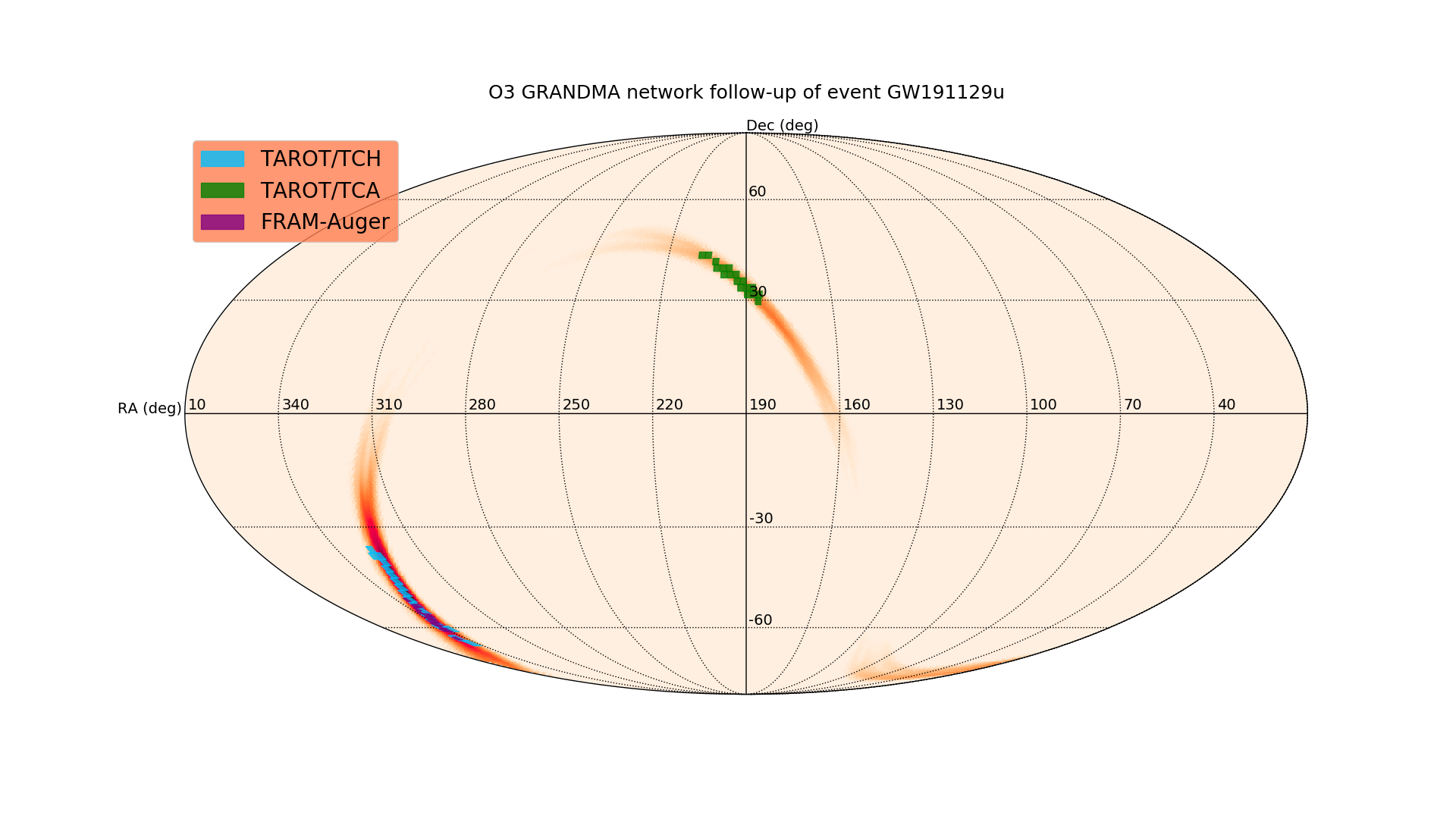}}
\subfloat[S191204r]{\includegraphics[width = 3.4in]{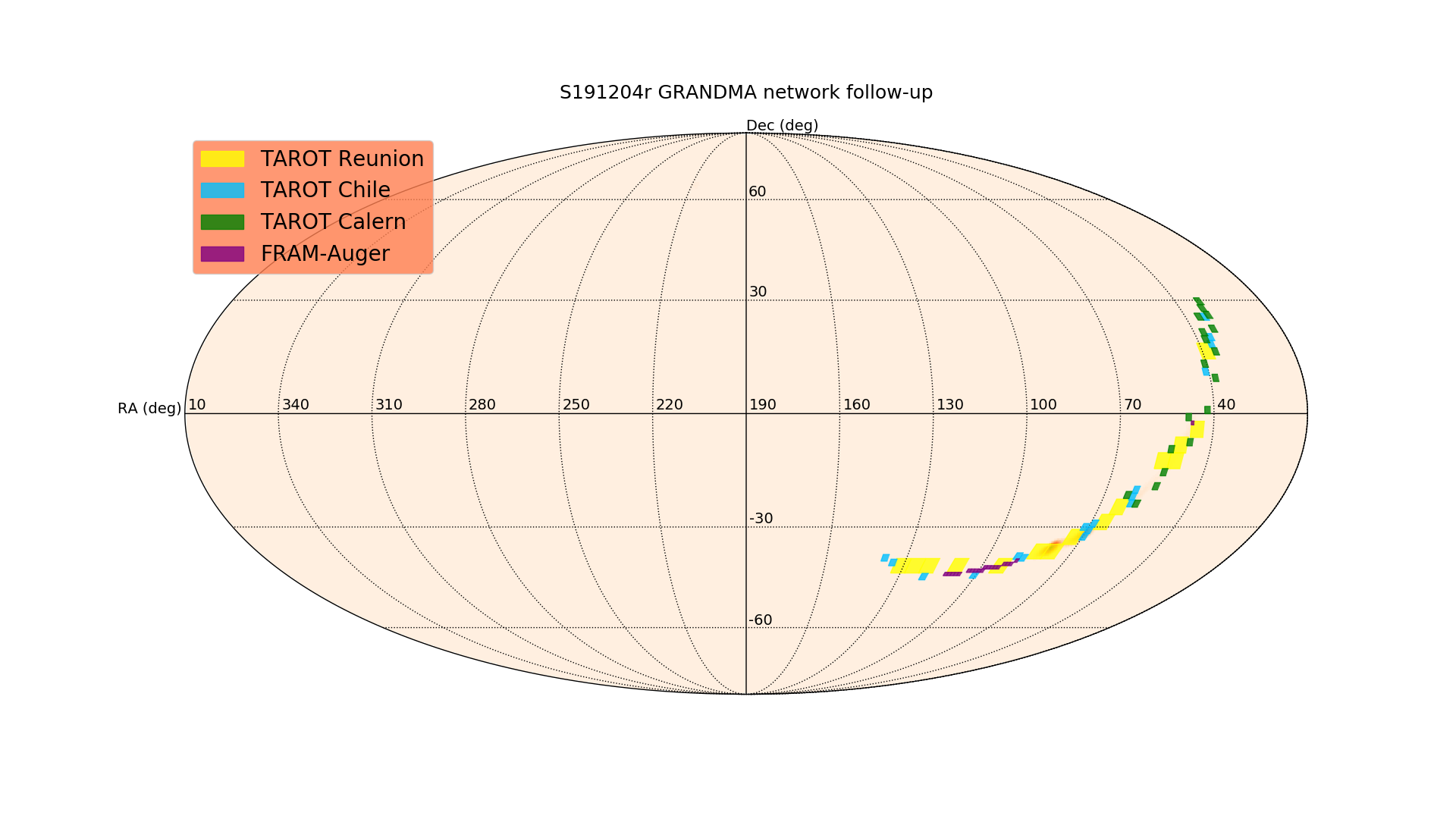}}\\
\subfloat[S191205ah]{\includegraphics[width = 3.4in]{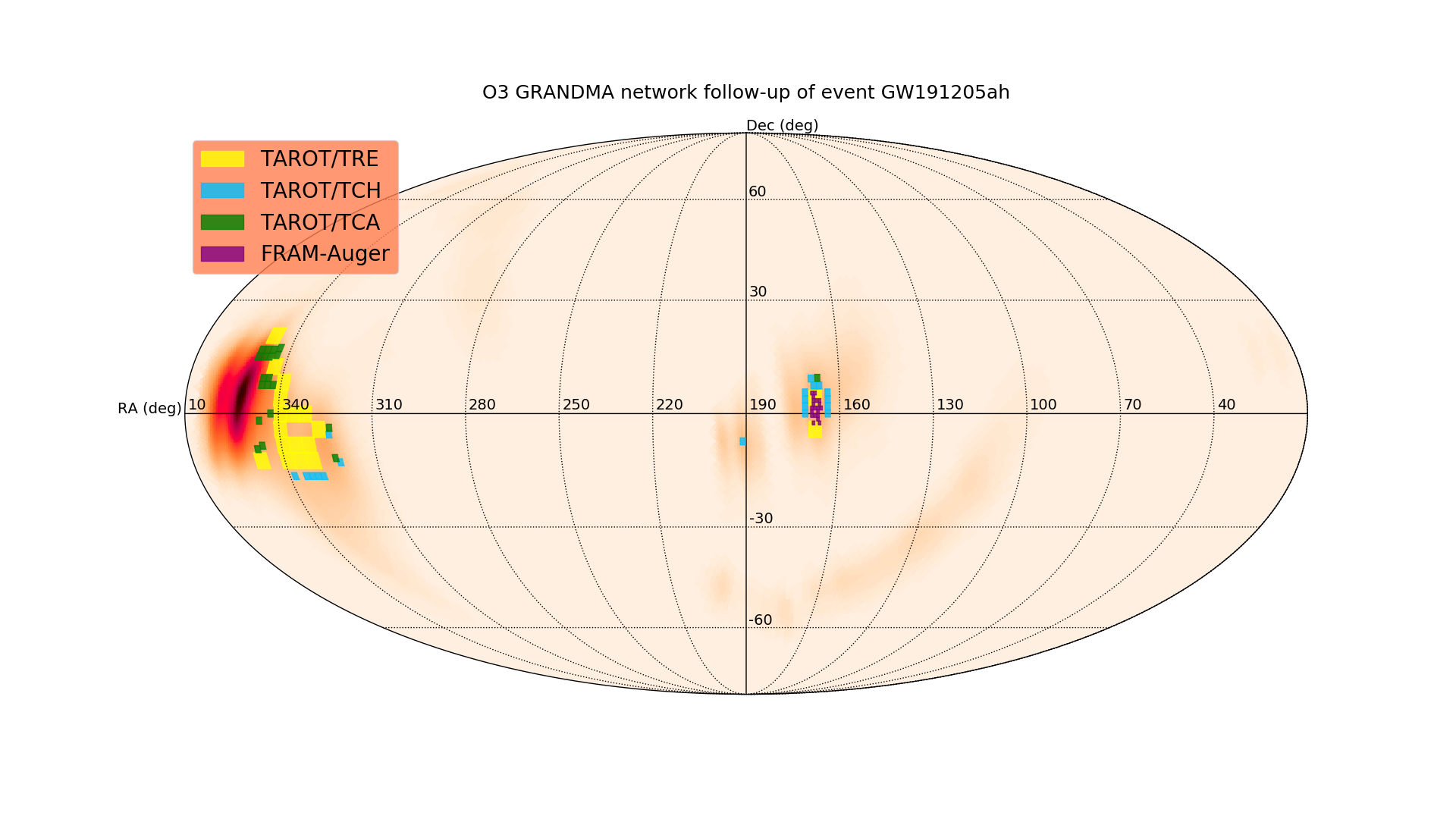}}
\subfloat[S191213g]{\includegraphics[width = 3.4in]{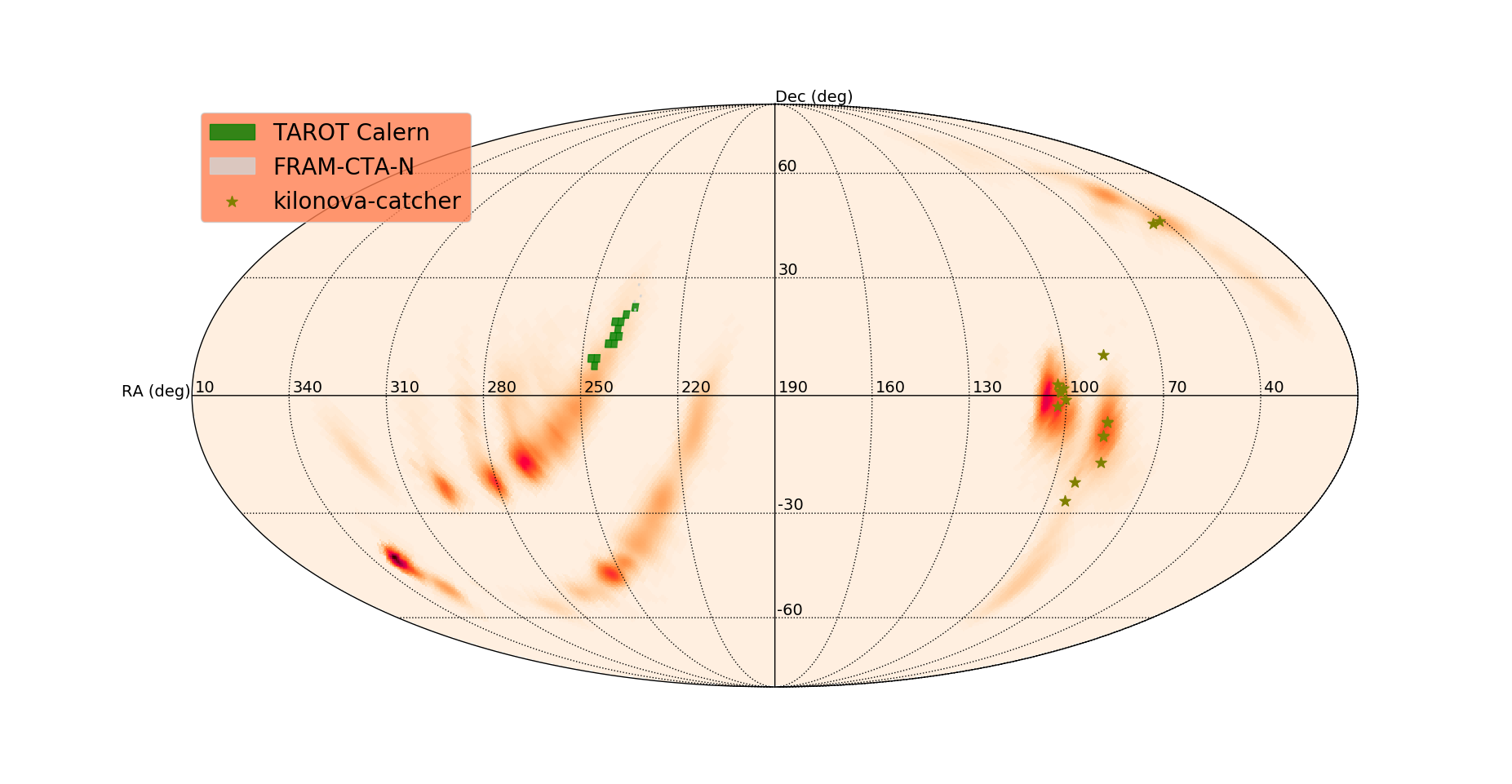}}\\
\subfloat[S191215w]{\includegraphics[width = 3.4in]{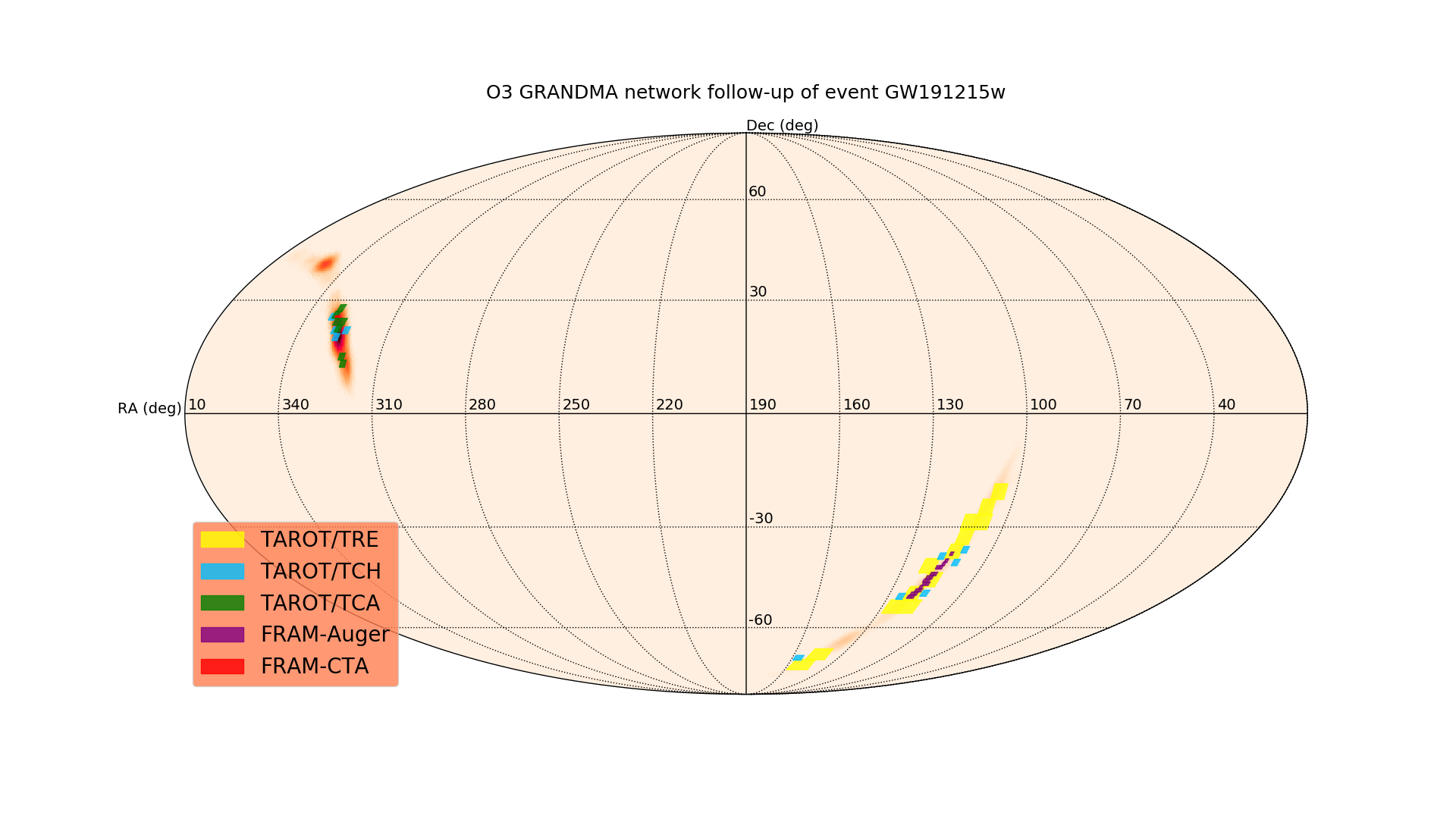}}
\subfloat[S191216ap]{\includegraphics[width = 3.4in]{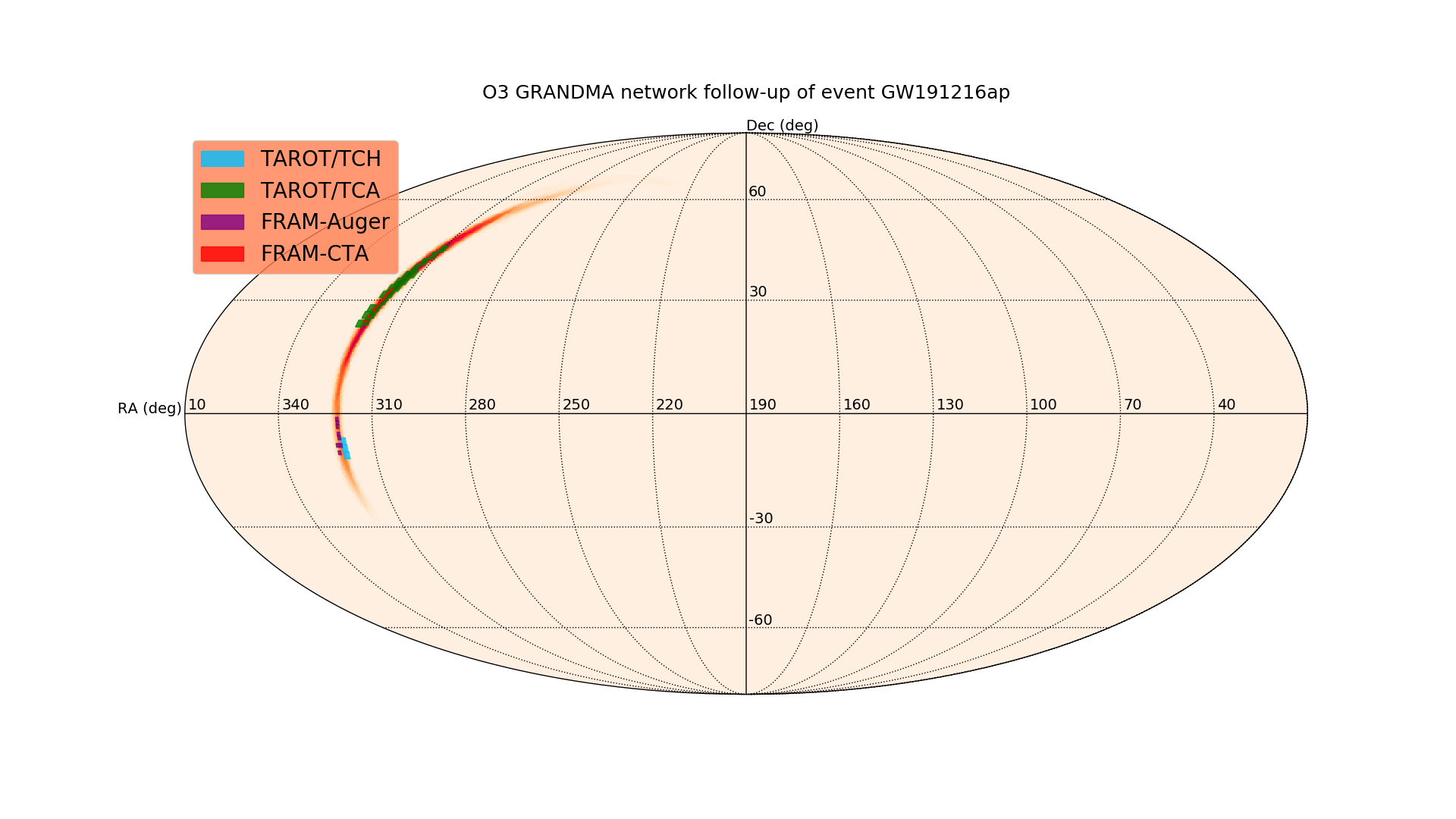}}\\
\label{followup_plots}
\caption{Sky coverage observations of the GRANDMA collaboration during the second five months of the O3 observational campaign.}
\end{figure*}

\begin{figure*}
\subfloat[S191222n]{\includegraphics[width = 3.4in]{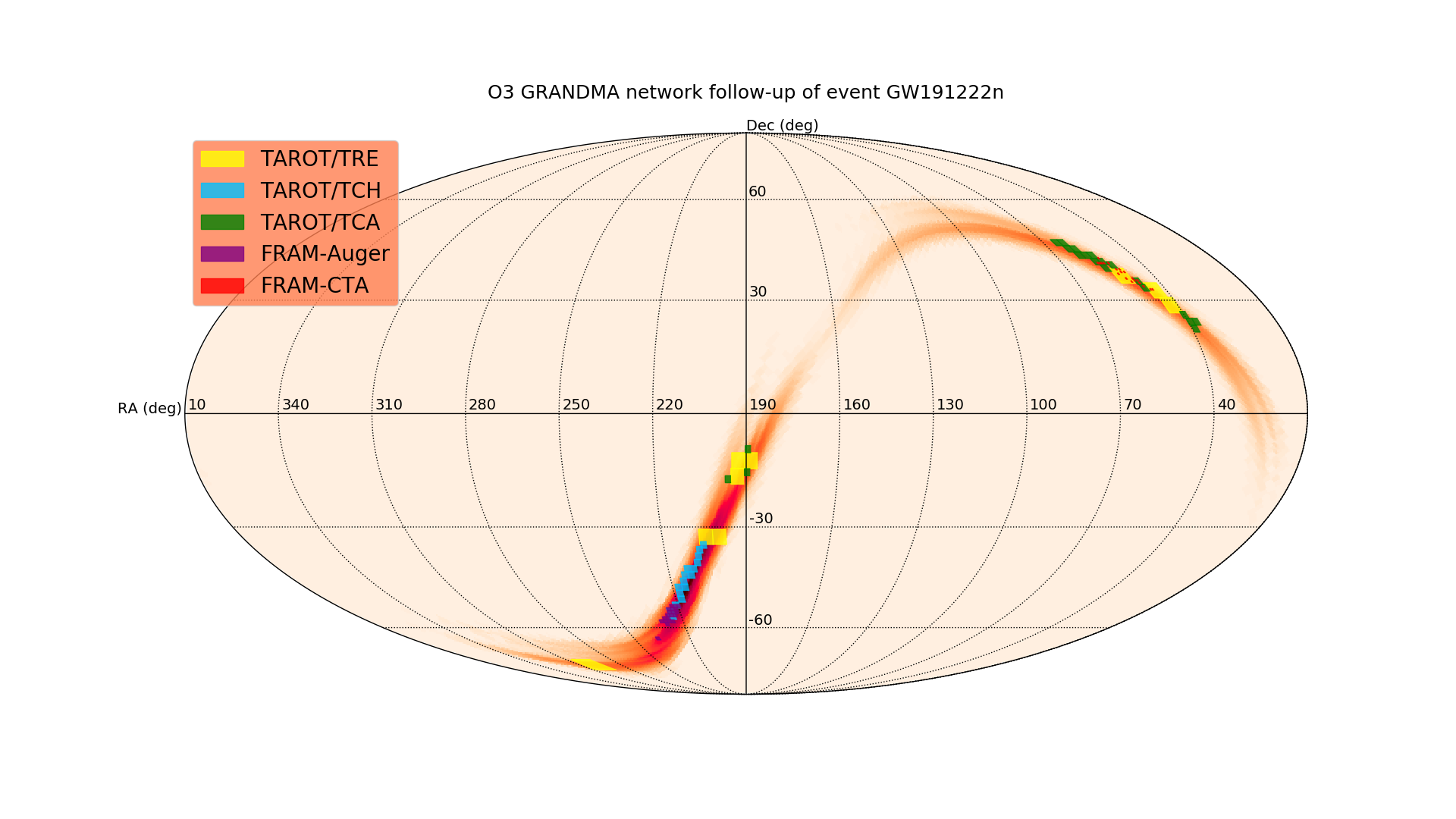}}
\subfloat[S200105ae]{\includegraphics[width = 3.4in]{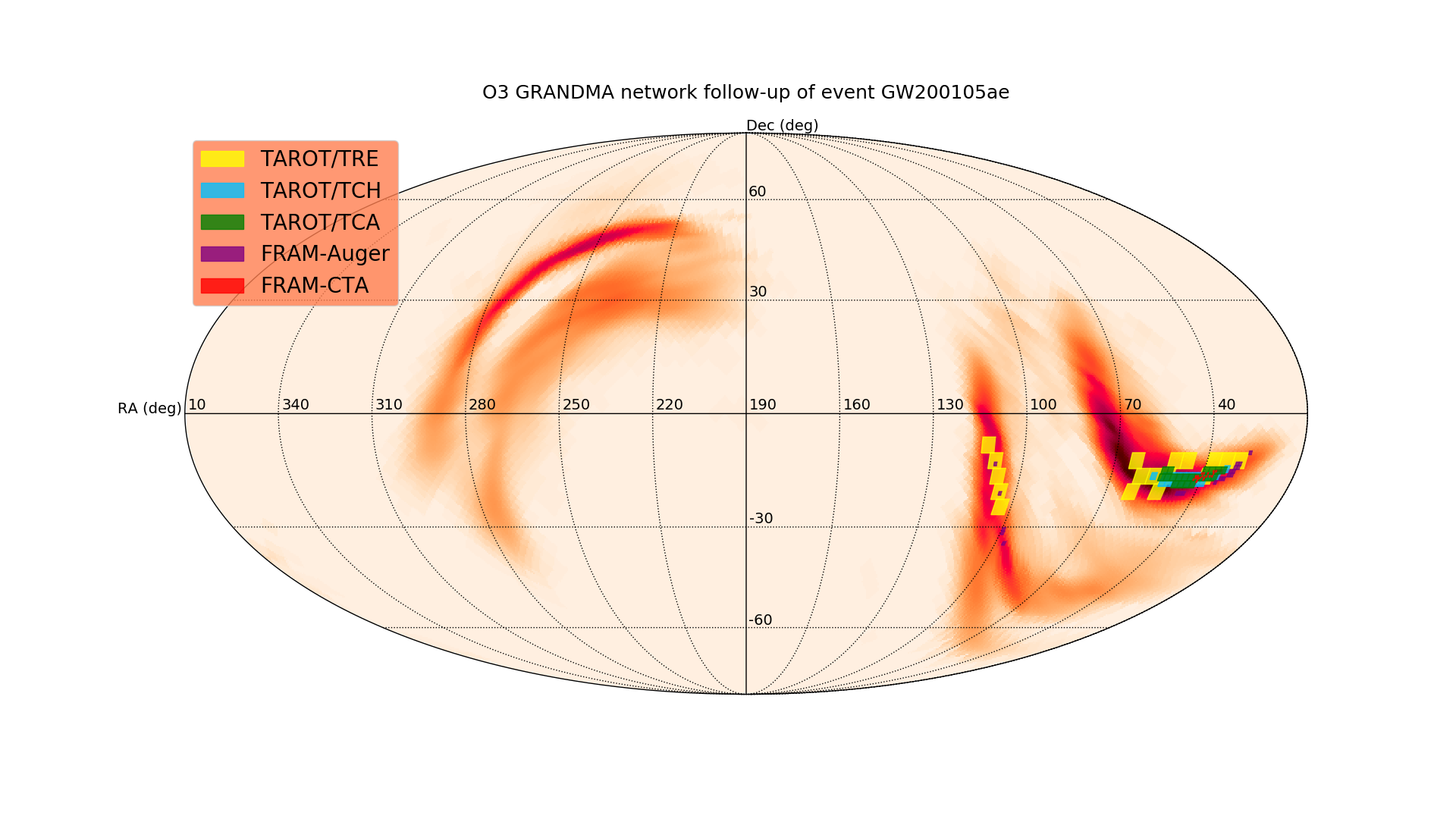}} \\
\subfloat[S200112r]{\includegraphics[width = 3.4in]{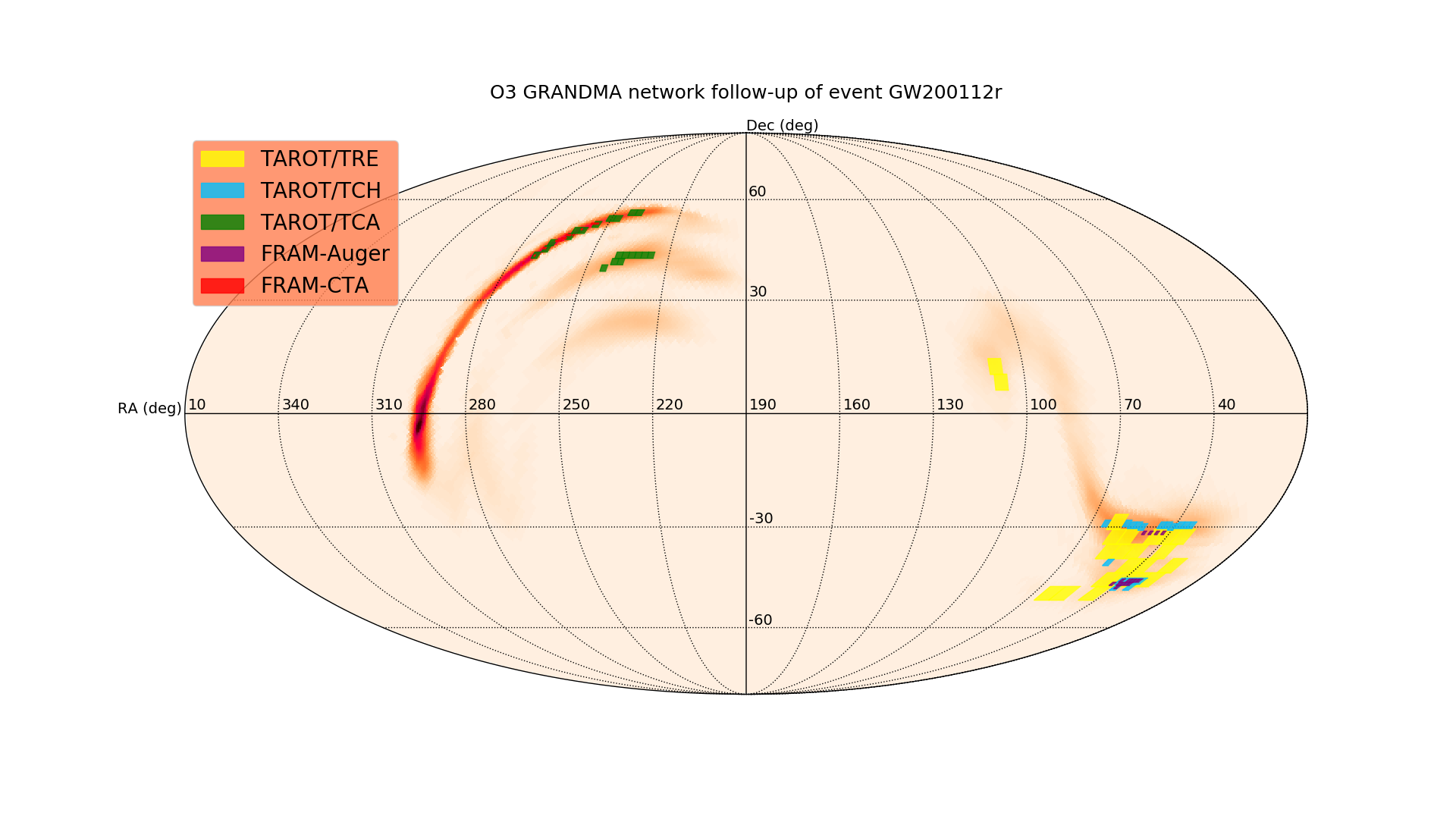}}
\subfloat[S200114f]{\includegraphics[width = 3.4in]{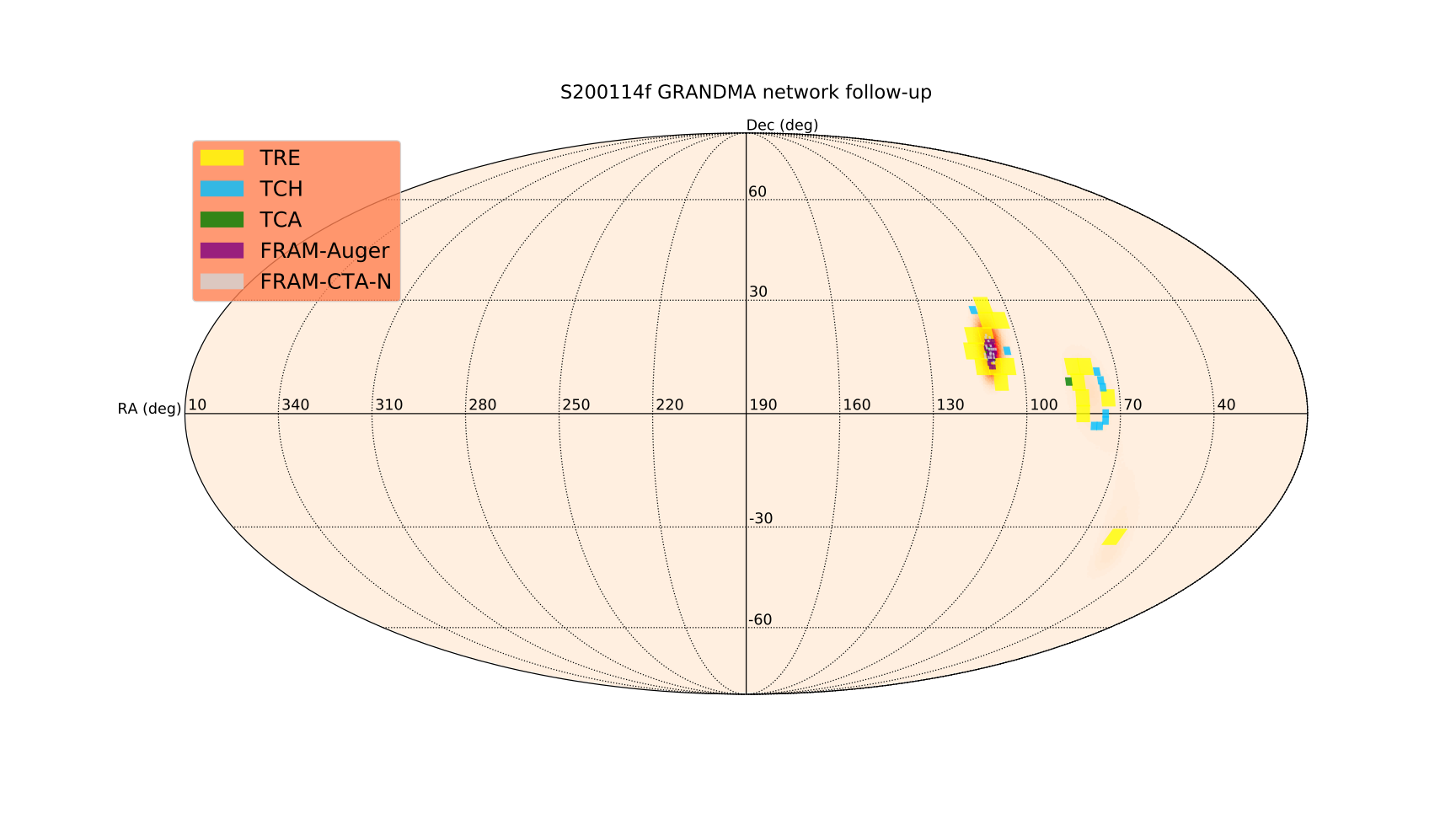}} \\
\subfloat[S200115j]{\includegraphics[width = 3.4in]{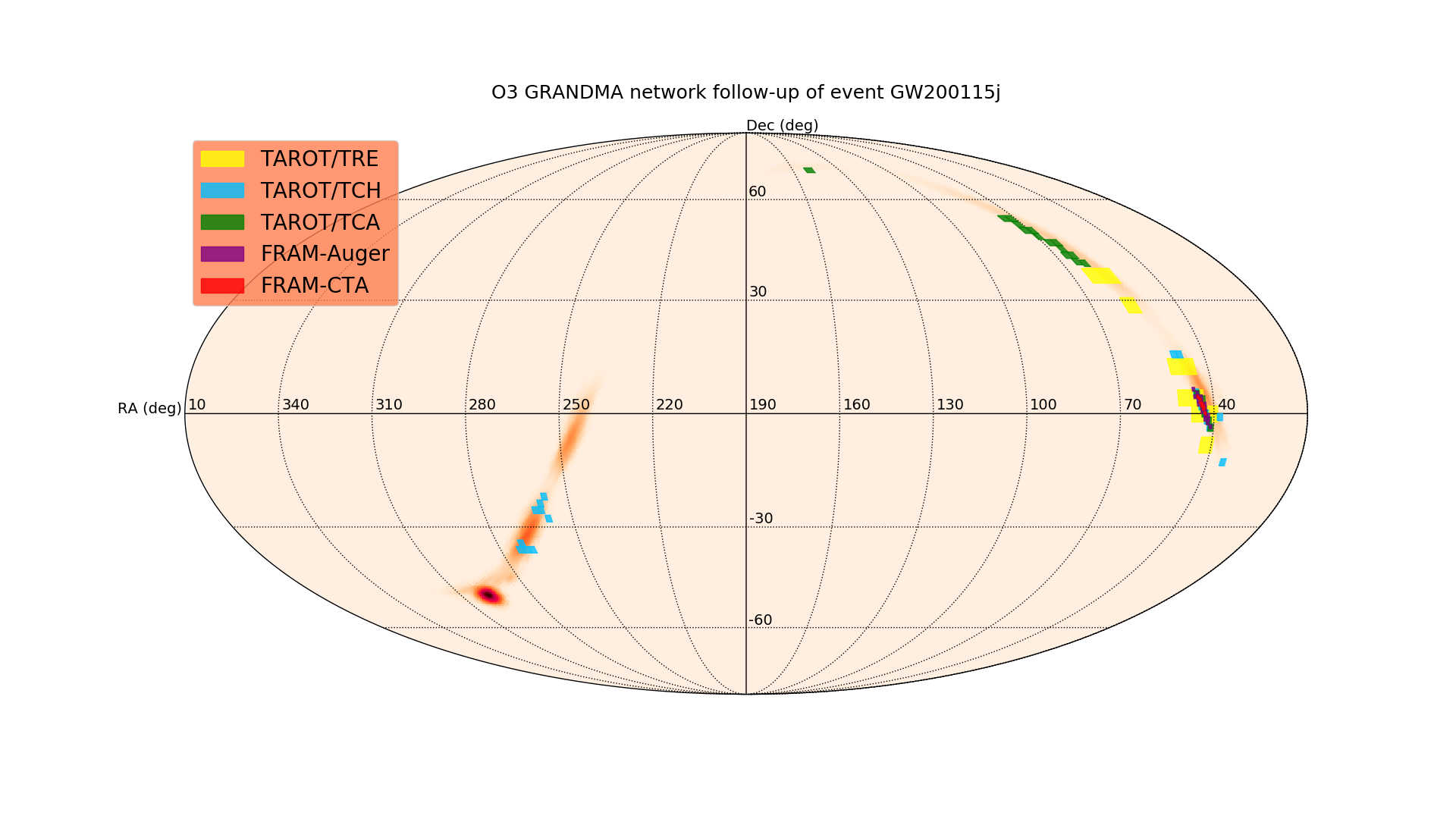}}
\subfloat[S200128d]{\includegraphics[width = 3.4in]{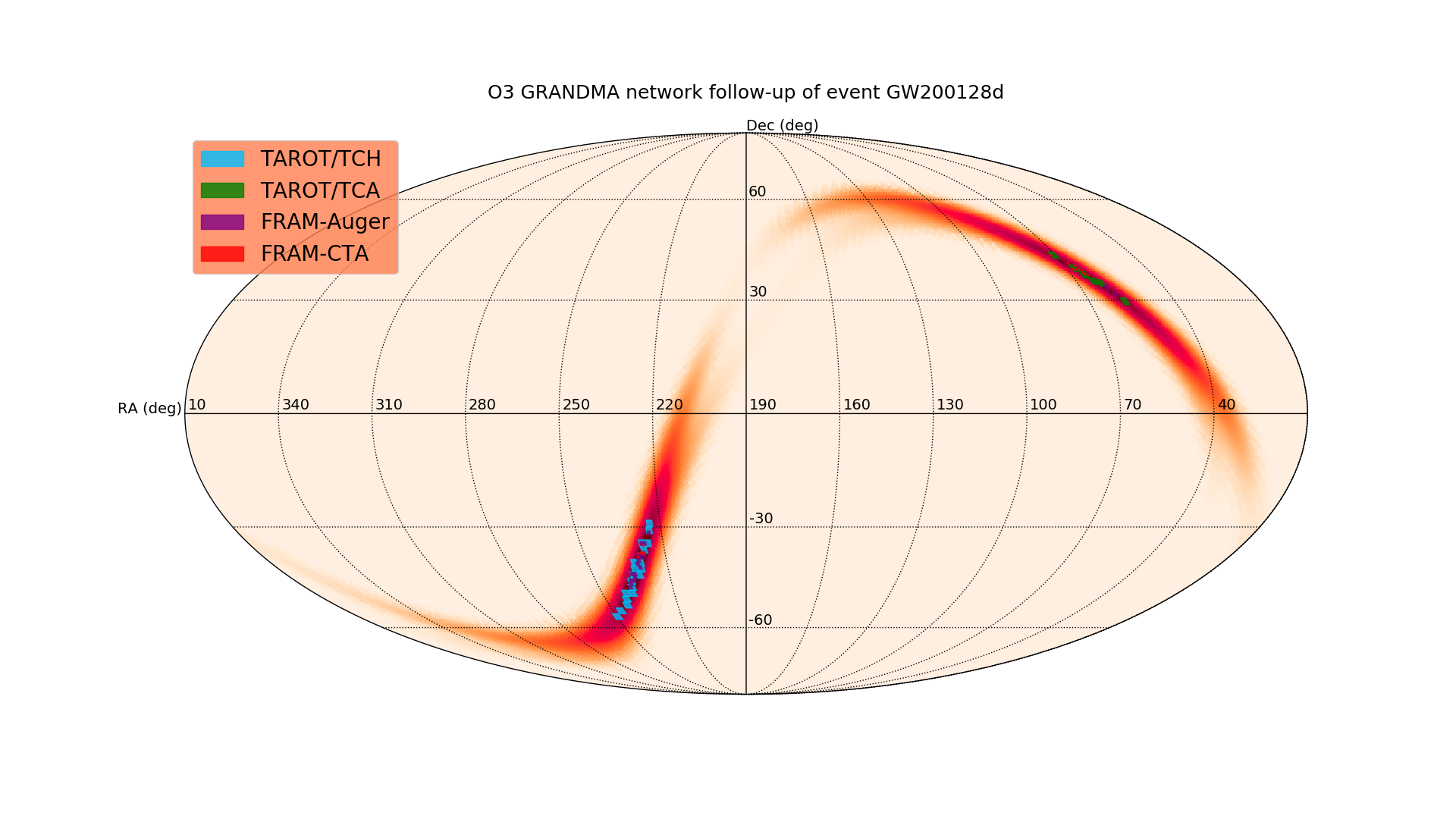}} \\
\subfloat[S200129m]{\includegraphics[width = 3.4in]{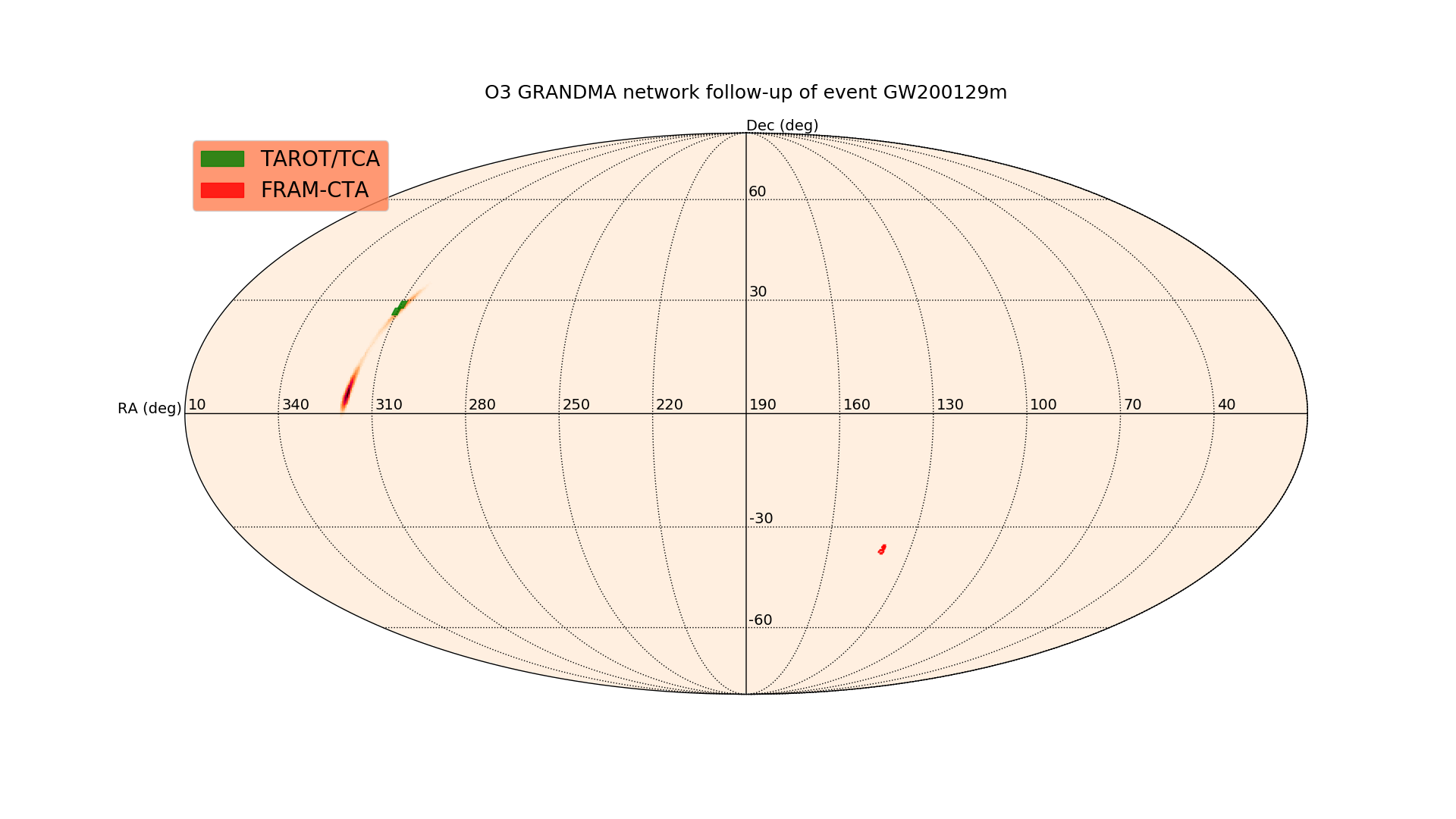}}
\subfloat[S200208q]{\includegraphics[width = 3.4in]{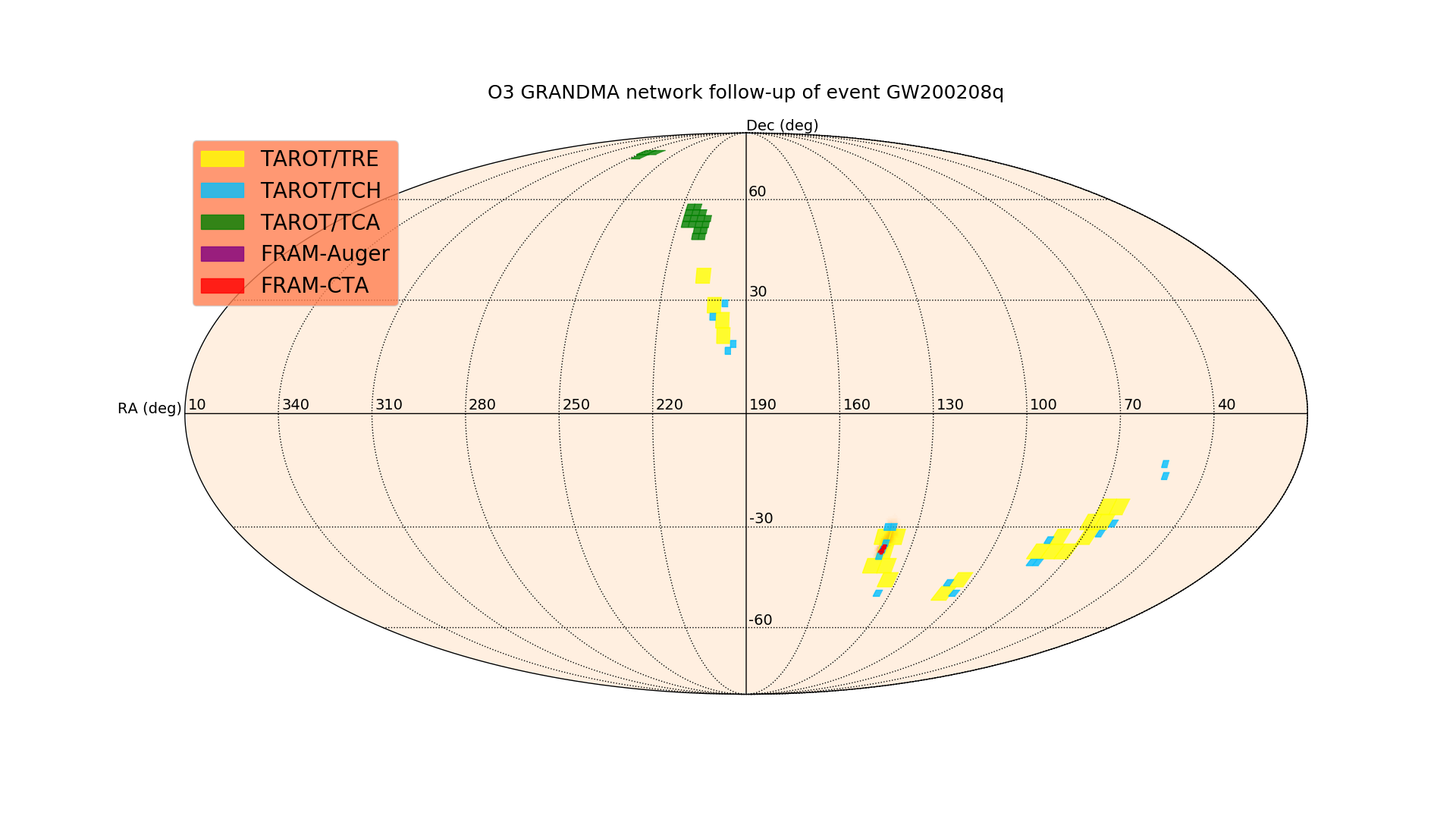}} \\
\contcaption{Sky coverage observations of the GRANDMA collaboration during the second five months of the O3 observational campaign.}
\end{figure*}

\begin{figure*}
\subfloat[S200213t]{\includegraphics[width = 3.4in]{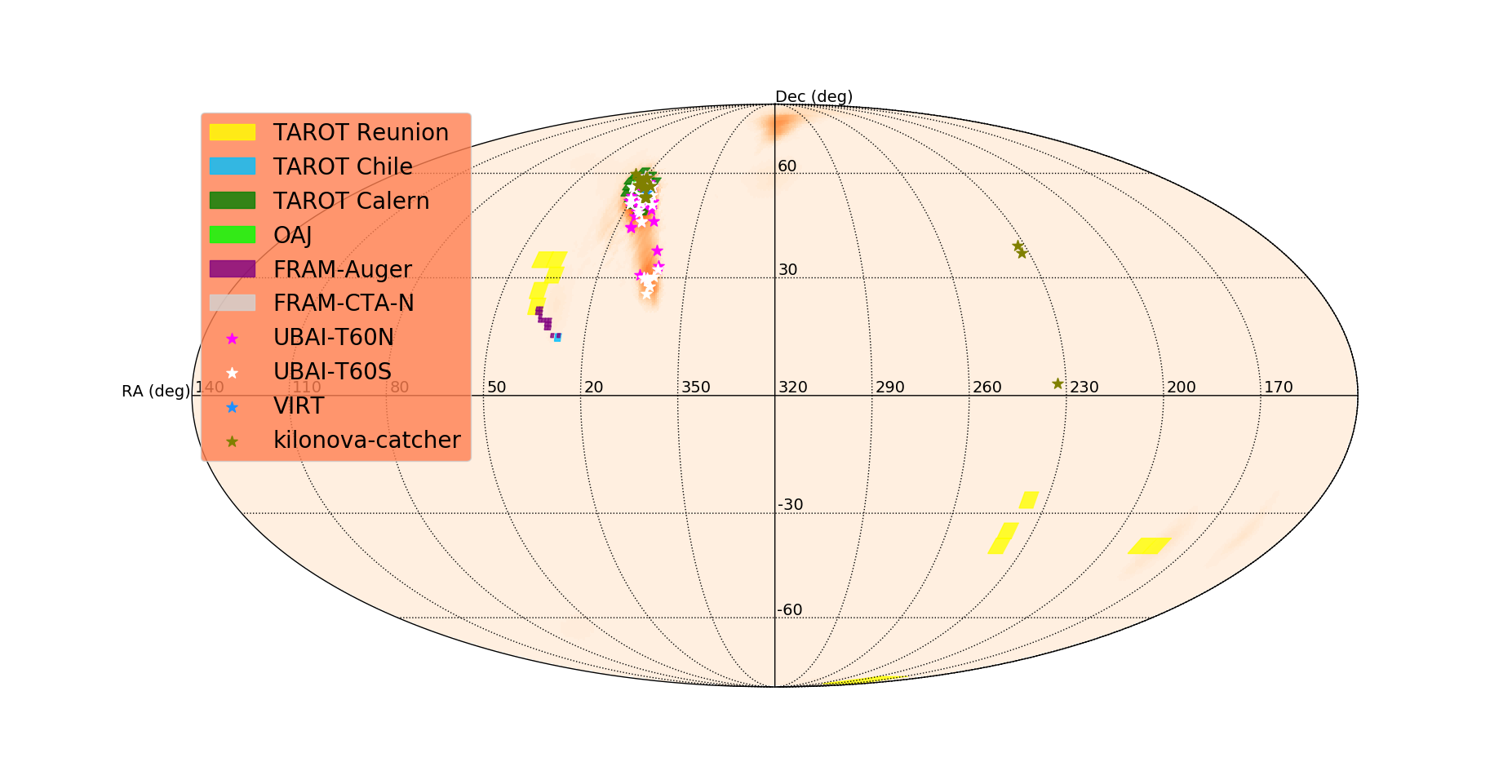}}
\subfloat[S200219ac]{\includegraphics[width = 3.4in]{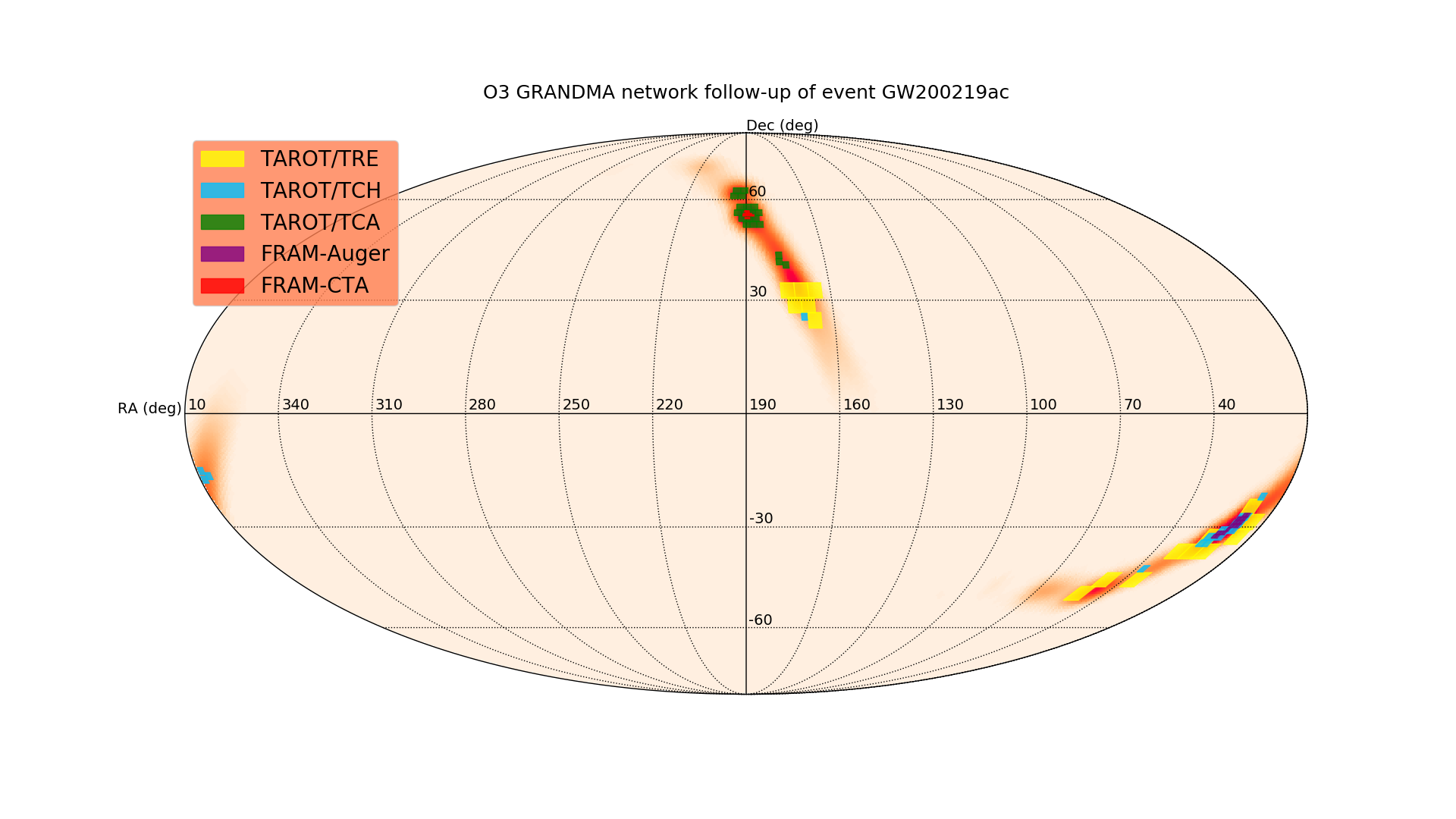}} \\
\subfloat[S200224ca]{\includegraphics[width = 3.4in]{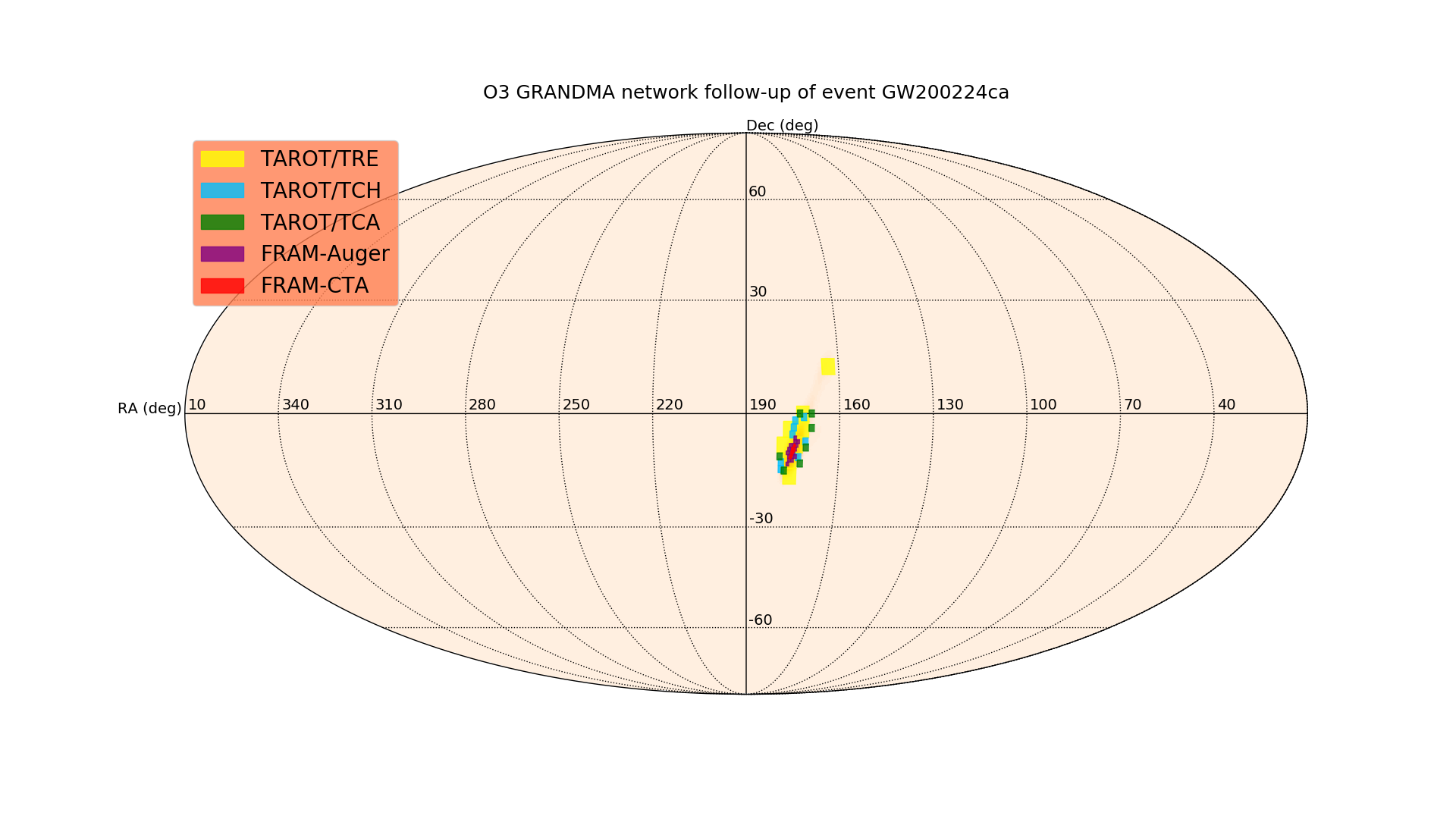}}
\subfloat[S200225q]{\includegraphics[width = 3.4in]{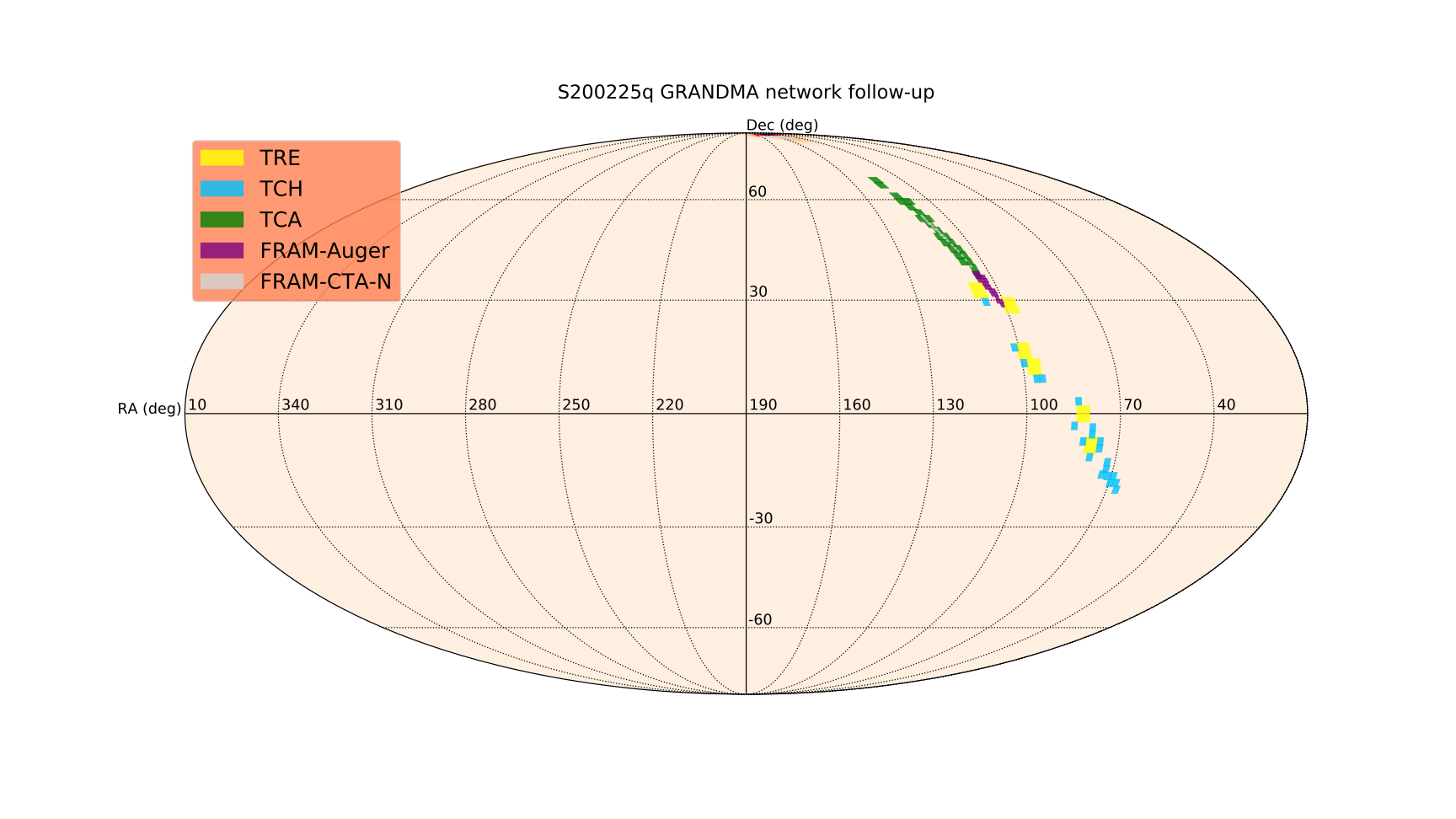}} \\
\subfloat[S200302c]{\includegraphics[width = 3.4in]{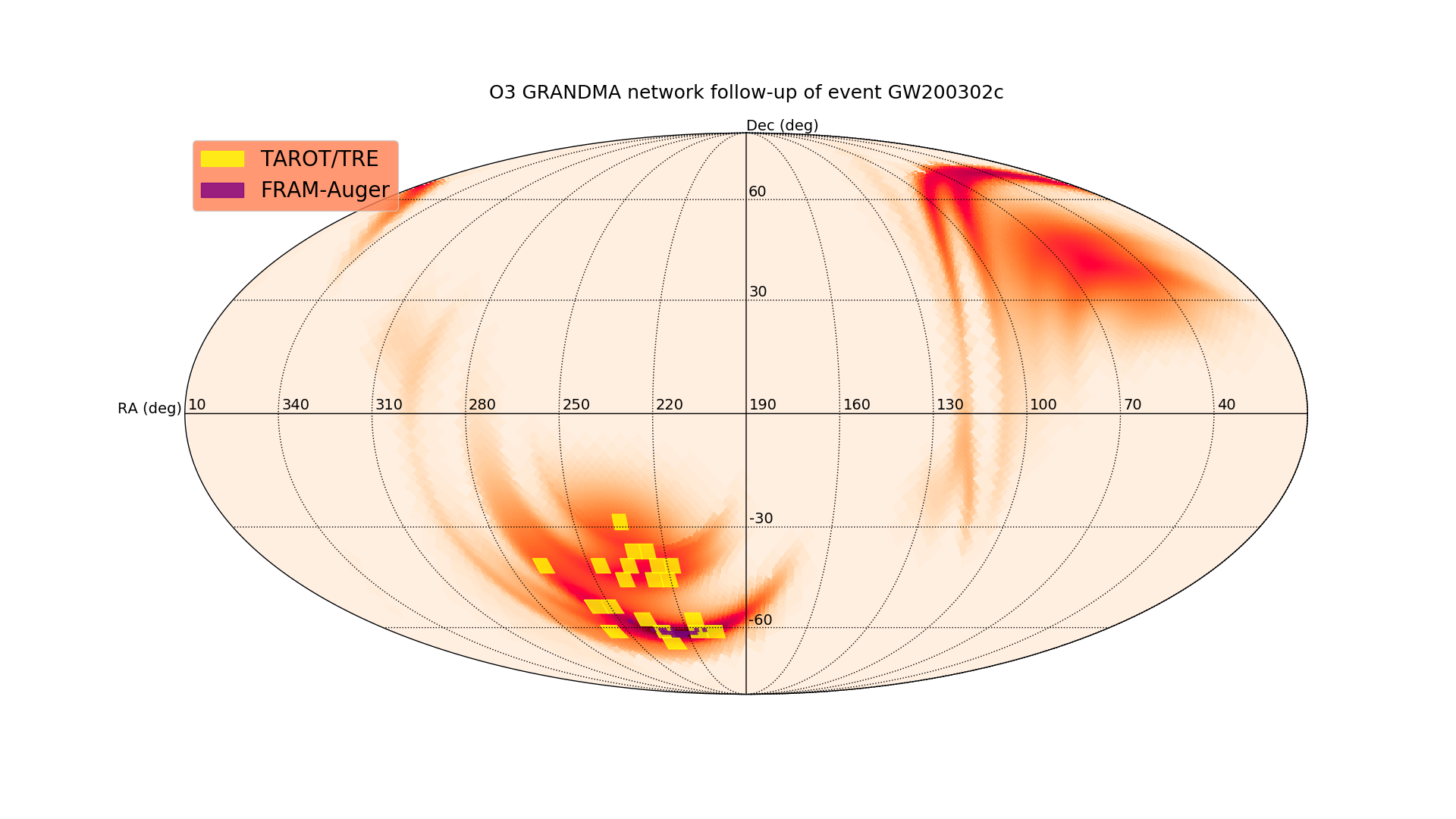}}
\subfloat[S200316bj]{\includegraphics[width = 3.4in]{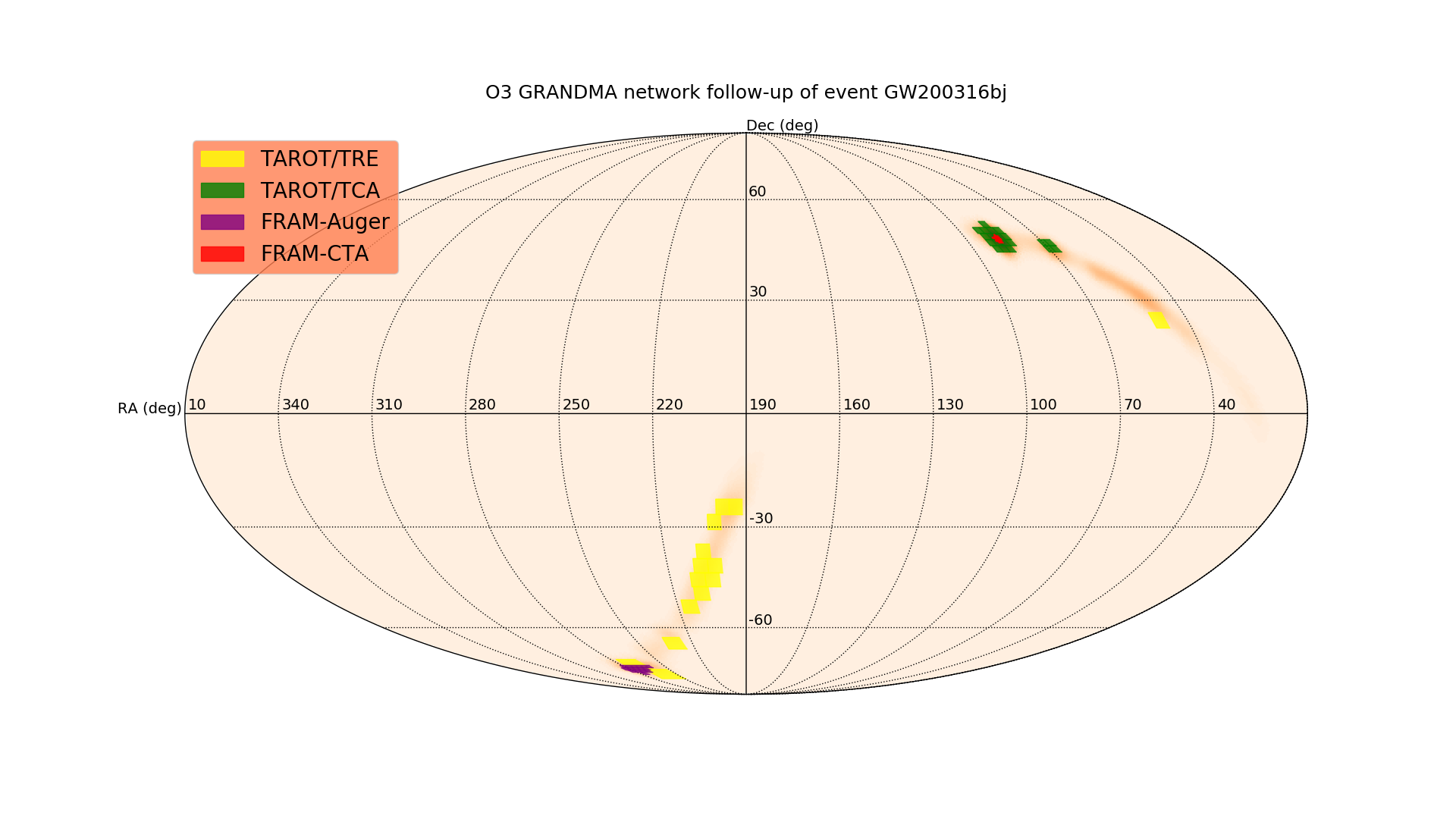}}\\
\contcaption{Sky coverage observations of the GRANDMA collaboration during the second five months of the O3 observational campaign.}
\end{figure*}

\section{Observations from the \textit{kilonova-catcher} program}

\begin{table*}
\caption{Summary of the GRANDMA \textit{kilonova-catcher} user observations of the BNS merger candidates detected during the last five months of the O3 runs. $\delta \mathrm{t}$ is the delay between the beginning of the observation and the trigger time. $\Delta \mathrm{t}$ is duration of the observations. The limiting magnitudes of each observational epoch are given at 5$\sigma$ confidence level in the AB system. The ``Lum.'' filter stands for ``Luminance'', which is almost the same as the clear filter (high transmission between between 400-700 nm) but with cuts at precise wavelengths. The design is made in order to remove emission lines of the local environment such as the sodium lines emitted by public illumination. (a) C11 f/5.8 telescope with a ST-8 camera at Observatoire Chante-Perdrix, France; (b) CDK 14'' PlaneWave 1100 GTO Astro-Physics telescope with a SBIG STL 1001e camera at Obs. Chaville Coregidora, Mexico; (c) iTelescope.net T5 with a SBIG ST-10 3 CCD Camera, in New Mexico, USA; (d) Celestron Edge HD14 à 2737mm (F/7.7) in Canada.}
\label{tab:KN_observations}
\begin{tabular}{ccccccccl}
\multicolumn{9}{c}{Observation log for \href{https://gracedb.ligo.org/superevents/S191213g/view/}{S191213g} (D = $201\pm81$ Mpc)}\\
\hline\hline
Galaxy & RA & Dec. & D & $\delta \mathrm{t}$ & $\Delta \mathrm{t}$ & Lim. mag & filter & Telescope/Observer\\
name & (deg) & (deg) & (Mpc) & (hr) & (hr) & (mag) & & \\
 \hline
UGC 02249 & 41.849 & 45.531 & 114.81 &  39.77  &  0.35  &  17.4  & Clear & (a)/M. Serrau\\
  &   &   &   & 44.83 & 0.27 & 17.0 & Lum & (b)/D.St-Gelais\\
2MASX 03011222+4454285 & 45.301 & 44.908 & 191.45 & 45.34 & 0.27 & 17.0 & Lum & (b)/D.St-Gelais\\
PGC 090026 & 86.831 & -6.836 & 95.14 &  48.05 & 0.27 & 16.6 & Lum & (b)/D.St-Gelais\\
PGC 147992 & 86.905 & -6.629 & 91.70 &  48.05 & 0.27 & 16.6 & Lum & (b)/D.St-Gelais\\
NGC 2076 & 86.698 & -16.782 & 33.70 &  48.53 & 0.27 & 16.7 & Lum & (b)/D.St-Gelais\\
NGC 2196 & 93.040 & -21.806 & 36.82 &  48.98 & 0.27 & 16.7 & Lum & (b)/D.St-Gelais\\
NGC 2206 & 93.999 & -26.765 & 85.85 &  49.44 & 0.27 & 16.8 & Lum & (b)/D.St-Gelais\\
2MASX 06435594+0148246 & 100.983 & 1.807 & 14.2994 & 50.38 & 0.27 & 16.7 & Lum & (b)/D.St-Gelais\\
2MASX 06472497+0110160 & 101.854 & 1.171 & 81.764 & 50.81 & 0.27 & 17.0 & Lum & (b)/D.St-Gelais\\
2MASX 06464860+0041324 & 101.703 & 0.692 & 176.15 & 51.26 & 0.27 & 16.7 & Lum & (b)/D.St-Gelais\\
2MASX 06404238-0059116 & 100.177 & -0.987 & 152.74 & 51.91 & 0.27 & 17.0 & Lum & (b)/D.St-Gelais\\
2MASX 06501768-0251397 & 102.574 & 2.861 & 32.24 & 52.39 & 0.27 & 17.0 & Lum & (b)/D.St-Gelais\\
2MASX 06502789-0244237 & 102.616 & -2.740 & 15.5784 & 52.39 & 0.27 & 17.0 & Lum & (b)/D.St-Gelais\\
PGC 981413 & 87.530 & -10.178 & 95.2902 & 52.87 & 0.27 & 16.9 & Lum & (b)/D.St-Gelais\\
PGC 982179 & 87.556 & 10.117 & 95.3345 & 52.87 & 0.27 & 16.9 & Lum & (b)/D.St-Gelais\\
PGC 983153 & 87.658 & -10.042 & 94.94 & 52.87 & 0.27 & 16.9 & Lum & (b)/D.St-Gelais\\
\hline
\multicolumn{9}{c}{Observation log for \href{https://gracedb.ligo.org/superevents/S200213t/view/}{S200213t} (D = $201\pm80$ Mpc)}\\
\hline\hline
Galaxy & RA & Dec. & D & $\delta \mathrm{t}$ & $\Delta \mathrm{t}$ & Lim. mag & filter & Telescope/Observer\\
name & (deg) & (deg) & (Mpc) & (hr) & (hr) & (mag) & & \\
 \hline
2MASX 00592750+5222015 & 14.865 & 52.3671 & 166.59 & 21.99 & 0.63 & 17.9 & Clear & (c)/H.-B. Eggenstein\\
LEDA 2418791 & 14.305 & 52.591 & 143.31 & 21.99 & 0.63 & 18.0 & Clear  & (c)/H.-B. Eggenstein\\
LEDA 2418926 & 14.692 & 52.598 & 176.99 & 21.99 & 0.63 & 18.0 & Clear  & (c)/H.-B. Eggenstein\\
     &        &        & & 45.29 & 0.27  & 16.8 & Lum.  & (b)/D. St-Gelais\\
2MASX 00570984+5222062 & 14.291 & 52.368 & 199.50 & 21.99 & 0.63 & 18.0 & Clear  & (c)/H.-B. Eggenstein\\
2MASX 00590114+5242317 & 14.755 & 52.709 & 218.72 & 21.99 & 0.63 & 18.0 & Clear  & (c)/H.-B. Eggenstein\\
2MASX 00562909+5224271 & 14.121 & 52.408 & 103.75 & 21.99 & 0.63 & 18.0 & Clear  & (c)/H.-B. Eggenstein\\
2MASX 00585906+5238557 & 14.746 & 52.649 & 103.03 & 21.99 & 0.63 & 18.0 & Clear  & (c)/H.-B. Eggenstein\\
2MASX 00585413+5236357 & 14.726 & 52.610 & 138.82 & 21.99 & 0.63 & 18.0 & Clear  & (c)/H.-B. Eggenstein\\
LEDA 2419200 & 14.726 & 52.6097& 141.38 & 21.99 & 0.63 & 18.0 & Clear  & (c)/H.-B. Eggenstein\\
2MASX 00585995+5235147 & 14.750 & 52.587 & 96.80 & 21.99 & 0.63 & 18.0 & Clear  & (c)/H.-B. Eggenstein\\
2MASX 00561754+5230539 & 14.073 & 52.515 & 257.46 & 21.99 & 0.63 & 18.0 & Clear  & (c)/H.-B. Eggenstein\\
2MASX 00585906+5238557 & 14.746 & 52.649 & 103.03 & 21.99 & 0.63 & 18.0 & Clear  & (c)/H.-B. Eggenstein\\
2MASX 01192705+5831151 & 19.863 & 58.521 & 154.80 & 116.64 & 0.40 & 17.4 & Lum. & (d)/R. M\'enard\\
2MASX 01192582+5824261 & 19.858 & 58.407 & 207.04 & 116.64 & 0.40 & 17.4 & Lum. & (d)/R. M\'enard\\
2MASX 01185204+5828404 & 19.717 & 58.478 & 231.19 & 116.64 & 0.40 & 17.4 & Lum. & (d)/R. M\'enard\\
2MASX 01180816+5834005 & 19.534 & 58.567 & 180.95 & 116.64 & 0.40 & 17.4 & Lum. & (d)/R. M\'enard\\
2MASX 01194891+5837391 & 19.954 & 58.628 & 173.69 & 116.64 & 0.40 & 17.4 & Lum. & (d)/R. M\'enard\\
2MASX 01014603+5556556 & 15.442 & 55.949 & 179.85 & 117.13 & 0.30 & 17.2 & Lum. & (d)/R. M\'enard\\
2MASX 01020132+5558266 & 15.506 & 55.974 & 176.43 & 117.13 & 0.30 & 17.2 & Lum. & (d)/R. M\'enard\\
2MASX 01300981+5750471 & 22.541 & 57.846 & 174.23 & 22.99 & 0.07  & 16.0 & Lum. & (b)/D. St-Gelais\\
 &        &        & & 45.61 & 0.27  & 16.9 & Lum. & (b)/D. St-Gelais\\
 &   &   &   & 117.52 & 0.30 & 17.1 & Lum. & (d)/R. M\'enard\\
2MASX 00592750+5222015 & 14.865 & 52.367 & 166.59 & 118.17 & 0.81 & 16.5 & Lum. & (d)/R. M\'enard\\
2MASX 01210956+5739281 & 20.290 & 57.658 & 196.26 & 49.53 & 0.27  & 16.7 & Lum. & (b)/D. St-Gelais\\
&        &        & & 119.06 & 0.30 & 16.5 & Lum. & (d)/R. M\'enard\\ 
2MASX 01103706+5536121 & 17.654 & 55.603 & 195.07 & 46.26 & 0.27  & 16.6 & Lum. & (b)/D. St-Gelais\\
&        &        & & 119.39 & 0.30 & 16.2 & Lum. & (d)/R. M\'enard\\
2MASX 01242969+5610556 & 21.124 & 56.182 & 83.78 & 38.04 & 0.10  & 18.2 & Clear & (a)/M. Serrau\\
&        &        & & 46.56 & 0.27  & 16.4 & Lum. & (b)/D. St-Gelais\\
\end{tabular}
\end{table*}

\begin{table*}
\addtocounter{table}{-1}
\caption{Continued.}
\begin{tabular}{ccccccccl}
\multicolumn{9}{c}{Observation log for \href{https://gracedb.ligo.org/superevents/S200213t/view/}{S200213t} (D = $201\pm80$ Mpc)}\\
\hline\hline
Galaxy & RA & Dec. & D & $\delta \mathrm{t}$ & $\Delta \mathrm{t}$ & Lim. mag & filter & Telescope/Observer\\
name & (deg) & (deg) & (Mpc) & (hr) & (hr) & (mag) & & \\
 \hline
2MASX 01240112+5609547 & 21.005 & 56.165 & 119.20 & 38.04 & 0.10 & 18.2 & Clear & (a)/M. Serrau\\
    &        &        & & 46.56 & 0.27 & 16.4  & Lum. & (b)/D. St-Gelais\\
2MASX 01283770+5755277 & 22.157 & 57.924 & 266.68 & 22.99 & 0.07 & 16.0 & Lum. & (b)/D. St-Gelais\\
    &        &        & & 45.61 & 0.27 & 16.9 & Lum. & (b)/D. St-Gelais\\
2MASX 01451454+5949020 & 26.311 & 59.817 & 294.78 & 24.00 & 0.27 & 15.8 & Lum. & (b)/D. St-Gelais\\
    &        &        & & 35.96 & 0.27 & 16.7 & Lum. & (b)/D. St-Gelais\\
2MASX 00592750+5222015 & 14.865 & 52.367 & 166.59 & 45.29 & 0.27 & 16.8 & Lum. & (b)/D. St-Gelais\\
2MASX 01113757+5526192 & 17.907 & 55.439 & 232.75 & 46.26 & 0.27 & 16.6 & Lum. & (b)/D. St-Gelais\\
\hline
\end{tabular}
\end{table*}

\section{Observations for AT2019wxt/PS19hgw}

\begin{table*}
\caption{Observation log for AT2019wxt/PS19hgw ($\textnormal{D}=144$ Mpc) with GRANDMA measurements and in addition measurements reported in various GCNs. The table is ordered by observation time. Magnitude upper limits are reported in AB magnitudes at $3\sigma$ confidence level for the GRANDMA measurements.}
\label{tab:2019wxt_observations}
\begin{tabular}{crcccc}
MJD & Magnitude & Error & Filter &
Telescope/Observer & Origin of data\\
(day) & (mag) & (mag) &  &  & \\
 \hline
58829.348  & $>21.00$ &  - & $z_P$ & \multirow{7}{*}{Pan-STARRS} & \multirow{7}{*}{\cite{gcn26485}}\\
58830.379 & $>20.30$ &  - & $z_P$ &  & \\
58830.379 & $>20.30$ &  - & $z_P$ &  & \\
58832.305 & $>19.40$ &  - & $i_P$ &  & \\
58833.335 & 19.37 & 0.07 & $i_P$ &  & \\
58833.320 & 19.32 & 0.07 & $i_P$ &  & \\
58833.305 & 19.38 & 0.05 & $i_P$ &  & \\
\hline
58835.703  & $>21.1$ & -  & $g^\prime$ & GRANDMA/TNT & This work \\
\hline
58835.790 & $>17.70$ & - & Lum. & GRANDMA/Serrau & This work\\
\hline
58835.80  & 19.60 & 0.1  & $J$ & GRAWITA/TNG    & \cite{gcn26499}\\
\hline
58835.830  & 18.99 & 0.02 & $i^\prime$ & NOT        & \cite{gcn26491}\\
\hline
$\sim$ 58835.853  & 18.52 & 0.34 & $v$ & \multirow{6}{*}{\textit{Swift}}      & \multirow{6}{*}{\cite{gcn26501}}\\
$\sim$ 58835.848 & 19.17 & 0.32 & $b$ &       & \\
$\sim$ 58835.848 & 19.47 & 0.24 & $u$ &      & \\
$\sim$ 58835.845 & 19.88 & 0.23 & $uvw1$ &       & \\
$\sim$ 58835.853 & 21.00 & 0.32 & $uvm2$ &       & \\
$\sim$ 58835.849 & 20.55 & 0.27 & $uvw2$ &     & \\
\hline
58835.860 & $>19.00$ & - & $R_C$ & GRANDMA/AZT-8 & This work\\
\hline
58835.920  & $>18.10$ & - & $I_C$ & GRANDMA/Eggenstein & This work\\
\hline
$\sim$ 58835.500  & 19.78 & 0.08  & $g^\prime$ & \multirow{3}{*}{Wendelstein-2m} & \multirow{3}{*}{\cite{gcn27057}}\\
$\sim$ 58835.500 & 20.10 & 0.09  & $i^\prime$ &  & \\
$\sim$ 58835.500 & 21.14 & 0.08  & $J$ &  & \\
\hline
58836.100  & 18.80 & 0.23 & Lum & GRANDMA/St-Gelais & This work\\
\hline
58836.190  & 19.62 & 0.10 & $g^\prime$ & \multirow{6}{*}{Palomar/P60-inch}       & \multirow{6}{*}{\cite{gcn26500}}\\
58836.190 & 19.61 & 0.11 & $g^\prime$ &        & \\
58836.190 & 19.38 & 0.12 & $i^\prime$ &        & \\
58836.190 & 19.40 & 0.11 & $i^\prime$ &         & \\
58836.190 & 19.41 & 0.11 & $r^\prime$ &        & \\
58836.190 & 19.40 & 0.18 & $r^\prime$ &        & \\
58836.190 &  $>17.10$ &  -   & $u^\prime$ &     & \\
58836.190 &  $>16.90$ &  -   & $u^\prime$ &   & \\
\hline
58836.384  & $>18.73$ &  -   & $B$   & \multirow{4}{*}{SAO/1-m}  & \multirow{4}{*}{\cite{gcn26520}}\\
58836.404 & $>18.83$ &  -   & $V$   &   & \\
58836.424 & 19.46 & 0.12 & $R_C$  &  & \\
58836.444 & 19.47 & 0.13 & $I_C$  &   & \\
\hline
58836.446  & 19.50 & 0.04 & $g^\prime$ & \multirow{3}{*}{Lulin/1-m}      & \multirow{3}{*}{\cite{gcn26503}}\\
58836.446 & 19.40 & 0.03 & $r^\prime$ &       & \\
58836.446 & 19.47 & 0.06 & $i^\prime$ &     & \\
\hline
58836.434  & 19.41 & 0.09  & $g_P$ & \multirow{5}{*}{Pan-STARRS} & \multirow{5}{*}{\cite{gcn26577}}\\
58836.436 & 19.32 & 0.07  & $r_P$ &  & \\
58836.438 & 19.30 & 0.07  & $i_P$ & & \\
58836.439 & 19.41 & 0.11  & $z_P$ &  & \\
58836.441 & 19.37 & 0.22  & $y_P$ &  & \\
\end{tabular}
\end{table*}

\begin{table*}
\addtocounter{table}{-1}
\caption{Continued.}
\begin{tabular}{cccccc}
\multicolumn{6}{c}{Observation log for AT2019wxt/PS19hgw ($\textnormal{D}=144$ Mpc)}\\
\hline\hline
MJD & Magnitude & Error & Filter &
Telescope/Observer & Origin of data\\
(day) & (mag) & (mag) &  &  & \\
 \hline
58836.465  & $>21.2$ & -  & $B$ & \multirow{5}{*}{GRANDMA/TNT} & \multirow{5}{*}{This work} \\
58836.473 & $>21.7$ & -  & $V$ &  & \\
58836.477 & $>21.7$ & -  & $g^\prime$ &  &  \\
58836.484 & $>21.5$ & -  & $r^\prime$ &  &  \\
58836.488 & $>20.9$ & -  & $i^\prime$ & &  \\
\hline
58836.620  & 19.59 & 0.07 & $I_C$ & YAHPT  & \cite{gcn26510}\\
\hline
$\sim 58836.500$  & 19.83 & 0.11  & $g^\prime$ & \multirow{3}{*}{Wendelstein/2-m} & \multirow{3}{*}{\cite{gcn27057}}\\
$\sim 58836.500$ & 19.84 & 0.09  & $i^\prime$ &  & \\
$\sim 58836.500$ & 20.06 & 0.11  & $J$ &  & \\
\hline
58837.114 & 19.43 & 0.02 & $r^\prime$ &  \multirow{4}{*}{DCT}      & \multirow{4}{*}{\cite{gcn26517}}\\
58837.120 & 19.62 & 0.01 & $i^\prime$ &       & \\
58837.126 & 19.58 & 0.03 & $z^\prime$ &       & \\
58837.126 & 19.86 & 0.02 & $g^\prime$ &       &  \\
\hline
58837.665  & 20.72 & 0.29 & $u^\prime$ & \multirow{5}{*}{GRANDMA/NARIT-TNT} & \multirow{5}{*}{This work} \\
58837.685 & 19.60 & 0.18 & $g^\prime$ &  & \\
58837.685 & 19.58 & 0.30 & $r^\prime$ &  & \\
58837.685 & 19.34 & 0.23 & $i^\prime$ &  & \\
58837.685 & 19.30 & 0.35 & $z^\prime$ &  & \\
\hline
58838.500 & 20.10 & 0.06  & $g^\prime$ & \multirow{3}{*}{Wendelstein/2-m} & \multirow{3}{*}{\cite{gcn27057}}\\
58838.500 & 19.83 & 0.06  & $i^\prime$ &  & \\
58838.500 & 19.84 & 0.10  & $J$ &  & \\
\hline
58841.211  & 20.43 & 0.07  & $g_P$ & \multirow{5}{*}{Pan-STARRS} & \multirow{5}{*}{\cite{gcn26577}}\\
58841.213 & 20.16 & 0.05  & $r_P$ & & \\
58841.214 & 19.97 & 0.04  & $i_P$ &  & \\
58841.216 & 19.82 & 0.05  & $z_P$ &  & \\
58841.218 & 19.74 & 0.12  & $y_P$ &  & \\
\hline
58843.523  & $>21.7$ & -  & $g^\prime$ & \multirow{3}{*}{GRANDMA/TNT} & \multirow{3}{*}{This work} \\
58843.531 & $>21.4$ & - & $r^\prime$ &  &  \\
58843.539 & $>20.8$ & -  & $i^\prime$ &  &  \\
\hline
58844.443  & 23.7 & 0.1  & $V$ & \multirow{2}{*}{GRANDMA/Xinglong-2.16} & \multirow{2}{*}{This work} \\
58844.454 & 20.4 & 0.1  & $R$ &  &  \\
\hline
$\sim 58846.500$ & 22.77 & 0.05  & $g^\prime$ & \multirow{9}{*}{Wendelstein/2-m} & \multirow{9}{*}{\cite{gcn27057}}\\
$\sim 58846.500$ & 20.92 & 0.06  & $i^\prime$ &  & \\
$\sim 58846.500$ & 20.66 & 0.11  & $J$ &  & \\
$\sim 58854.500$ & 23.68 & 0.08  & $g^\prime$ &  & \\
$\sim 58854.500$ & 22.64 & 0.07  & $i^\prime$ &  & \\
$\sim 58854.500$ & 21.60 & 0.12  & $J$ &  & \\\
$\sim 58862.500$ & $>24.00$ &   -   & $g^\prime$ &  & \\
$\sim 58862.500$ & $>22.70$ &   -   & $i^\prime$ &  & \\
$\sim 58862.500$ & $>21.91$ &   -   & $J$ &  & \\
\hline
\end{tabular}
\end{table*}

\section{Affiliations}

$^{1}$Universit\'e de Paris, CNRS, Astroparticule et Cosmologie, F-75013 Paris, France\\
$^{2}$N.Tusi Shamakhy astrophysical Observatory\ Azerbaijan National Academy of Sciences, settl.Mamedaliyev, AZ 5626, Shamakhy, Azerbaijan\\
$^{3}$American University of Sharjah, Physics Department, PO Box 26666, Sharjah, UAE\\
$^{4}$National Astronomical Research Institute of Thailand, 260, Moo 4, T. Donkaew, A. Mae Rim, Chiang Mai, 50180, Thailand\\
$^{5}$Astronomical Observatory\ Taras Shevshenko National University of Kyiv, Observatorna str. 3, Kyiv, 04053, Ukraine\\
$^{6}$Nuclear Physics Department\ Taras Shevchenko National University of Kyiv, 60 Volodymyrska str., Kyiv, 01601, Ukraine\\
$^{7}$E. Kharadze Georgian National Astrophysical Observatory, Mt.Kanobili, Abastumani, 0301, Adigeni, Georgia\\
$^{8}$Samtskhe-Javakheti  State  University, Rustaveli Str. 113,  Akhaltsikhe, 0080,  Georgia\\
$^{9}$Instituto de Astrof\'isica de Andaluc\'ia (IAA-CSIC), Glorieta de la Astronom\'ia s/n, 18008 Granada, Spain\\
$^{10}$ARTEMIS UMR 7250 UCA CNRS OCA, boulevard de l'Observatoire, CS 34229, 06304 Nice CEDEX 04, France\\
$^{11}$Ulugh Beg Astronomical Institute, Uzbekistan Academy of Sciences, Astronomy str. 33, Tashkent 100052, Uzbekistan\\
$^{12}$IJCLab, Univ Paris-Saclay, CNRS/IN2P3, Orsay, France\\
$^{13}$School of Physics and Astronomy, University of Minnesota, Minneapolis, Minnesota 55455, USA\\
$^{14}$OzGrav-UWA, University of Western Australia, School of Physics, M013, 35 Stirling Highway, Crawley, WA 6009, Australia\\
$^{15}$Institut für Physik und Astronomie, Universität Potsdam, Haus 28, Karl-Liebknecht-Str. 24/25, 14476, Potsdam, Germany
$^{16}$University of the Virgin Islands, United States Virgin Islands 00802\\
$^{17}$Volkssternwarte Paderborn e.V.
Im Schlosspark 13, 33104 Paderborn - Schloss Neuhaus, Germany\\
$^{18}$FZU -- Institute of Physics of the Czech Academy of Sciences, Prague, Czech Republic\\
$^{19}$IRAP, Universit\'e de Toulouse, CNRS, UPS, 14 Avenue Edouard Belin, F-31400 Toulouse, France\\
$^{20}$Universit\'e Paul Sabatier Toulouse III, Universit\'e de Toulouse, 118 route de Narbonne, 31400 Toulouse, France\\
$^{21}$Physics Department and Astronomy Department, Tsinghua University, Beijing, 100084, China\\
$^{22}$OrangeWave Innovative Science, LLC, Moncks Corner, SC 29461, USA\\
$^{23}$Department of Physics and Materials Science, Faculty of Science, Chiang Mai University, Chiang Mai, 50200, Thailand\\
$^{24}$Universit\'e Paris-Saclay, CNRS, CEA, D\'epartement d'Astrophysique, Astrophysique, Instrumentation et Mod\'elisation de Paris-Saclay, 91191, Gif-sur-Yvette, France.\\
$^{25}$Astronomy and Space Physics Department\ Taras Shevchenko National University of Kyiv, 60 Volodymyrska str., Kyiv, 01601, Ukraine\\
$^{26}$National University of Uzbekistan, 4 University str., Tashkent 100174, Uzbekistan\\
$^{27}$Laboratoire d’Annecy de Physique des Particules (LAPP), Univ. Grenoble
Alpes, Université Savoie Mont Blanc, CNRS/IN2P3, F-74941 Annecy, France\\
$^{28}$Yunnan Astronomical Observatories/Chinese Academy of Science, Kunming, 650011, China \\
$^{29}$National Astronomical Observatories/Chinese Academy of Science 20A Datun Road, Beijing, 100012, China\\


\label{lastpage}

\end{document}